\documentclass[nolinenumber]{aastex631}

\usepackage{amsmath}

\begin{document}

\title{Nucleosynthetic Analysis of Three-Dimensional Core-Collapse Supernova Simulations}

\author[0000-0002-0042-9873]{Tianshu Wang}
\affiliation{Department of Astrophysical Sciences, Princeton University}

\correspondingauthor{Tianshu Wang}
\email{tianshuw@princeton.edu}

\author[0000-0002-3099-5024]{Adam Burrows}
\affiliation{Department of Astrophysical Sciences, Princeton University}




\begin{abstract}
We study in detail the ejecta conditions and theoretical nucleosynthetic results for 18 three-dimensional core-collapse supernova (CCSN) simulations done by F{\sc ornax}. {Most simulations are carried out to at least 3 seconds after bounce, which allows us to follow their longer-term behaviors.} We find that multi-dimensional effects introduce many complexities into ejecta conditions. We see stochastic electron fraction evolution, complex peak temperature distributions and histories, and long-tail distributions of the time spent within nucleosynthetic temperature ranges. These all lead to substantial variation in CCSN nucleosynthetic yields and differences with 1D results.  We discuss the production of lighter $\alpha$-nuclei, radioactive isotopes, heavier elements, and a few isotopes of special interest. Comparing pre-CCSN and CCSN contributions, we find that a significant fraction of elements between roughly Si and Ge are generically produced in CCSNe. We find that $^{44}$Ti exhibits an extended production timescale compared to $^{56}$Ni, which may explain its different distribution and higher than previously predicted abundances in supernova remnants such as Cas A and SN1987A. We also discuss the morphology of the ejected elements. This study highlights the high-level diversity of ejecta conditions and nucleosynthetic results in 3D CCSN simulations and emphasizes the need for additional long-term {($\sim$10 seconds)} 3D simulations to properly address such complexities.
\end{abstract}



\section{Introduction}
Core-collapse supernovae (CCSNe) are the final acts of massive stars ($M_{\text{ZAMS}}\gtrsim 8M_\odot$), and these strong explosions each inject into the interstellar medium fresh nucleosynthesis of a few solar masses, including elements newly synthesized during the explosion and those produced during stellar evolution. Thus, CCSNe play an important role in the chemical evolution of the universe, and calculating the ejecta elemental composition remains a fundamental issue in CCSN theory \citep{woosley2002,janka2012,nomoto2013,burrows2021,arcones2023}.

CCSN ejecta are a mixture of multiple components. After core bounce, a shock forms and moves outward. If the matter behind the shock gains enough energy from neutrinos to overcome the shock ram pressure \citep{wang2022}, the shock expands unstoppably and leads to a successful CCSN explosion. Matter in the outer envelope is pushed out and heated by this now exploding shock, and depending upon the peak temperature it achieves, the matter may experience explosive burning (as opposed to nuclear burning in hydrostatic equilibrium) \citep{woosley2002,nomoto2013,arcones2023}. These shells also include the nucleosynthetic contribution of the previous stellar evolution stages. Behind the shock there are high-entropy, low-density bubbles that can only be seen in multi-dimensional simulations, which is neutrino-heated matter that moves faster than its surroundings (for 3D examples, see \citet{lentz2015,roberts2016,muller2017,oconnor2018,vartanyan2018,ott2018,summa2018,burrows2019,burrows2020,kuroda2020,stockinger2020,bollig2021}). At later times, transonic outflows starting from the surface of the proto-neutron star (PNS), known as the neutrino-driven winds, emerge \citep{duncan1986,burrows1987,burrows1995,qian1996,wang2023}, and a wind termination shock is formed when it catches up with the more slowly moving primary ejecta. Several different nucleosynthetic processes occur in different components of the ejecta, in particular explosive nucleosynthesis and $\alpha$-rich freeze-out. Since the electron fractions ($Y_e$) in the ejecta deviate from the initial values due to charged-current interactions with neutrinos, some heavy elements ($Z>26$) can be produced via processes such as the (weak) r-process and $\gamma$-process \citep{woosley1978,janka2012,burrows2021}. Aided by the neutrino luminosity emitted from the proto-neutron star (PNS), nucleosynthesis processes involving neutrino interactions such as the $\nu$- and $\nu p$- processes are also thought to exist \citep{woosley1990,frohlich2006,Pruet2006}.  

There has been much effort to determine the overall elemental composition of the CCSNe ejecta, as well as to map out and quantify the effects of each individual nucleosynthetic process. 
Thanks to their relative simplicity, hundreds to thousands of spherically symmetric (1D) models have been analyzed in order to study the systematic dependence of overall nucleosynthesis on progenitor properties such as ZAMS mass, metallicity, and rotation (e.g., \citet{thielemann1996,rauscher2002,woosley2002,heger2010,nomoto2013,sukhbold2016,limongi2018,curtis2019}). Since first-principles 1D radiation hydrodynamic simulations don't explode except for very low mass progenitors, such models either add extra parameters to mimic multi-dimensional effects or apply piston or thermal-bomb methods to force the models to explode with a given prescribed certain energy. Although these models provide useful insight into the elemental composition of CCSN ejecta, they predict inaccurate inner ejecta conditions and the corresponding results compare unfavorably to multi-dimensional studies (\citet{lentz2015,wongwathanarat2015,harris2017,wanajo2018,sieverding2020,sieverding2023,wang2023}). In addition to the different explosion properties  between 1D and 3D simulations (e.g., ejecta masses and explosion energies), many features are missed by 1D models that have significant influence on the ejecta abundances. Long-lasting post-explosion accretion that occurs only in multi-dimensional simulations maintains the neutrino luminosity at a relatively high level even after the explosion commences \citep{muller2017,burrows2020,bollig2021}, and this influences both the dynamical and thermal histories of the ejecta. Interactions between accreting and out-going matter may slow down the outflow and extend the time it spends in interesting temperature ranges, leading to significantly different nucleosynthetic results \citep{sieverding2023}. The outflow velocity is also loosely related to its electron fraction, which is crucial for nucleosynthesis. In addition, one-dimensional models enter the spherical neutrino-driven wind phase much earlier, and the wind properties are generally different from those found in 3D simulations \citep{wang2023}. 

To capture all these processes and effects accurately and obtain a quantitatively correct understanding of the nucleosynthesis in CCSNe, three dimensional simulations carried out to late times are needed. Some effort has been made in this direction ( \citet{wongwathanarat2015,harris2017,muller2017,vartanyan2019,stockinger2020,sieverding2020,sieverding2023}), but the number of 3D simulations with nucleosynthetic results is still far from sufficient\footnote{There are many 3D simulations without any nucleosynthetic analysis.}. In addition, most long-term 3D models don't continue the radiation transport calculation after about one second after bounce. Instead, they cut out the central object and apply a parametric spherical symmetric neutrino-driven wind boundary condition \citep{stockinger2020}, or employ neutrino transport calculation results from spherical symmetric simulations \citep{bollig2021}. These methods reduce the computational cost significantly and have little effect on the primary ejecta launched earlier. However, these approaches significantly change the dynamical and thermal histories of the matter ejected at later times, especially the neutrino-driven winds. Such effects are generally larger in more massive progenitors, because they maintain stronger asymmetrical accretion at late times, whose consequences can't be reproduced properly by spherical symmetric methods. Although the total mass of later, secondary ejecta and neutrino-driven winds is small, it is proven to have a strong influence on isotopes such as $^{44}$Ti \citep{sieverding2023} and heavy elements with $Z>26$ \citep{wang2023}. Therefore, three-dimensional models with radiation transport turned on throughout the entirety of a simulation are preferred in order to obtain the overall ejecta abundance pattern.

In this paper, we present a detailed the nucleosynthetic analysis of 18 long-term 3D CCSN simulations spanning a large progenitor ZAMS mass range from 9 to 60 $M_\odot$. Most of these simulations are carried out to more than 3 seconds after bounce. The 3D radiation transport calculation is never turned off, and this allows us to obtain more realistic dynamical and thermal histories for the ejected matter to later times. With this unprecedentedly large 3D long-term CCSN simulation set, this paper aims to improve our understanding of the nucleosynthetic conditions in the context of CCSN explosions. 

The paper is arranged as follows. In Section \ref{sec:method}, we describe details of the progenitor models, our simulations, and the methods employed. We discuss the ejecta conditions and the complexities of the 3D simulations in Section \ref{sec:conditions}. The nucleosynthetic results are provided in Section \ref{sec:results}. Here, we also provide the final elemental abundances and the fractional yields from the explosive wind phases. We also discuss the production of some isotopes that are frequently the subject of observations and study the roles of a few interesting nucleosynthetic processes. In addition, we discuss the spatial distribution of ejected elements. Finally, in Section \ref{sec:conclusion} we summarize the paper and our conclusions, and provide possible caveats. 

\section{Method}\label{sec:method}
\subsection{Summary of Simulations}
All simulations used in this study were generated by the multi-group multi-dimensional radiation/hydrodynamics code F{\sc{ornax}} \citep{skinner2019,vartanyan2019,burrows2019,burrows2020}. The simulations employ the SFHo equation of state (EOS) of \citet{steiner2013}, which is broadly consistent with extant theoretical and experimental constraints \citep{tews2017,Reed2021,Lattimer2023}. There are 12 logarithmically-distributed energy groups for each of our three neutrino species (electron-type, anti-electron type, and the rest are bundled as ``$\mu$-type'').

The ZAMS masses of the simulations are 9, 9.25, 9.5, 11, 15.01, 16, 17, 18, 18.5, 19, 19.56, 20, 23, 24, 25, 40, and 60 $M_\odot$. {The total masses left at collapse are 8.74, 8.98, 9.21, 10.69, 12.57, 13.09, 13.70, 14.27, 14.67, 14.37, 13.88, 15.03, 15.08, 14.69, 15.90, 15.34, 7.29 $M_\odot$, respectively.}. The 9, 9.25, 9.5, 11, 40, and 60 $M_\odot$ progenitors come from \citet{sukhbold2016} (S16), while all the others come from \citet{sukhbold2018} (S18). All progenitors have solar metallicity initially. Although the S16 progenitors have more detailed pre-CCSN nucleosynthesis results, we have included the S18 progenitors because (a) they are a much finer resolution set, and more importantly (b) the authors have fixed a major coding error in the stellar evolution code Kepler which influenced all models with ZAMS mass above $\sim$14 $M_\odot$ \citep{sukhbold2018}.

\begin{figure}
    \centering
    \includegraphics[width=0.48\textwidth]{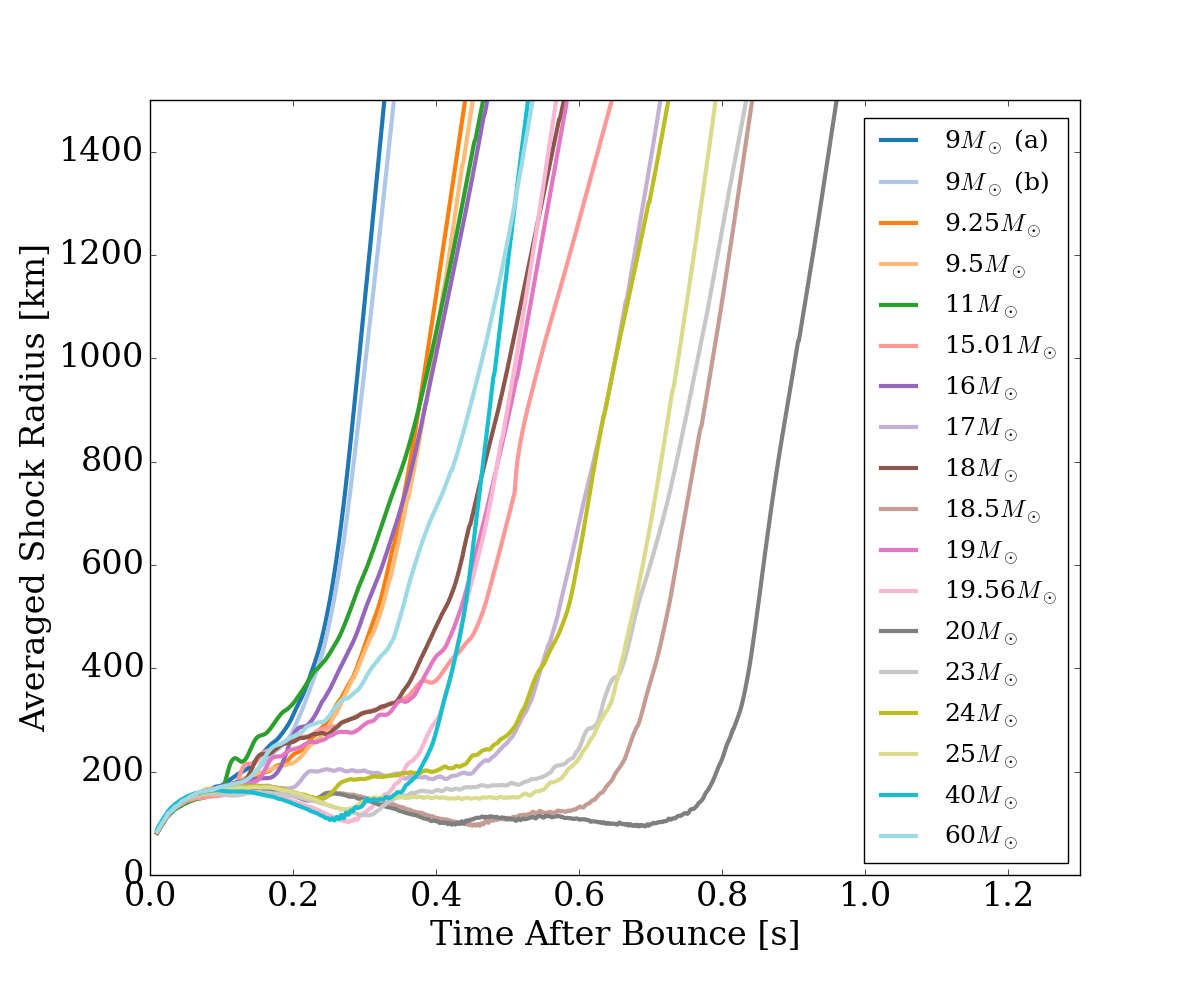}
    \includegraphics[width=0.48\textwidth]{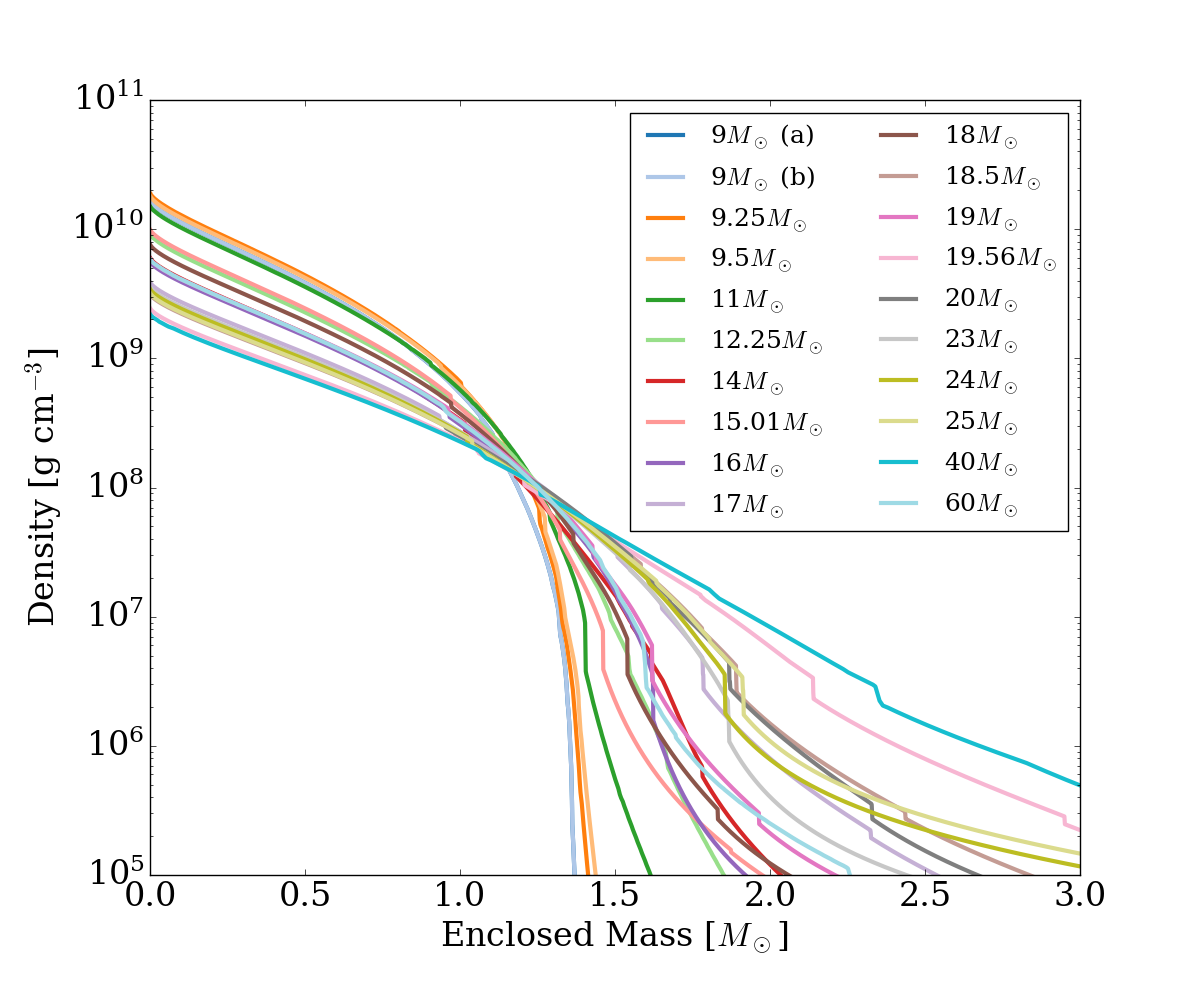}
    \caption{Average shock radii as a function of time and the initial progenitor density profiles for all models. {\bf Left:} The temporal evolution of the shock radii. The explosion time varies from about 160 ms to about 800 ms after bounce. {\bf Right:} Progenitor density profiles. The density jump at Si/O interfaces at about $10^6$ to $10^7$ g cm$^{-3}$ can be clearly seen. The accretion of the Si/O interfaces will often lead to shock revival after the shock stalls. Although the density jump at the interface can inaugurate explosion \citep{wang2022}, the accretion of the interface may not directly lead to unstoppable shock expansion, since some models (usually the more massive ones) can start to explode even hundreds of milliseconds its accretion. This means that a significant amount of additional mass exterior to the Si/O interface will accrete onto the PNS. This is demonstrated more clearly in Figure \ref{fig:r-T}.}
    \label{fig:rshock}
\end{figure}

Some of the CCSN simulations used in this work have been published in \citet{coleman2022,vartanyan2023,wang2023,burrows2023}. Each simulation is done with a spherical grid of 1024(r)$\times$128($\theta$)$\times$256($\phi$), and the outer boundary radius varies from 30,000 to 100,000 kilometers (km). Each model is first simulated spherically until 10 milliseconds after bounce, after which it is continued in full 3D.
Figure \ref{fig:rshock} shows the temporal evolution of the shock mean radii and the mass density profiles for all the progenitor models, while Table \ref{tab:simulation_summary} summarizes some basic properties of the models. The explosion time (defined as the time after which the shock never shrinks again) varies from about 160 milliseconds (ms) to about 800 ms after bounce.

\begin{table}
    \centering
    \begin{tabular}{c|cccccc}
    ZAMS Mass [$M_\odot$]  &Progenitor Set  &Duration [s] &Explosion Time$^\dagger$ [s] &$M_{0.6\text{MeV}}$ [$M_\odot$] 
    &$M_{0.1\text{MeV}}$ [$M_\odot$] &{Ejecta Mass$^\dagger$}\\
    \hline
    9(a)  &S16  &1.775    &0.162   &0.0128  &0.0478 &7.40\\ 
    9(b)  &S16  &1.950    &0.162   &0.0115  &0.0439 &7.40\\ 
    9.25  &S16  &3.532    &0.239   &0.0198  &0.0678 &7.60\\ 
    9.5   &S16  &2.375    &0.240   &0.0262  &0.0861 &7.81\\ 
    11    &S16  &4.492    &0.189   &0.0618  &0.198  &9.18\\ 
    40    &S16  &1.760    &0.364   &0.215   &2.79   &12.92\\ 
    60    &S16  &4.545    &0.274   &0.131   &1.12   &5.51\\ 
    \hline
    15.01 &S18  &4.383    &0.382   &0.0703  &0.611  &10.94\\
    16    &S18  &4.184    &0.217   &0.0823  &0.625  &11.42\\
    17    &S18  &2.037    &0.481   &0.0831  &0.693  &11.67\\ 
    18    &S18  &4.328    &0.316   &0.115   &0.906  &12.59\\ 
    18.5  &S18  &3.954    &0.647   &0.155   &1.42   &12.53\\
    19    &S18  &4.075    &0.346   &0.112   &1.01   &12.58\\
    19.56 &S18  &3.890    &0.368   &0.320   &2.08   &11.61\\
    20    &S18  &3.323    &0.779   &0.108   &1.24   &12.89\\ 
    23    &S18  &6.228    &0.598   &0.122   &1.36   &13.20\\ 
    24    &S18  &3.919    &0.489   &0.159   &2.20   &12.67\\
    25    &S18  &3.119    &0.599   &0.142   &1.77   &13.81\\ 
    \end{tabular}
    \caption{This table summarizes some basic properties of our models. S16 means that the progenitor is taken from \citet{sukhbold2016}, while S18 means that it's taken from \citet{sukhbold2018}. Initial progenitor models from both sets contain the elemental abundances calculated by a small 19-isotope network, while the S16 progenitors have extra co-processed nucleosynthesis results calculated by an adaptive $\sim$2000 isotope network \citep{sukhbold2016}. The mass of the ejecta that have at any time reached 0.6 MeV (the NSE criterion) is indicated by $M_{0.6\text{MeV}}$. The mass of the freeze-out component is only a few percent of the total ejecta mass, but this component plays a very important role in ejecta abundance calculations. The mass of the ejecta parcels that have reached 0.1 MeV is also shown. 
    Differences between the 9(a) and 9(b) models are due mostly to the presence of imposed perturbations in the initial model in model 9(a). This led to a slightly more vigorous explosion launch. The 19.56 and 40 $M_\odot$ models are special because they form black holes a few seconds after the successful explosion \citep{burrows2023}, and they have higher explosion energies than other models. {$^\dagger$We assume that everything exterior to the shock at the end of each simulation will be ejected.}
    }
    \label{tab:simulation_summary}
\end{table}

Although all other models don't have initial perturbations\footnote{{There are relative numerical errors at about the $10^{-14}-10^{-15}$ level due to machine truncation errors, which serve as seeds for turbulence and convection. Numerical tests have shown that numerical perturbations at this level or even $\sim$$10^4$ times larger are unable to influence the explosion properties.}}, an initial perturbation is added to the 9(a) model at 10 milliseconds after bounce. The initial perturbation is chosen to be an $n=4$, $l=10$, $v_{max}=100$ km/s spherical harmonic velocity field between 200 and 1000 km, following the prescription of \citet{muller2015}. Compared to the non-perturbed 9(b) model, 9(a) explodes slightly earlier and a bit more vigorously. This leads to significant lower electron fractions in 9(a) in the early phase of the explosion. {Since explosion properties are only weakly influenced ($<10\%$ deviation in shock velocity, explosion energy, and ejecta mass), this may indicate that nucleosynthesis might be more sensitive to initial perturbations than explosion properties.} To study the effects of initial perturbations, we also carried out a 18 $M_\odot$ simulation starting from a 3D progenitor model provided by \citet{muller2016}. This 3D progenitor model is the same as its 1D counterpart in the S18 set until 293.5s before the onset of collapse, after which oxygen shell burning is simulated in 3D. We compare the ejecta conditions in this 3D progenitor 18 $M_\odot$ simulation with those in the 1D case, and we see almost no difference. This means that the strong effects seen in the 9(a) model may not be as strong in other models. This issue is discussed in more detail in Section \ref{sec:conditions}. 
Two other special cases are the 19.56 and 40 M$_\odot$ models. In addition to their higher explosion energies, these two models experience the general relativistic instability and form black holes \citep{burrows2023} at $\sim$1.76 and $\sim$3.9 seconds after bounce, respectively. Though forming black holes, these models too eject fresh nucleosynthesis. 

\subsection{Tracer Method}
Tracers are passive mass elements advected according to their fluid vector velocities. We have tested three tracer methods similar to those compared in \citet{sieverding2022}: (a) the inline method, in which tracers are updated during the F{\sc{ornax}} calculation at each hydrodynamic timestep ($\sim$10$^{-6}$s), (b) a forward method, in which tracers are post-processed and integrated forward in time starting at their initial positions, and (c) a backward method, in which tracers are post-processed and integrated backward in time starting at their final positions. The inline method is the most accurate, but it's also the most time-consuming. Updating tracer positions is a fast operation itself, but the work-load balance at late times is very expensive because a large number of injected tracers fall onto the PNS. To keep track of the neutrino-driven winds, none of these tracers can be discarded because at late times the whole PNS becomes convective and the possibility for some of these tracers to be ejected in winds is nonzero. The inline method takes negligible time at early phases, but it slows down the calculation by more than 70\% at about 2 seconds after bounce and worsens thereafter. Therefore, the inline method is only applied to a few models in this study to check the accuracy of post-processed methods.

The fluid velocity fields of the simulations are saved every millisecond. We divide the 1-ms time interval into $N_{sub}$ equal length substeps in the post-processed methods (b) and (c) to avoid allowing the tracer particles to bypass multiple grid cells in a timestep, with the velocity field linearly interpolated in time and space to the positions of the tracer particles. We use adaptive $N_{sub}$ to ensure each tracer moves no more than half a grid cell size along each direction per substep. This adaptive method saves computation time because tracers spend most of their time at large radii, where $N_{sub}$ is much smaller than for those close to the PNS.

We find that the backward method provides more accurate ejecta conditions for nucleosynthesis calculations, which comports with the findings of \citet{sieverding2022}. Due to the lower time resolution of the post-processing methods, they are unable to follow the correct tracer trajectory once the tracer enters the chaotic convective region close to the PNS. For tracers that are pushed out by the shock, both forward and backward methods give accurate trajectories. For those that reach a radius below a few hundred kilometers, the infalling phase of a tracer is better tracked by the forward method, while the outmoving phase is better tracked by the backward method. Since any tracer that has entered the convective region will reach nuclear statistic equilibrium (NSE) and forget its previous composition, it is the ejection phase that is most important for the nucleosynthesis calculations and, thus, the backward method is preferred. Furthermore, we find that the forward method is unable to follow the behavior of the neutrino-driven winds. The winds are fast-moving and low-density with a low total mass, so the number of forward tracers representing this component is small unless the initial tracer positions are chosen extremely carefully. The backward method doesn't have this problem, as the neutrino-driven wind regions are easily identified at the end of each simulation. 

We add about 320,000 post-processed tracers to each simulation and apply the backward method to them. {All nucleosynthesis results shown in this paper are based on such backward tracers.} The tracers are placed logarithmically along the r-direction above 500 km and uniformly along the $\theta$- and $\phi$-directions at the end of each simulation. More massive models generally have heavier tracers because there is more mass above the PNS on the simulation grid. The tracers are backwardly evolved to 10 milliseconds after bounce and are then mapped onto the progenitor model using their mass coordinates. This is because the earlier collapse phase is evolved in spherical symmetry and the Lagrangian mass coordinate does not change during this phase.

\subsection{Pre-CCSN Nucleosynthesis}
The nucleosynthesis results of the pre-CCSN stellar evolution phases are taken from the progenitor models. These are our initial abundances. Both \citet{sukhbold2016} (S16) and \citet{sukhbold2018} (S18) progenitors contain a standard 19-isotope network which includes only $^{1}$H, $\alpha-$nuclei, and a few iron-group isotopes, while S16 has another co-processed adaptive network which includes $\sim$2000 isotopes \citep{sukhbold2016}. Using the S16 models, we have compared the differences in the final nucleosynthetic yields due to differences between the small and large networks. We find that the final abundances of $\alpha-$nuclei lighter than $^{44}$Ti and isotopes from mostly NSE freeze-out (many isotopes between $A=44$ and $A=60$) are almost independent of the choice of pre-CCSN network. The net yield of $^{44}$Ti is also insensitive to the choice of network, but the small network has a non-negligible ``pseudo-$^{44}$Ti" contribution in the envelope exterior to the shock, which artificially enhances the apparent final $^{44}$Ti mass abundance by a factor of a few. This envelope contribution of $^{44}$Ti is not seen in the large network. 

The large network is used as the starting point of the CCSN nucleosynthesis calculation for the S16 progenitors. For these models, we are able to study the CCSN contribution per se to the final ejecta for all isotopes in which we are interested. The S18 progenitors have only the 19-isotope network, which limits our study to a small subset of isotopes ($\alpha-$nuclei, $^{44}$Ti, $^{56}$Ni and $^{57}$Ni). Carrying out detailed pre-CCSN nucleosynthesis calculations for the S18 progenitors is beyond the scope of this work, and we leave it to future work by others. 

\subsection{Nucleosynthesis Calculation}
All nucleosynthesis calculations are done with SkyNet \citep{lippuner2017} with a 1530-isotope network including elements up to $A=100$. The reaction rates are taken from the JINA Reaclib \citep{cyburt2010} database, and we include neutrino interactions with protons and neutrons. Reactions for the $\nu-$process \citep{woosley1990} are not included. For the purpose of interfacing with SkyNet, the detailed neutrino spectra extracted from F{\sc ornax} are fit to Fermi-Dirac functions in a way that preserves the particle number ($\int E^2f(E)dE$), the average energy ($\int E^3f(E)dE$), and the average reaction rates ($\int E^4f(E)dE$). These fit parameters are fed into SkyNet to address the effects of neutrino-matter interactions on the nucleosynthetic yields. The NSE criterion is set at 0.6 MeV ($\sim$7 GK), and SkyNet switches to its NSE evolution mode if the temperature is above this mark and the strong interaction timescale is shorter than the timescale of density changes \citep{lippuner2017}. Table \ref{tab:simulation_summary} shows the mass of the ejecta that have at any time reached this NSE criterion. The mass of the freeze-out component is only a few percent of the total ejecta mass, but this component plays a very important role in ejecta abundance calculations. The mass of the ejecta parcels that have reached 0.1 MeV is also shown in the Table \ref{tab:simulation_summary}. The electron fractions ($Y_e$) are calculated by F{\sc{ornax}} when the temperature is above the NSE threshold, which allows the neutrino spectra to be appropriately non-thermal. The nucleosynthesis calculation starts from the point after which the tracers never reach NSE again. 
In this work, we use the \citet{lodders2021} solar abundances to calculate production factors, which are ratios between the mass abundances of the ejecta and the Sun. 

{We emphasize that elemental contributions of pre-SN winds are not included in this work. We refer readers to \citet{sukhbold2016} for a better understanding of such winds.}

\section{Ejecta Conditions}\label{sec:conditions}

\begin{figure}
    \centering
    \includegraphics[width=0.48\textwidth]{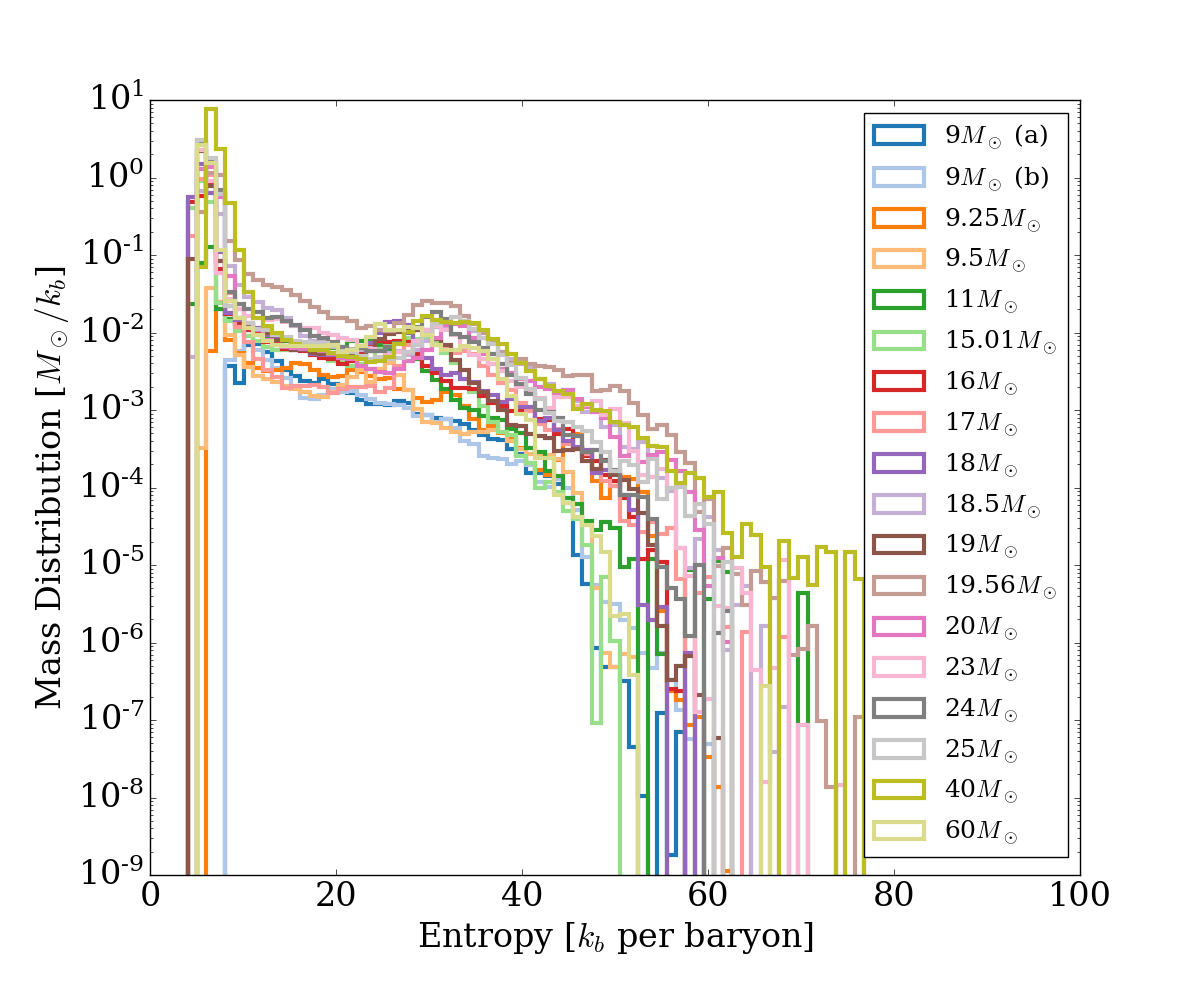}
    \includegraphics[width=0.48\textwidth]{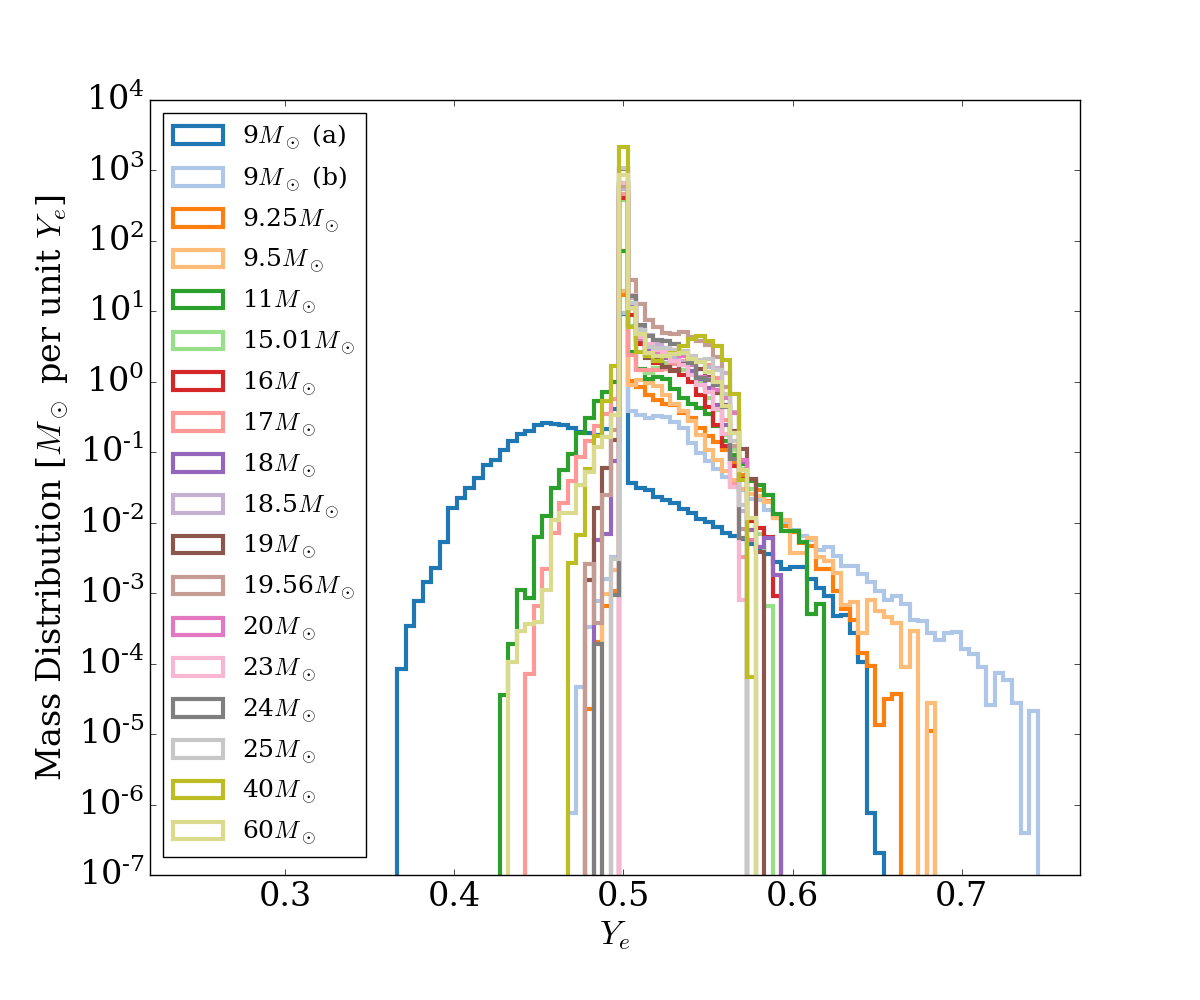}
    \caption{Ejecta electron fraction and entropy distributions. Both variables are measured either at NSE freeze-out (for tracers that have reached NSE) or at the peak temperature (for those that never reach NSE). {\bf Left:} Distributions of ejecta entropy. Although matter ejected later has higher entropy, none of our models reach $S>80$ $k_b$ baryon$^{-1}$. {\bf Right:} Distributions of ejecta electron fractions. Most models have modest electron fractions roughly between $\sim$0.45 and $\sim$0.60. The 9(a) model is exceptional because the initial perturbations make it explode a bit more vigorously and earlier than the 9(b) model, which leads to a faster ejection. As a result, the 9(a) model has more neutron-rich ejecta than other models. The effects of initial perturbations may not be as strong for more massive models with stronger accretion and later explosions.  }
    \label{fig:tracer-ye-S}
\end{figure}

\begin{figure}
    \centering
    \includegraphics[width=0.48\textwidth]{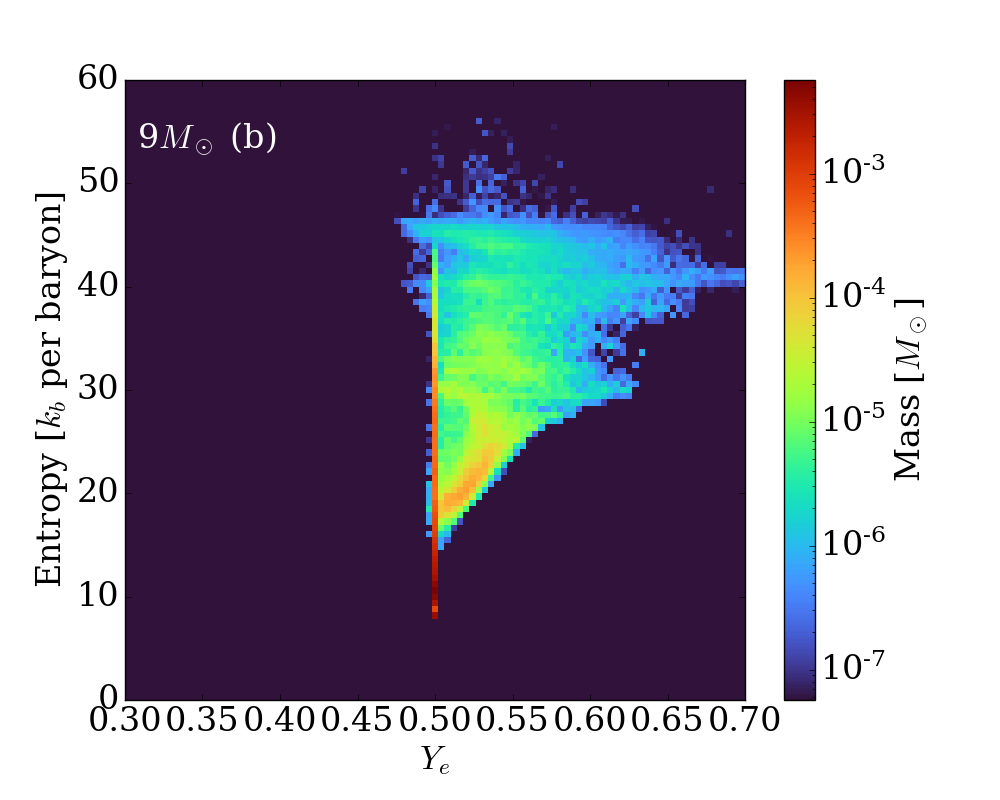}
    \includegraphics[width=0.48\textwidth]{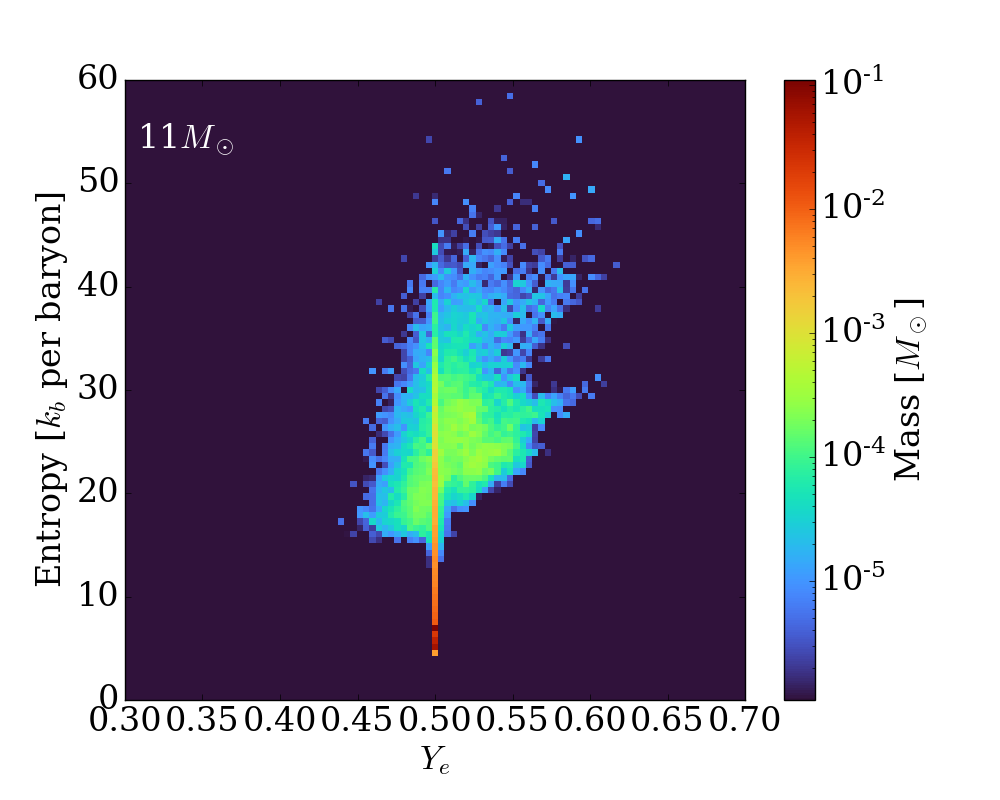}
    \includegraphics[width=0.48\textwidth]{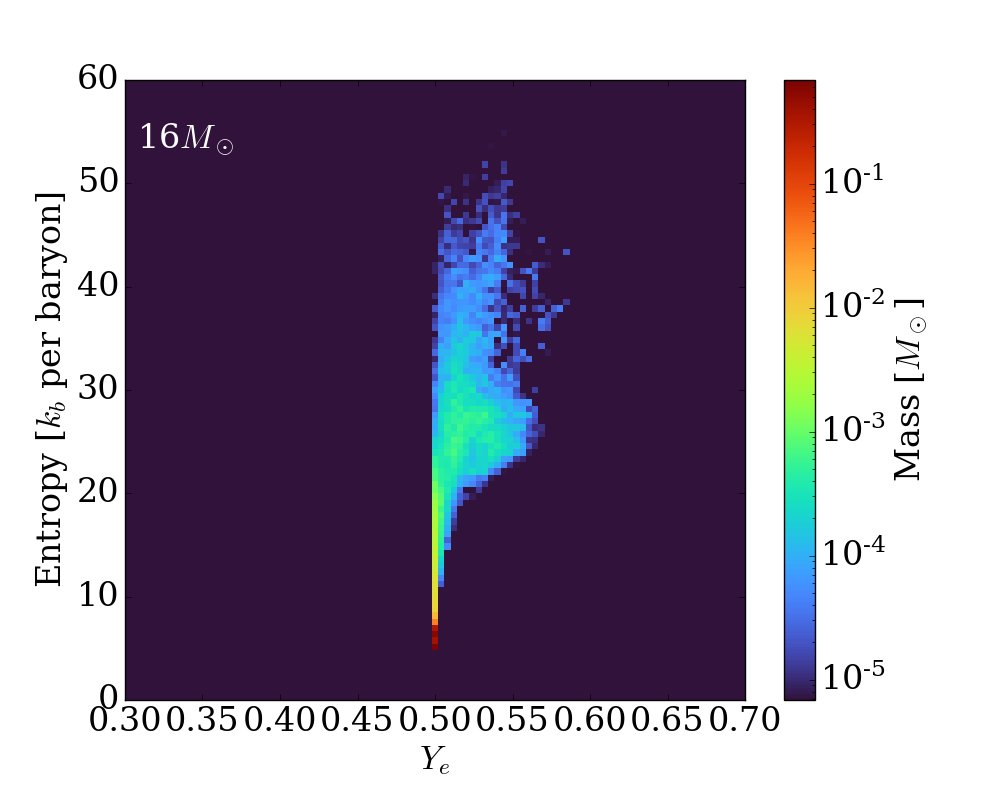}
    \includegraphics[width=0.48\textwidth]{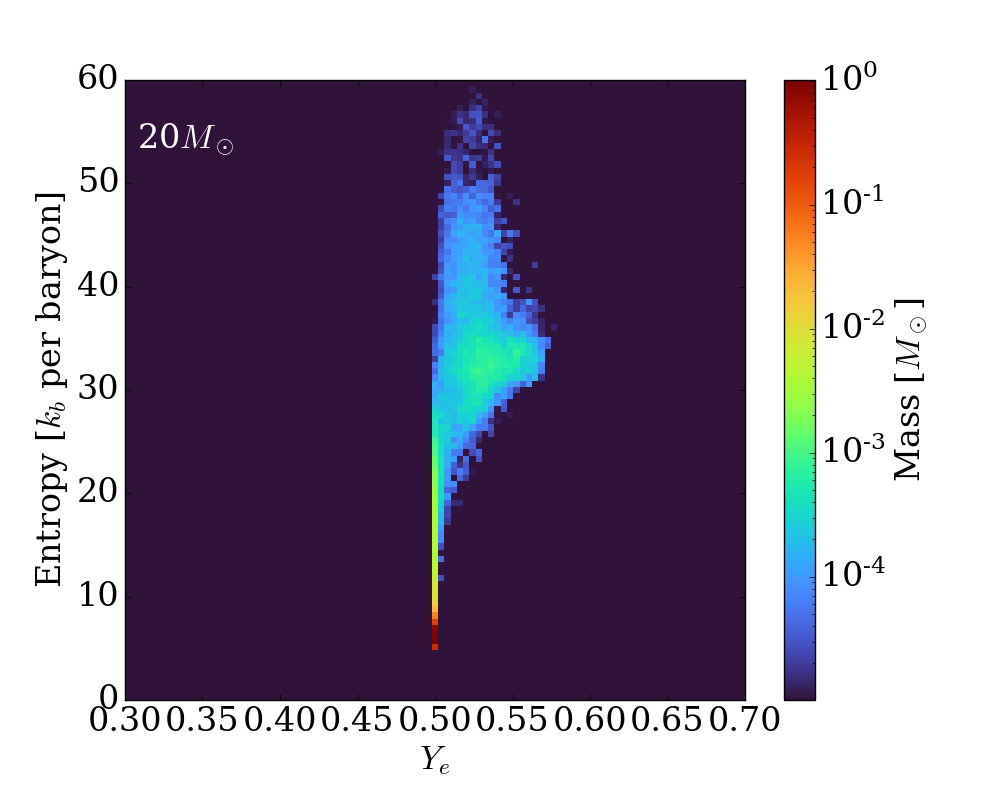}
    \includegraphics[width=0.48\textwidth]{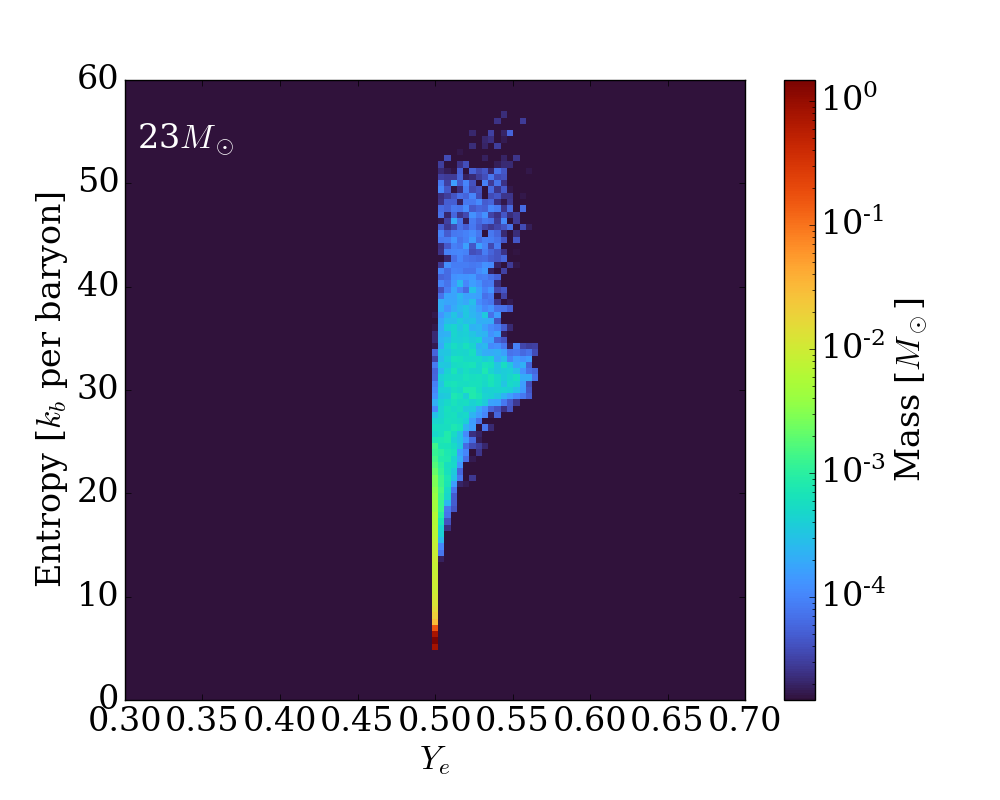}
    \includegraphics[width=0.48\textwidth]{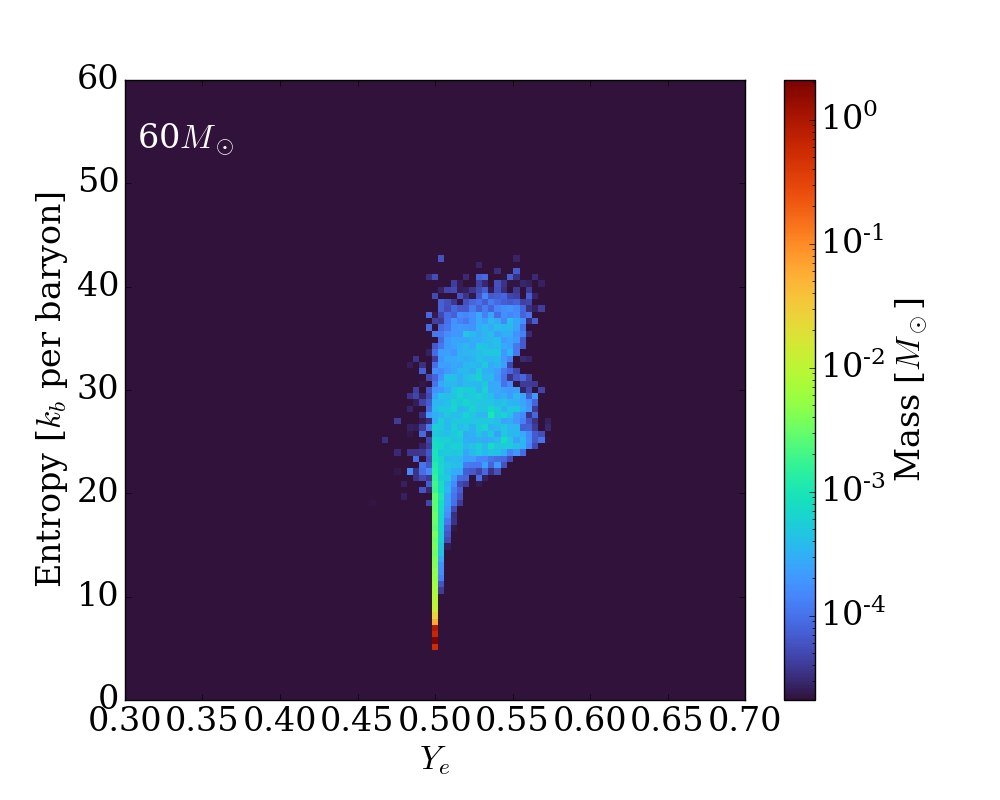}
    \caption{Joint distributions of the electron fraction and entropy of models 9 $M_\odot$ (b), 11 $M_\odot$, 16 $M_\odot$, 20 $M_\odot$, 23 $M_\odot$, and 60 $M_\odot$. A flag-like shape can be seen in all models. The ejecta are mostly proton-rich, but there can be some neutron-rich components. Figure \ref{fig:ye-t} shows the actual temporal evolution of the electron fraction and the occurrence of neutron-rich phases seems stochastic.}
    \label{fig:ye-S}
\end{figure}

\begin{figure}
    \centering
    \includegraphics[width=0.48\textwidth]{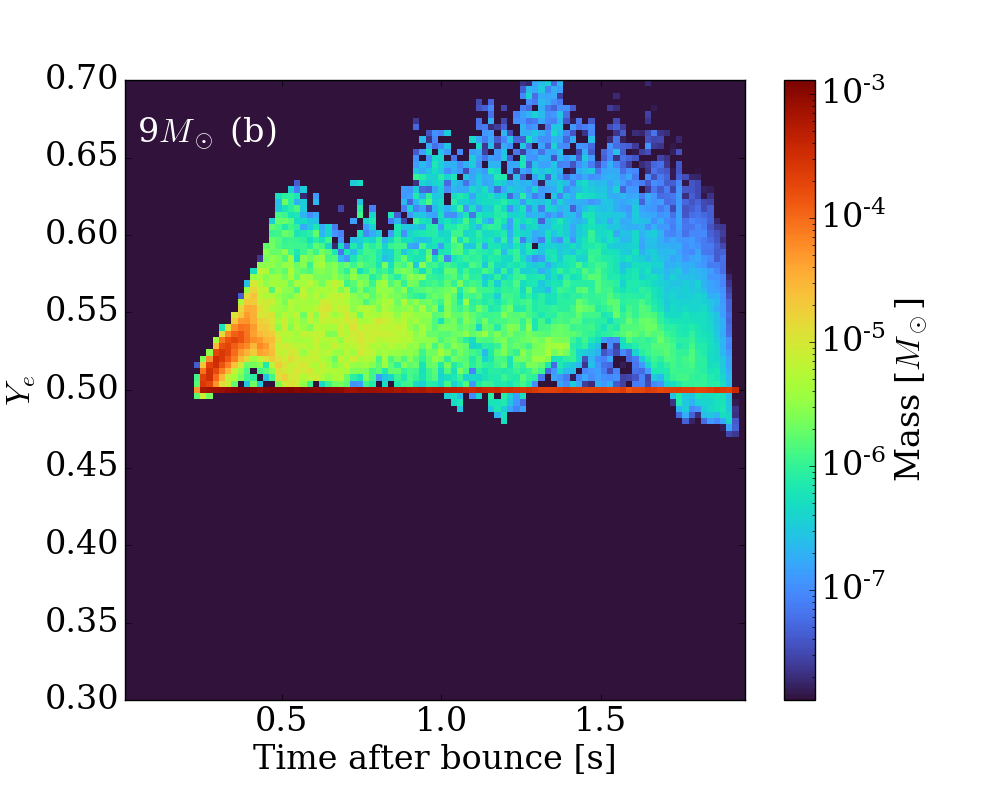}
    \includegraphics[width=0.48\textwidth]{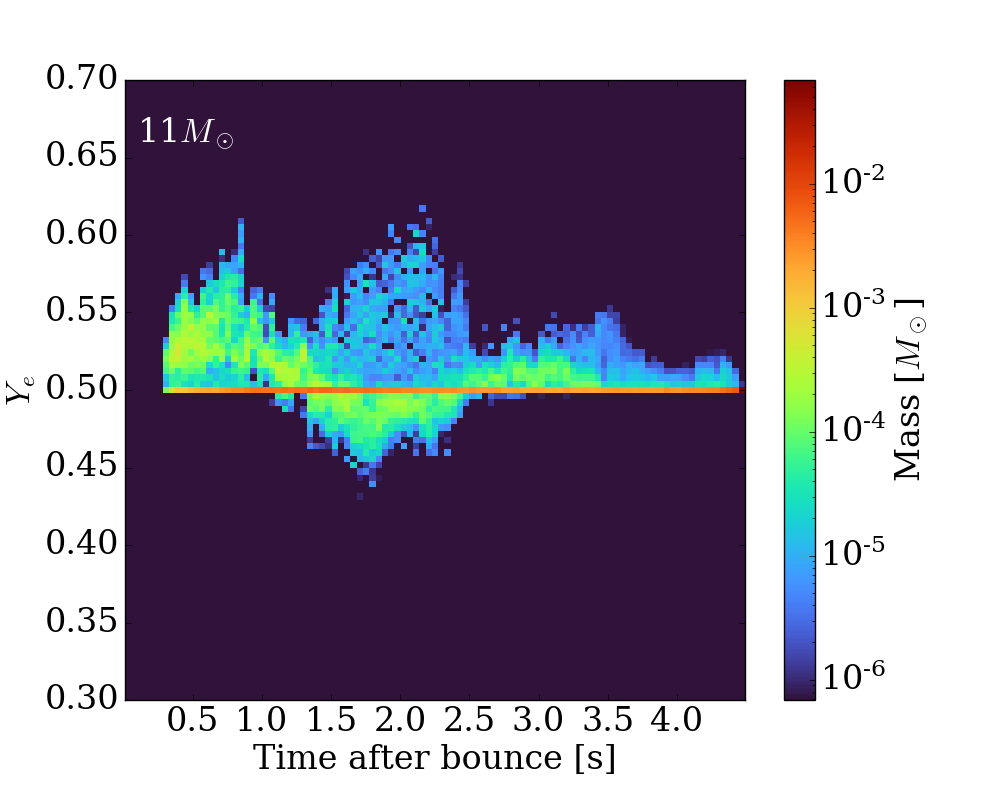}
    \includegraphics[width=0.48\textwidth]{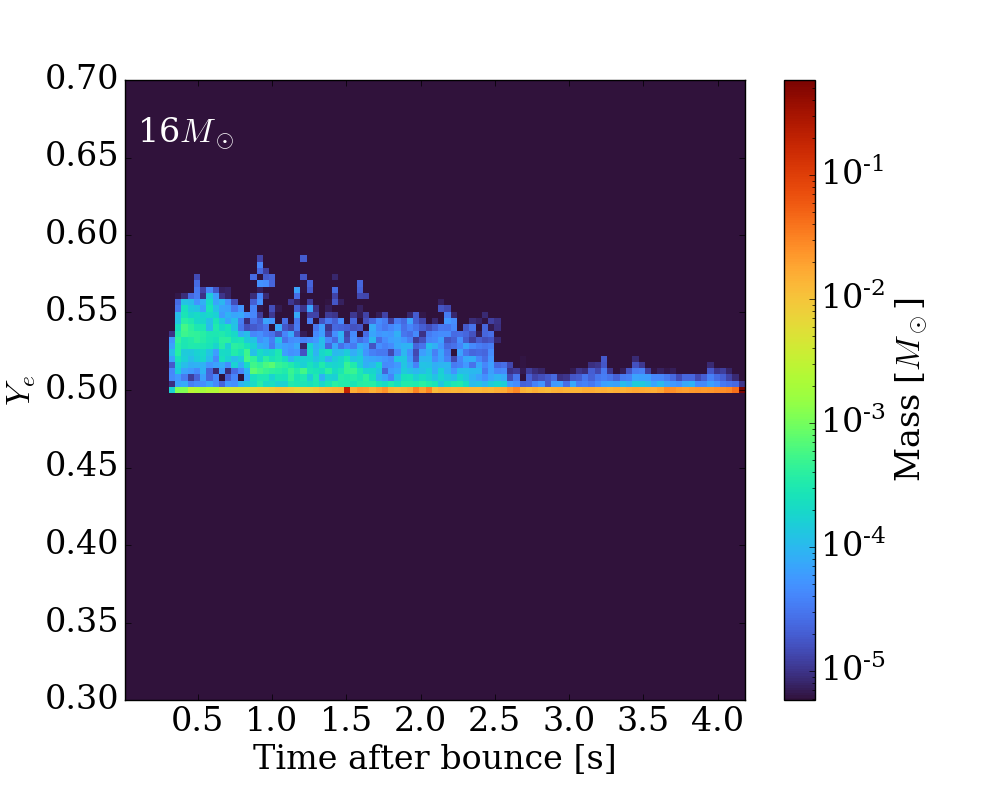}
    \includegraphics[width=0.48\textwidth]{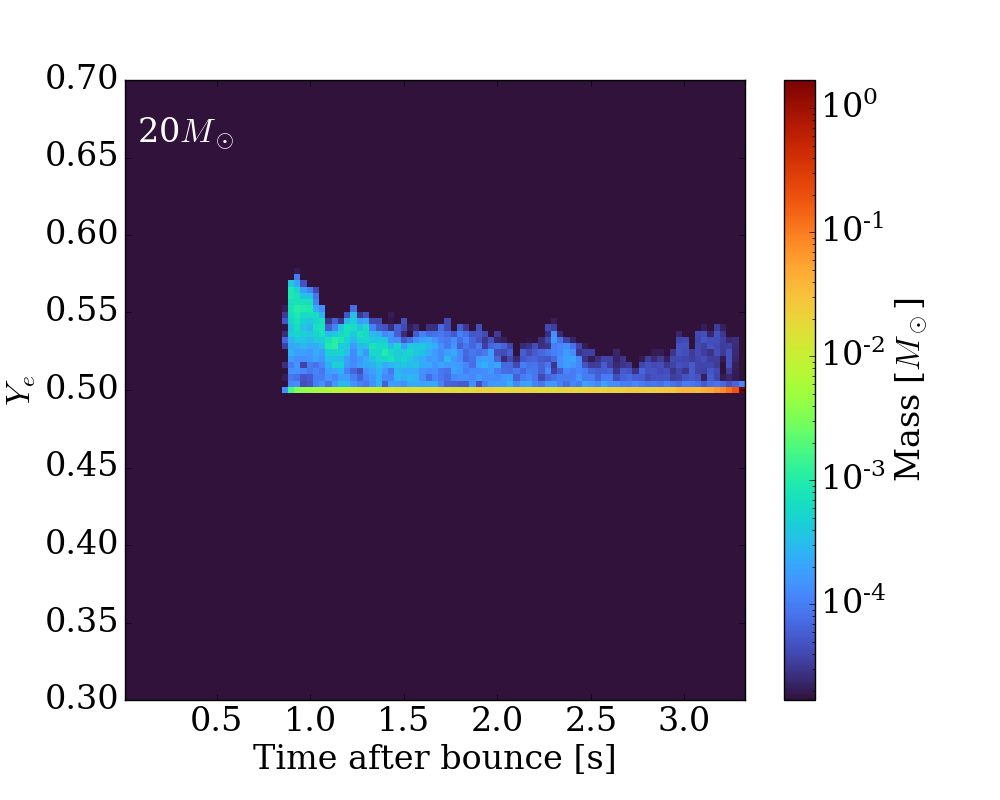}
    \includegraphics[width=0.48\textwidth]{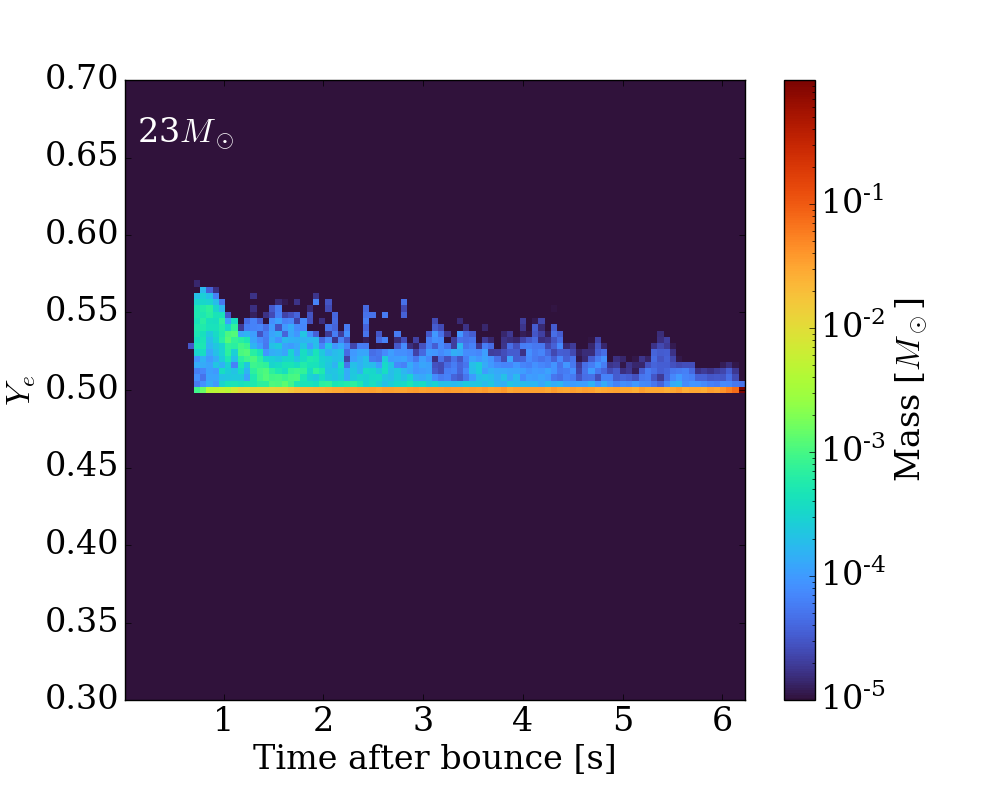}
    \includegraphics[width=0.48\textwidth]{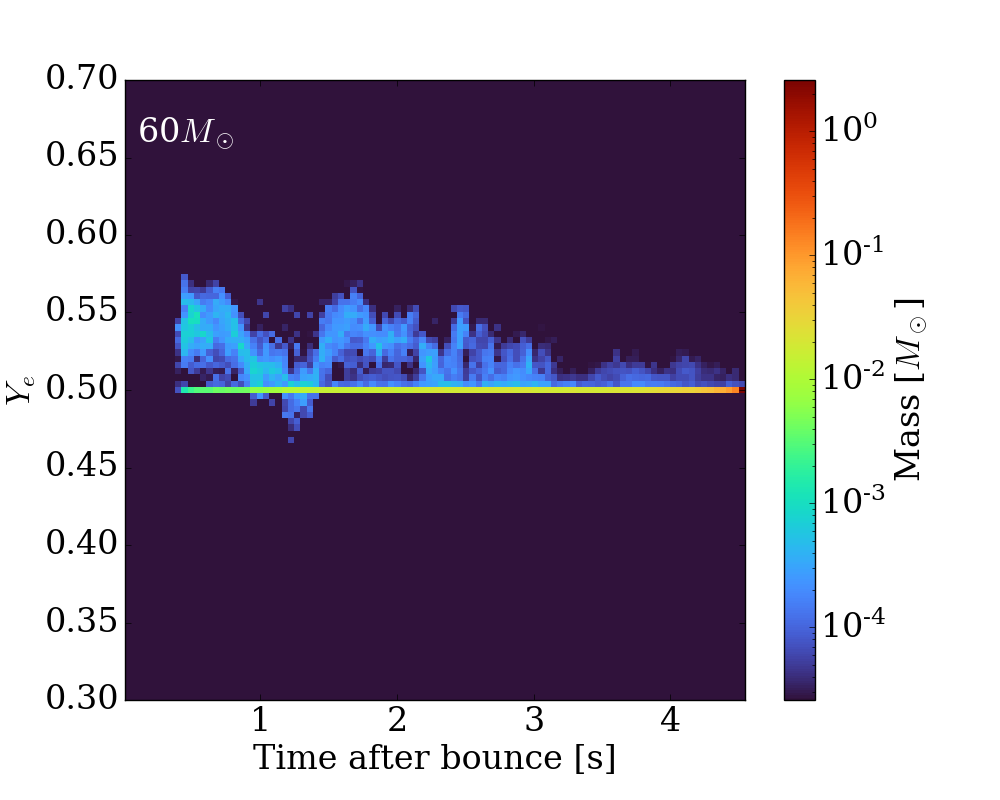}
    \caption{Ejecta electron fractions as a function of ejected time. The 9 $M_\odot$ (b), 11 $M_\odot$, 16 $M_\odot$, 20 $M_\odot$, 23 $M_\odot$ and 60 $M_\odot$ models are shown here. The horizontal line at $Y_e=0.5$ represents the matter ejected directly by the explosive shock. Such matter has never been close enough to the PNS to interact with neutrinos and change its electron fraction. The other component of the ejecta comes from the PNS and it has varying electron fractions.  This component is mostly proton-rich, but the $Y_e$ evolution seems stochastic, and there can be neutron-rich phases. }
    \label{fig:ye-t}
\end{figure}

In this section, we summarize and discuss the thermal conditions of the ejecta. Figure \ref{fig:tracer-ye-S} shows the mass distributions of the ejecta electron fractions and entropies. Both quantities are measured at either the time of NSE freeze-out (for tracers that have reached NSE) or at the time when the peak temperature is achieved (for tracers that never reach NSE). 
Although matter ejected later has on average a higher entropy, none of our models has reached S$>$80 $k_b$ baryon$^{-1}$ by the end of the simulation. Most models have modest electron fractions roughly between $\sim$0.45 and $\sim$0.60. Figure \ref{fig:ye-S} shows the joint distribution of the ejecta electron fractions and entropies in a few representative simulations. The vertical line at $Y_e=0.5$ represents the matter ejected directly by the explosive shock. This component never falls close enough to the PNS to interact with neutrinos. Thus, its electron fraction remains unchanged during the explosion. In addition to the prominent vertical line, most simulations show a similar flag-like shape in the $Y_e-S$ plane, and manifest a skew to proton-rich conditions at higher entropies. However, there can be some neutron-rich parcels as well, whose occurrence depends upon the details of the explosion and appears slightly stochastic. Figure \ref{fig:ye-t} shows the temporal evolution of the ejecta electron fraction. It is clear that there is a varying-$Y_e$ component, which is associated with neutrino-heated matter and neutrino-driven winds. For this component, the electron fractions remain proton-rich for most of the time, but neutron-rich phases can also be seen in a few models. However, we haven't yet found a reasonable way to easily predict the detailed temporal evolution of ejecta electron fractions.

\begin{figure}
    \centering
    \includegraphics[width=0.48\textwidth]{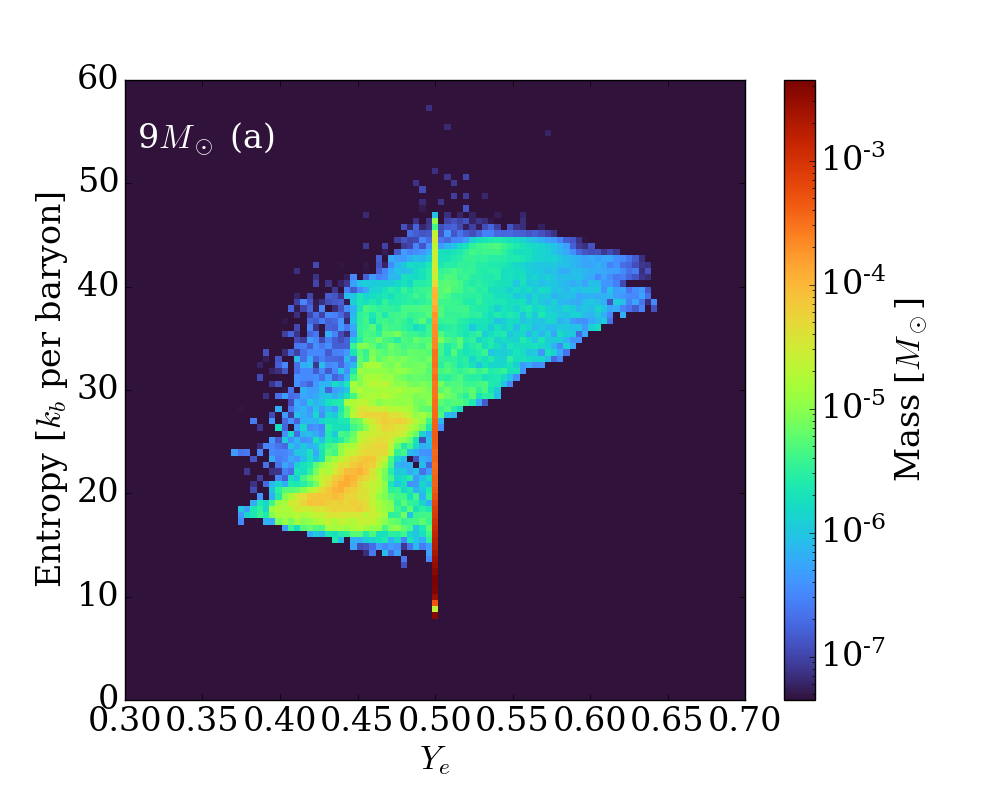}
    \includegraphics[width=0.48\textwidth]{figures/ye-S/9b_tracer_ye_S.png}
    \includegraphics[width=0.48\textwidth]{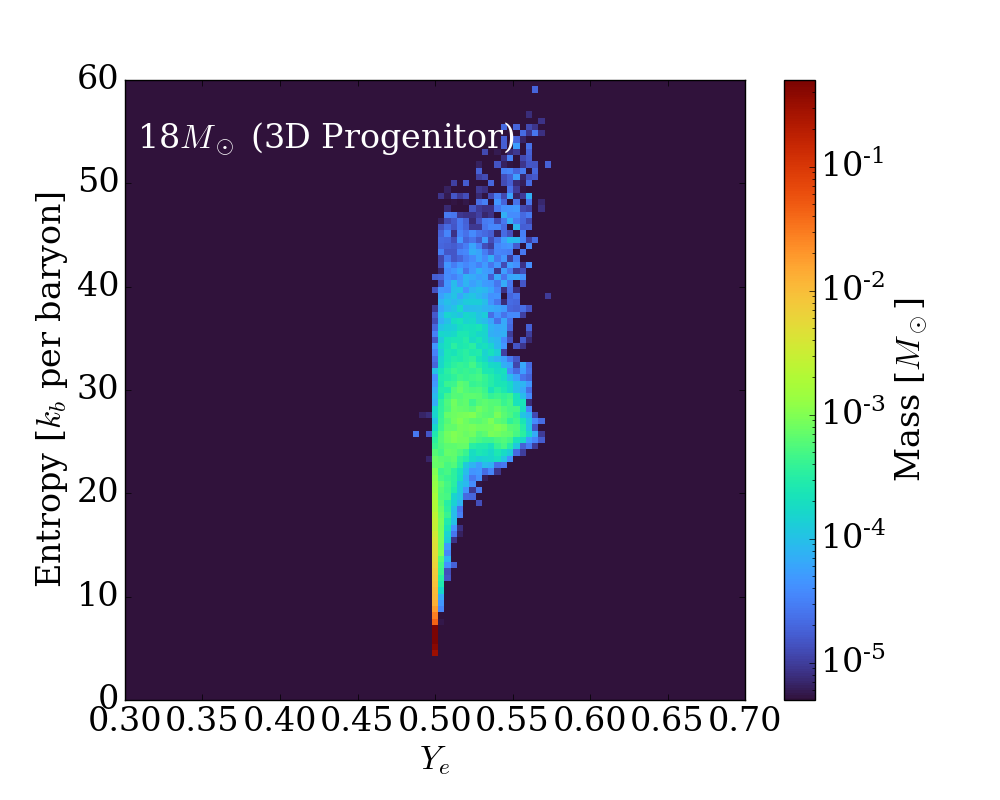}
    \includegraphics[width=0.48\textwidth]{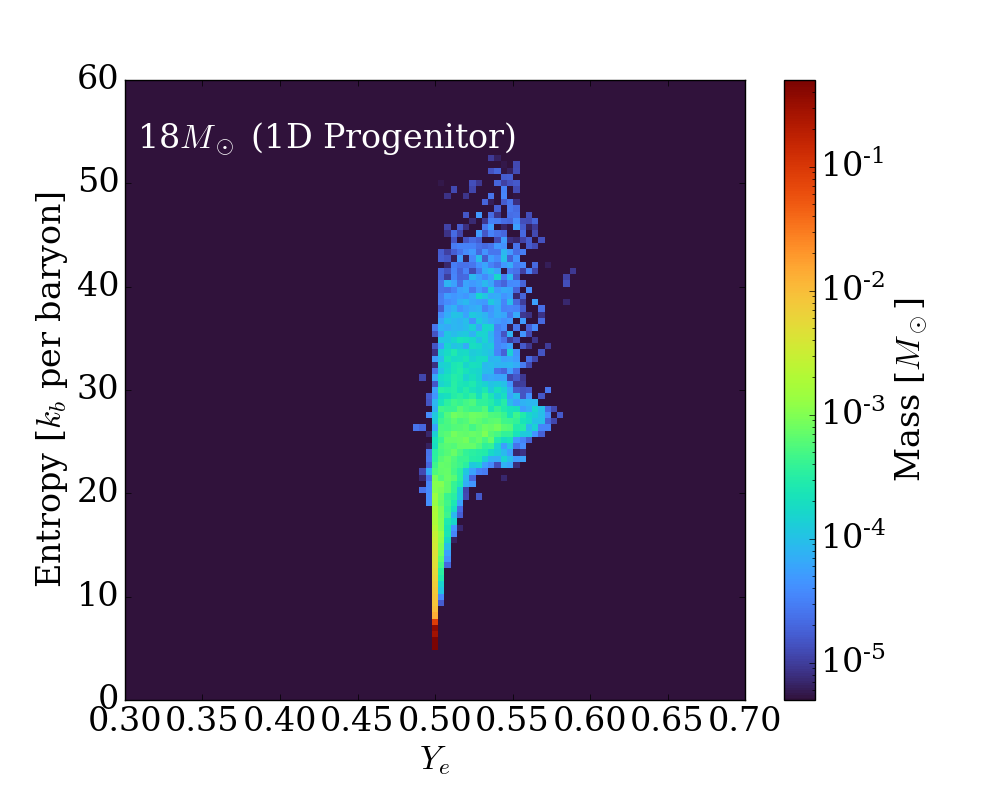}
    \caption{Ejecta condition comparisons for simulations with and without initial perturbations. {\bf Top Left:} The perturbed 9 $M_\odot$ model (referred to as model 9(a) in this paper). The initial perturbation is chosen to be an $n=4$, $l=10$, $v_{max}=100$ km s$^{-1}$ spherical harmonic velocity field between 200 and 1000 km following \citet{muller2015}. {\bf Top Right:} The non-perturbed 9 $M_\odot$ model (referred to as model 9(b) in this paper). {\bf Bottom Left:} The 18 $M_\odot$ model started from a 3D progenitor which has realistic initial perturbations. {\bf Bottom Right:} The 18 $M_\odot$ model started from the corresponding 1D progenitor. Although the initial perturbations in the 9(a) model significantly change the ejecta electron fractions, the two 18 $M_\odot$ simulations show almost identical distributions. {Assuming a weak impact of initial perturbation on the overall explosion properties, the effects of initial perturbations on $Y_e$ may not be as strong for more massive models with stronger accretion and later explosions.} 
    However, the number of such comparisons is too small to draw a final conclusion, so the effects of initial perturbations are still to be firmly demonstrated. We plan to study this topic in more detail in future work.}
    \label{fig:1d-3d}
\end{figure}

Not only does the late-time $Y_e$ evolution seem unpredictable, we also see that some of the matter parcels of the 9(a) model achieve low $Y_e$s early during its post-bounce evolution. 
This is due to the initial perturbations we added to that progenitor model. During collapse, matter compression leads to electron capture, making the matter neutron-rich. Later, the net effect of charge-current interactions with neutrinos and anti-neutrinos increases the electron fraction and the matter becomes proton-rich if it swells long enough in the turbulent region above the PNS. However, the ejected matter remains neutron-rich if it is ejected quickly and there is not enough time to interact with the neutrinos. Initial perturbations make the 9(a) model explode slightly earlier and more vigorously compared to the non-perturbed 9(b) model, which is why there is more neutron-rich ejecta in that model. {As stated earlier, since the explosion properties of the 9(a) model are only weakly influenced, this may indicate that nucleosynthesis might be more sensitive to initial perturbations than explosion properties.} However, such effects may not be as strong in other models. For models that explode later and a bit more slowly, there seems always enough time to boost $Y_e$ above 0.5.  In this case, even with initial perturbations in the convective progenitor, the effect of such perturbations on the final $Y_e$ distribution can be minor. To explore this, we simulated two 18 $M_\odot$ models with 1D and 3D progenitors, for which the latter boasts realistic initial perturbations in velocity and density typical of more massive models. The 1D progenitor is taken from the S18 set which is evolved to core collapse using the stellar evolution code Kepler. The 3D progenitor is the same model as the 1D up until 293.5 seconds (s) before the onset of collapse, after which the oxygen-shell burning calculations are done in 3D by the hydrodynamics code P{\sc ROMETHEUS} \citep{muller2016}. Figure \ref{fig:1d-3d} shows the $Y_e-S$ distributions of the 18 $M_\odot$ models with and without initial perturbations, compared to the corresponding quantities for the 9(a) and 9(b) models. For these 18 $M_\odot$ models, we see almost no difference between ejecta conditions. This means that the effects of realistic initial perturbations may not influence the nucleosynthesis for such more massive models, in contrast with what we see for the 9(a) and 9(b) models. However, we have not done an exhaustive study of this difference and it is premature to draw definitive conclusions at this stage, though our preliminary results are suggestive. We plan to study this issue in more detail in future work. {We emphasize that we are focusing here only on the role of relatively weak initial perturbations. It is thought that stronger initial perturbations are able to significantly change explosion properties (such as shock speed, explosion energy, ejecta mass) \citep{couch2013,muller2015,bollig2021}. However, it is still unclear whether weak perturbations, which don't change the explosion properties, may have a strong influence on nucleosynthesis results.}

\begin{figure}
    \centering
    \includegraphics[width=0.48\textwidth]{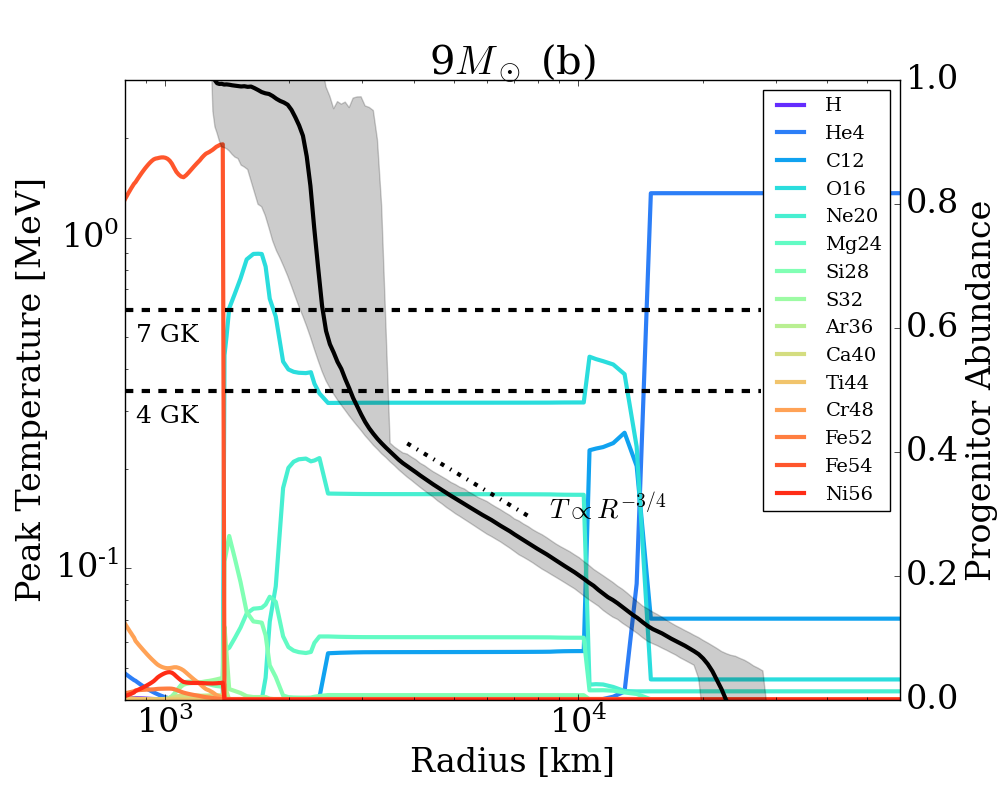}
    \includegraphics[width=0.48\textwidth]{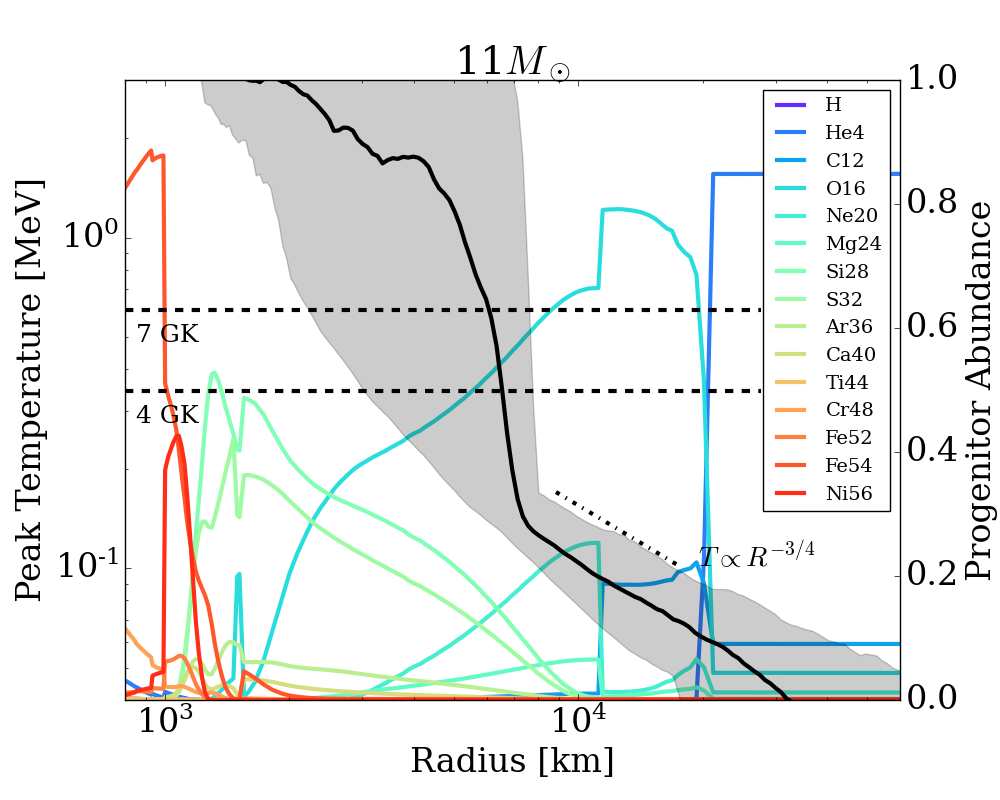}
    \includegraphics[width=0.48\textwidth]{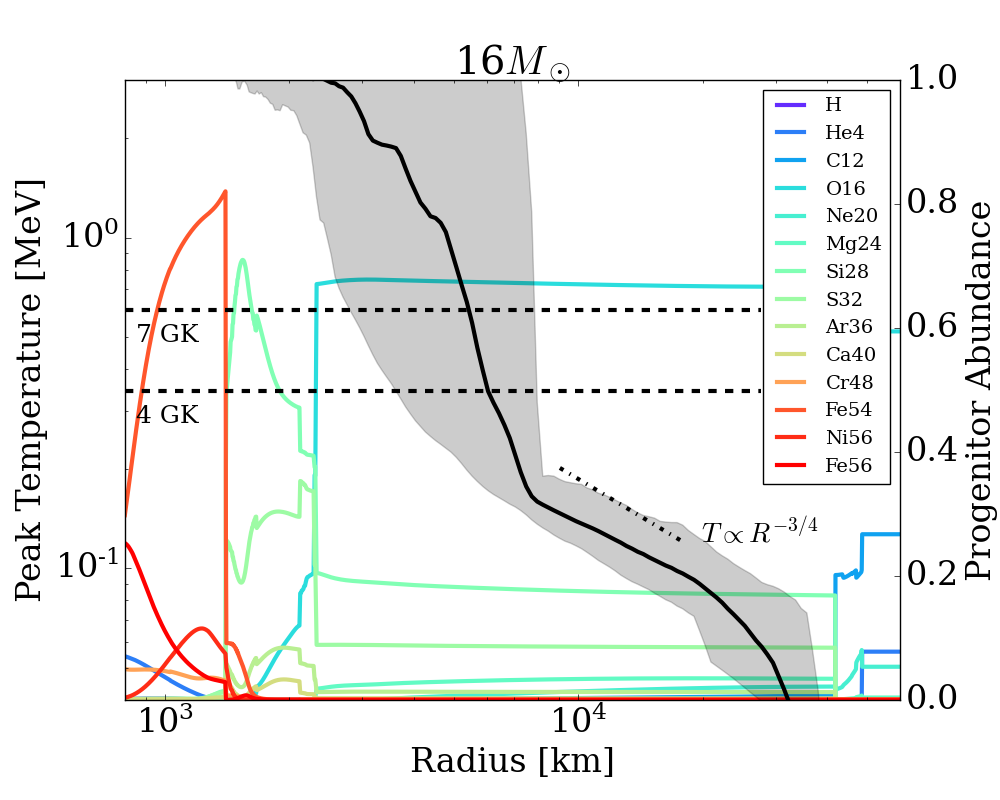}
    \includegraphics[width=0.48\textwidth]{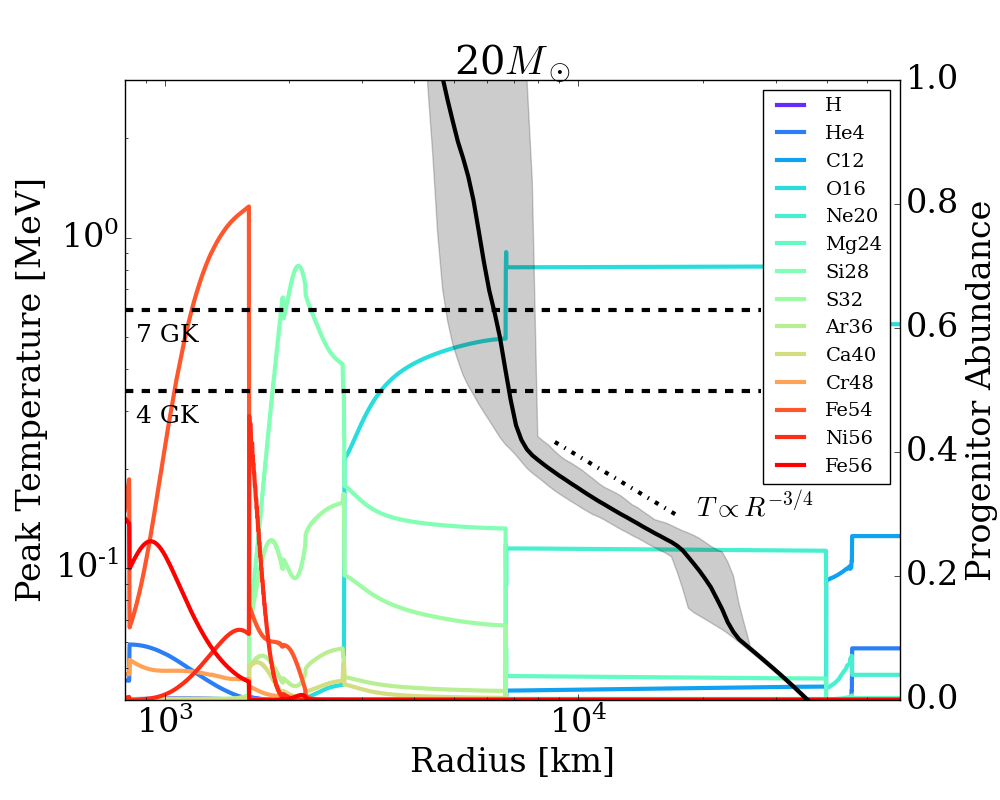}
    \includegraphics[width=0.48\textwidth]{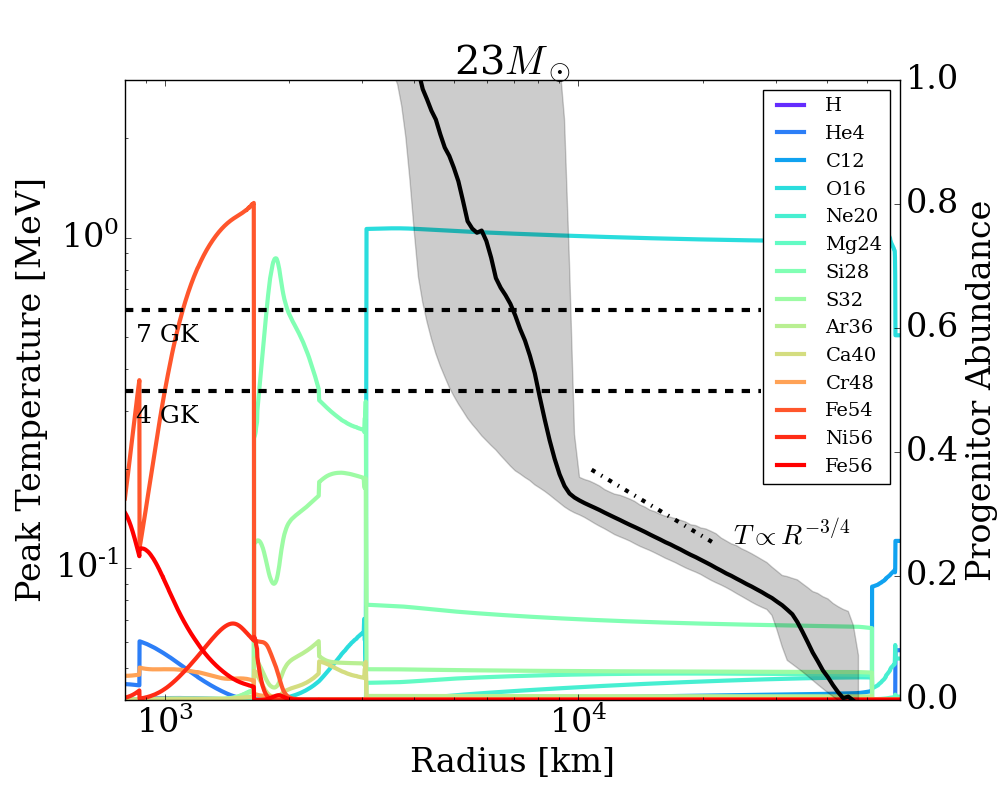}
    \includegraphics[width=0.48\textwidth]{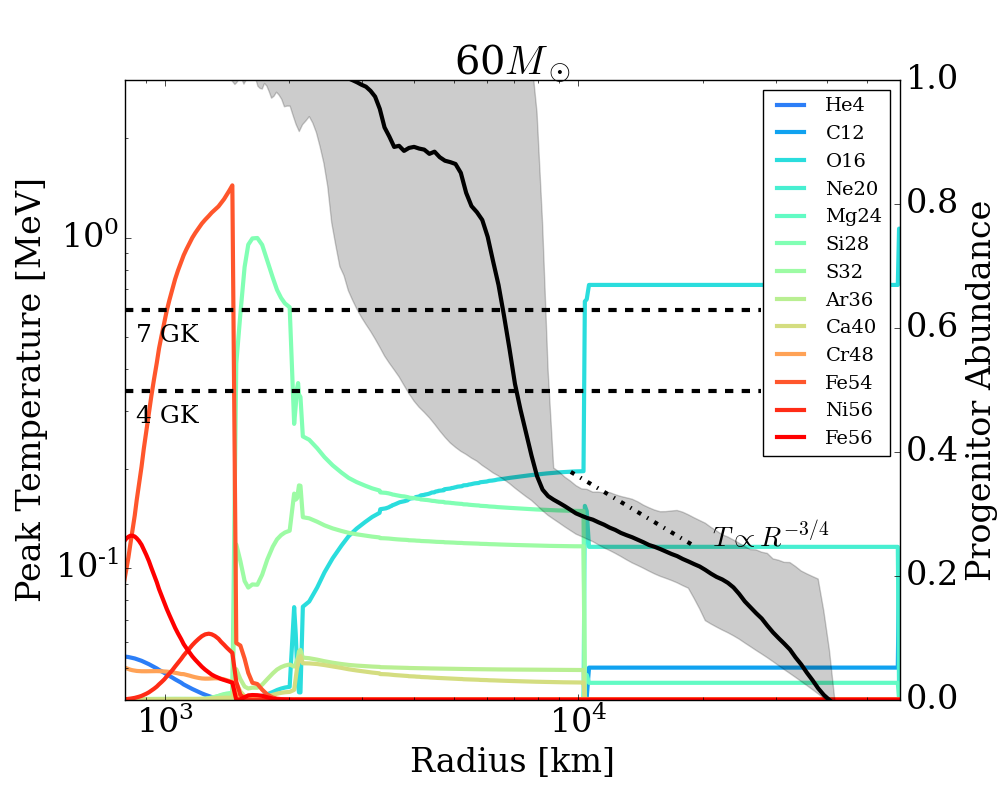}
    \caption{Peak temperatures and initial abundances as a function of initial radius for the 9 $M_\odot$ (b), 11 $M_\odot$, 16 $M_\odot$, 20 $M_\odot$, 23 $M_\odot$, and 60 $M_\odot$ models. In this plot, we include only matter that is eventually ejected. The black curve shows the mass averaged peak temperature experienced by matter at the given radius of the progenitor and the gray shadow shows the range of peak temperatures. The width of this gray shadow is related to the asymmetry of the explosion.
    The colored lines depict the initial abundances in the progenitor. The complete silicon burning temperature ($\sim$4 GK) and the NSE threshold ($\sim$7 GK) are marked as horizontal black dashed lines. The dot-dashed line shows the power-law relation $T\propto R^{-3/4}$. The peak temperatures in the less massive models follow this power-law relation quite well when the temperature drops below $\sim$0.2 MeV ($\sim$2.3 GK). In this region, most of the matter is ejected directly by the primary explosion. In the more massive models, the deviations from this power law are larger because the explosion energy is deposited during a longer period of time, which leads to shallower slopes. See Section \ref{sec:conditions} for more details.
    }
    \label{fig:r-T}
\end{figure}

Figure \ref{fig:r-T} depicts the peak temperatures achieved by each layer of the progenitor model. In this plot, we include only matter that is eventually ejected. The black solid line is the mass-averaged peak temperature, while the gray shadow region shows the range of peak temperatures. {The width of the shaded band is directly related to the asymmetry of the explosion.} Two regions can be seen clearly in this plot from small to large radii. The first region has the highest temperatures, steepest slopes, and largest ranges of peak temperatures. Matter in this region reaches higher temperatures because it (at least partly) enters the turbulent region above the PNS where neutrino heating is vigorous. The large range of peak temperatures in this region is due to the complex convective motions and the different ejection/dwell times. Higher temperatures are achieved if the matter reaches a smaller radius and is later ejected. Therefore, the minimum peak temperature curve roughly indicates the behavior of the earliest ejected matter. Because the explosion is often triggered by the accretion of the Si/O interface \citep{wang2022}, there is a correlation between the decrease of the minimum peak temperature and the oxygen shell mass of the progenitors. However, there is sometimes a delay between the accretion of the Si/O interface and the explosion, which can also be seen in the shock radius evolution depicted in Figure \ref{fig:rshock}. It is worth noting that not all matter in this region will eventually be ejected. A significant fraction will accrete onto the PNS and the exact amount of accretion depends upon the details of the 3D simulations. 

The peak temperature will suddenly drop when the blast wave is able to eject all the matter in a layer. This marks the starting point of the second region on Figure \ref{fig:r-T}. In models with simultaneous explosion and accretion, matter in this second region may still fall onto the PNS and a small fraction will move out with the wind. This effect may shift the maximum peak temperature curve upward in longer-duration simulations, but the mass-averaged curve is not much influenced. The peak temperature behavior in this region is similar to that seen in 1D studies, and the width of the gray shadow which echoes the asymmetry of the explosion is small. The relation between the peak temperature and the radius in the less massive models can be well-described by the $T\propto R^{-3/4}$ relation (shown as the dot-dashed line), which assumes the explosion energy is deposited instantly. This assumption doesn't hold in some more massive models, because long-lasting accretion can power the neutrino heating which continues to pump energy into the ejecta. This is why the more massive progenitor models always show shallower slopes (higher temperatures) and deviate from the $T\propto R^{-3/4}$ relation. The second region extends to the shock positions at the end of our simulations, where the peak temperature drops to that of the unshocked matter. 

\begin{figure}
    \centering
    \includegraphics[width=0.48\textwidth]{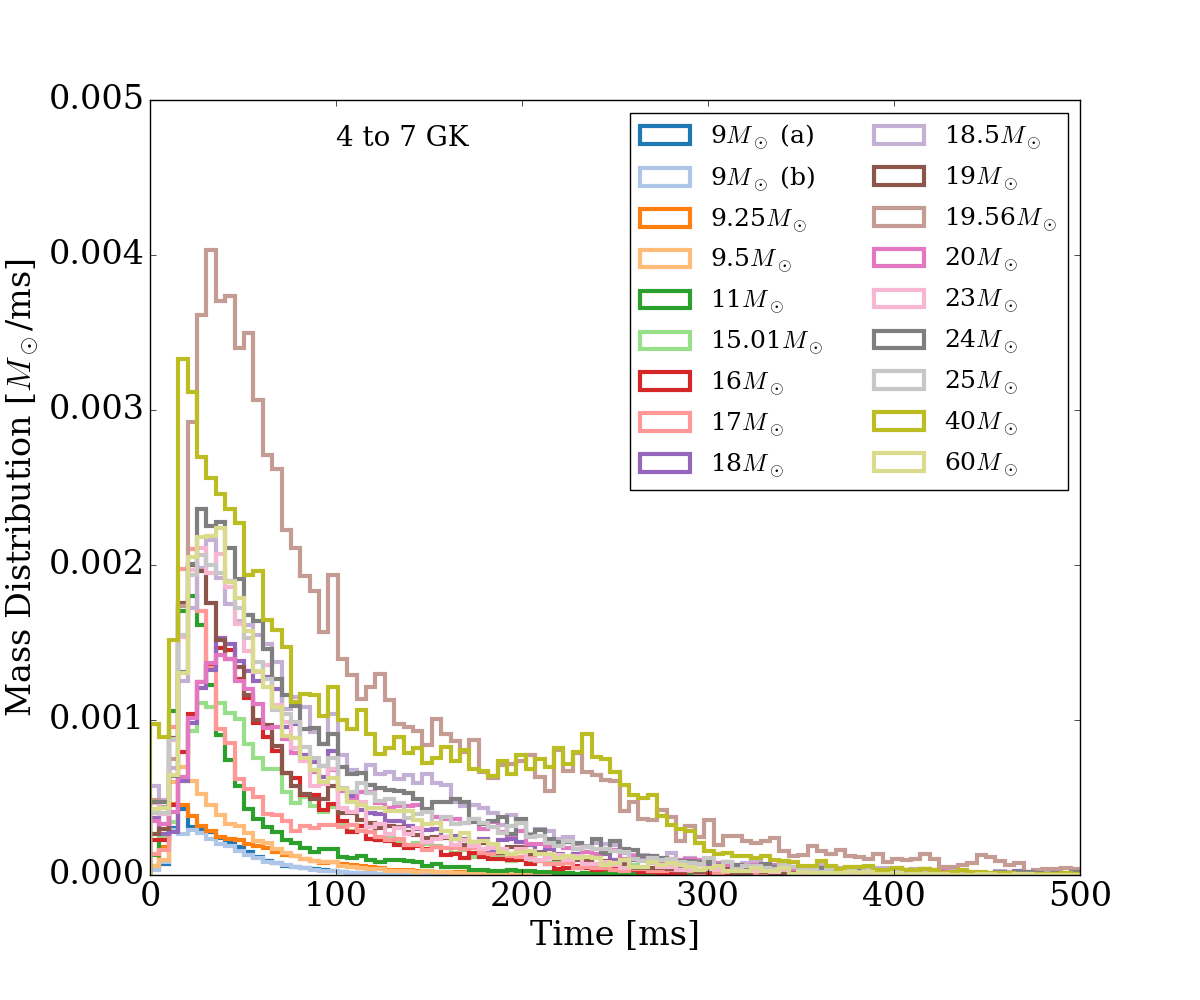}
    \includegraphics[width=0.48\textwidth]{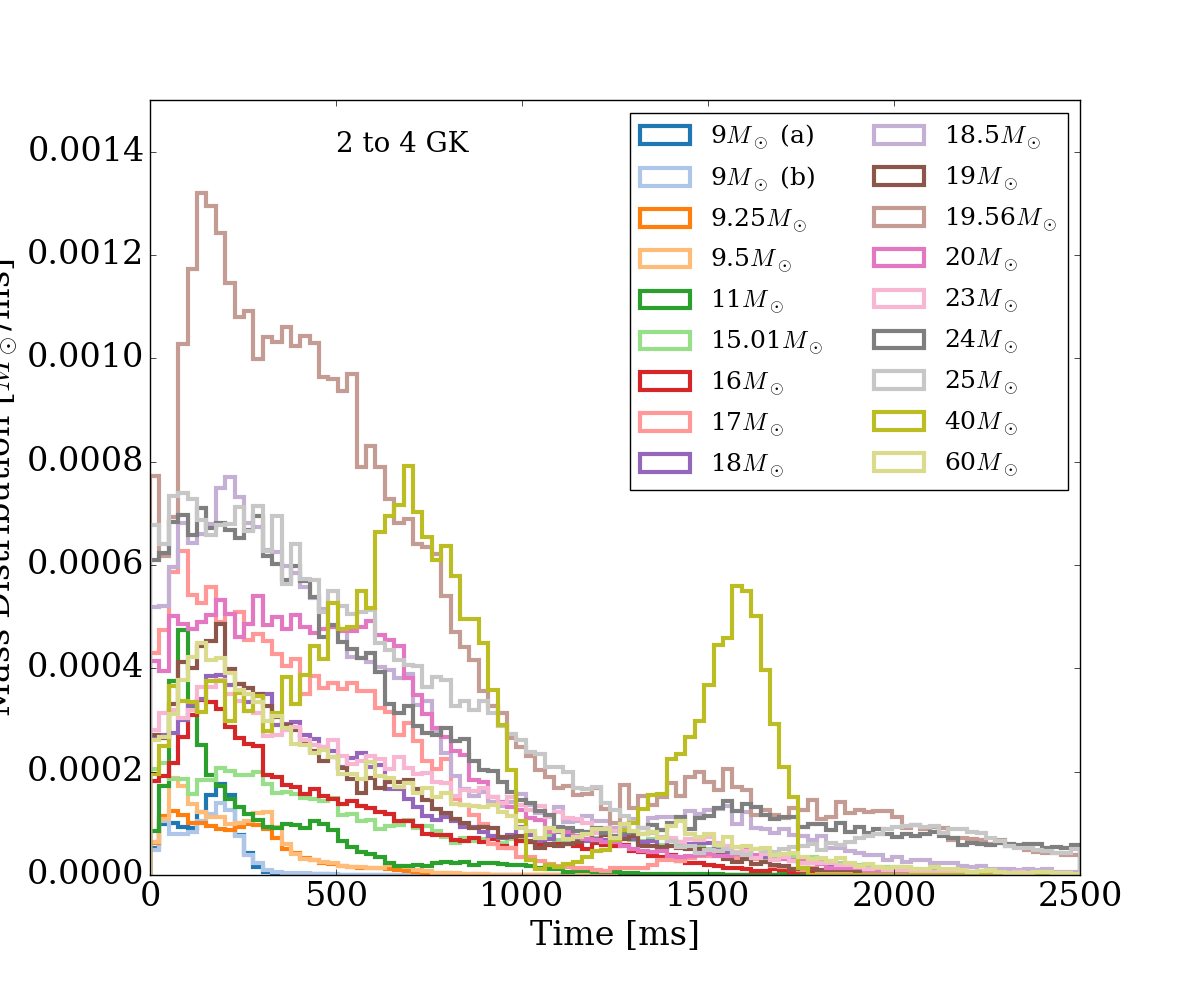}
    \caption{Distributions of the time the tracers spend in the $2-4$ and $4-7$ GK intervals. Although less massive models show similar distribution shapes in the $4-7$ GK temperature range, more massive models have distinct long-tail or even multi-peak distributions in both temperature ranges. This is due to long-lasting accretion which interacts with the ejecta and delays their ejection, which will further change the results of the nucleosynthetic calculations.}
    \label{fig:hist-dt}
\end{figure}

\begin{figure}
    \centering
    \includegraphics[width=0.48\textwidth]{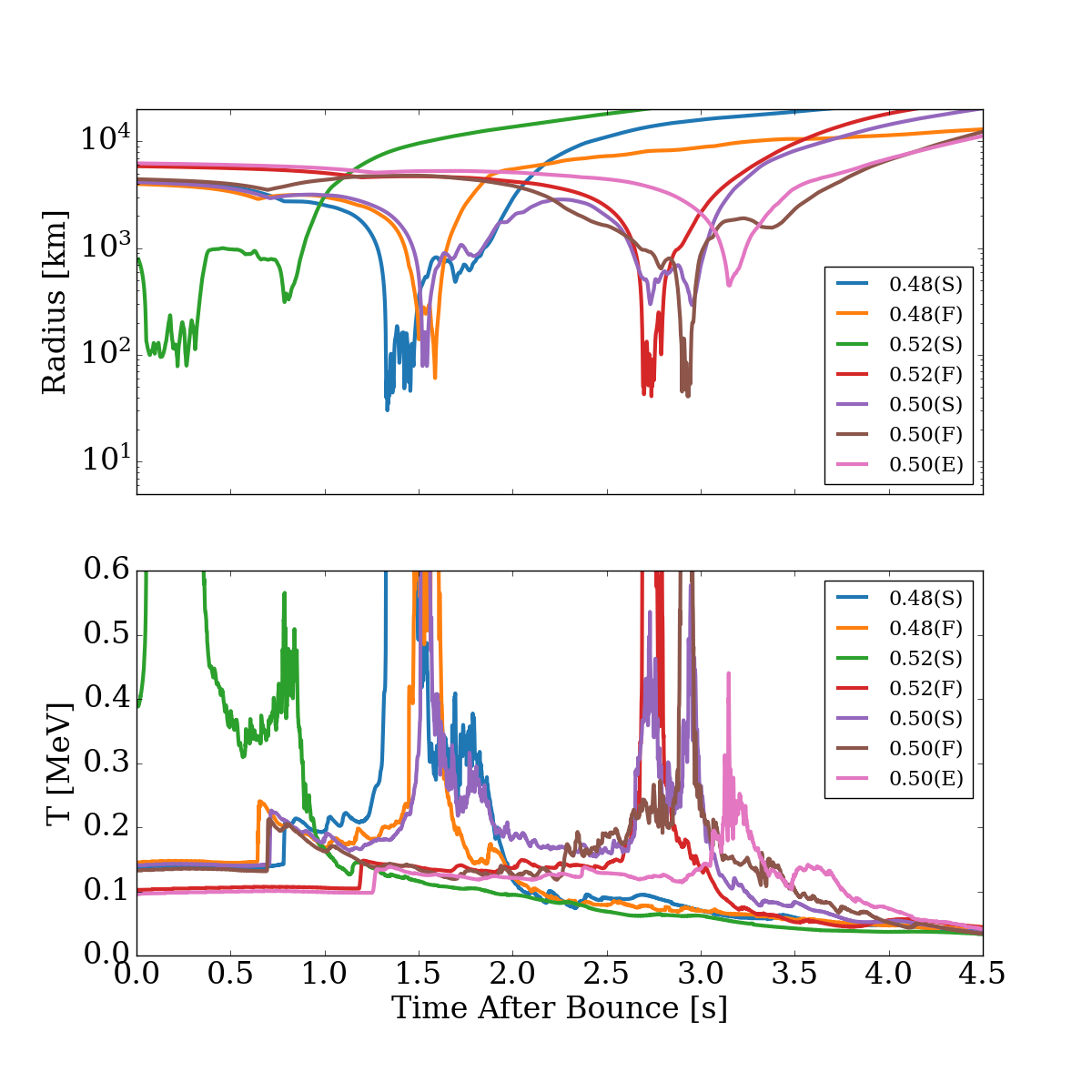}
    \includegraphics[width=0.48\textwidth]{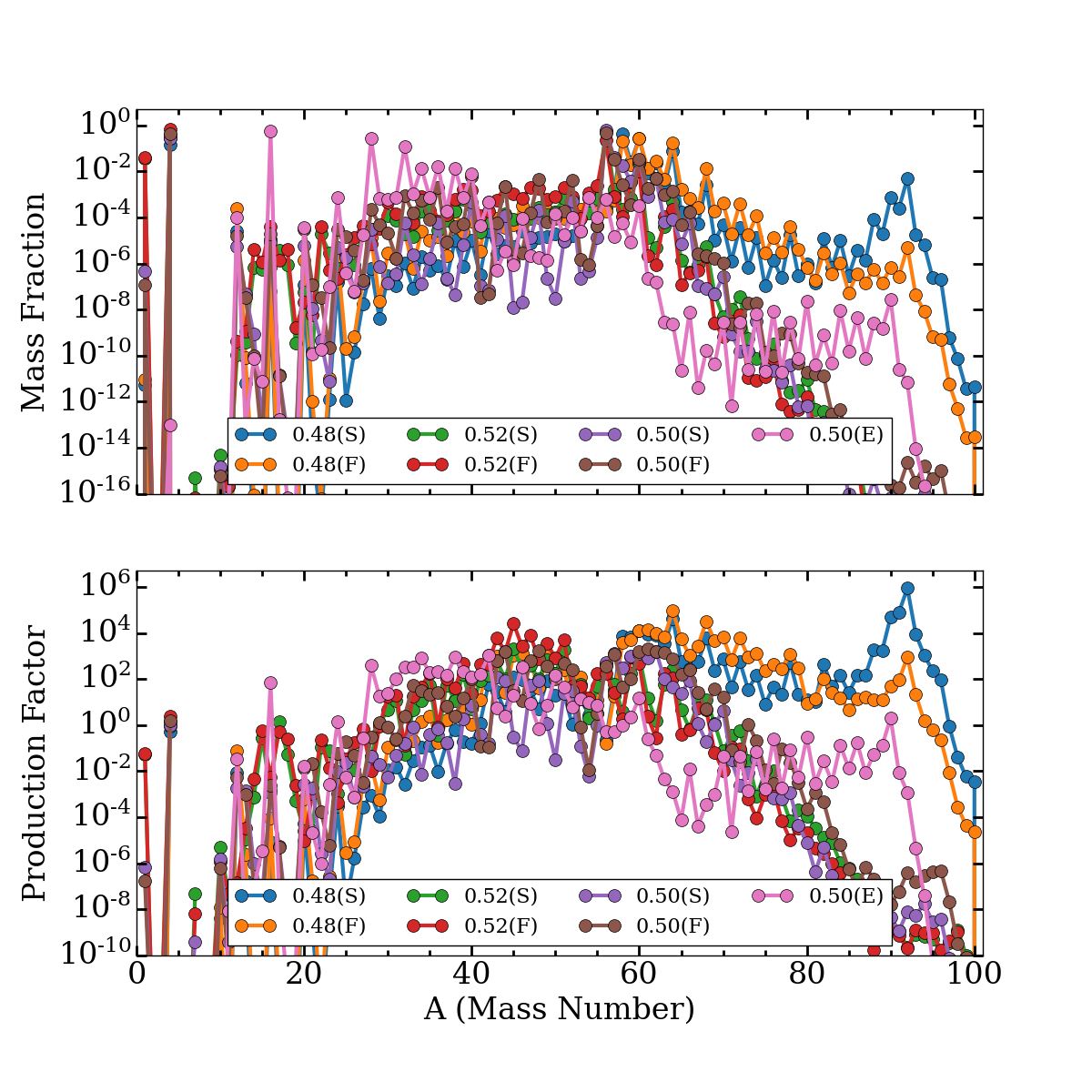}
    \caption{Histories and nucleosynthesis results of six selected tracers with different $Y_e$s. Numbers in the tracer names show the $Y_e$ of each tracer, while ``S" and ``F" represent relatively slow and fast ejection. ``E'' means that the tracer has never reached NSE. {\bf Top Left:} Tracer radii as a function of time. The ejection phases of most tracers are not monotonic due to interactions with infalling material, and they experience non-monotonic thermal histories. {\bf Bottom Left:} Temperature evolution of the selected tracers. The non-monotonic temperature evolution can significantly influence the nucleosynthesis results because it allows multiple transitions between local equilibria of nuclear reactions and non-equilibrium phases \citep{magkotsios2011,sieverding2023}. {\bf Top Right:} Mass fractions of the final compositions of the tracers. {\bf Bottom Right:} Production factors of the selected tracers. It is clearly seen that the element abundances are easily changed by more than one order of magnitude due solely to the differences in average ejection velocities. See Section \ref{sec:conditions} for more details.
    }
    \label{fig:selected-tracer}
\end{figure}

In addition to the electron fraction and peak temperature, another important factor in nucleosynthesis calculations is the time ejecta parcels spend in certain temperature ranges. Figure \ref{fig:hist-dt} shows the distribution of time spent in $4-7$ GK and $2-4$ GK intervals. Although less massive models show similar distribution shapes in the $4-7$ GK temperature range, more massive models have distinct long-tail or even multi-peak distributions in both temperature ranges. This is due to the long-lasting accretion which interacts with the ejecta and delays its ejection, and this further alters the nucleosynthesis. As an example, Figure \ref{fig:selected-tracer} shows six selected tracers with different $Y_e$s. The numbers in the tracer names identify the $Y_e$ of each tracer, while ``S" and ``F" represent relatively slow and fast ejection. ``E'' gives an example of tracers that have never reached NSE. The top left panel shows the tracer radius as a function of time. Due to the interaction with infalling matter, the ejection phases of most tracers are not simple, and they can manifest non-monotonic thermal histories. The bottom left panel shows the temperature evolution of the selected tracers. The non-monotonic temperature evolution can significantly influence the nucleosynthesis results because it allows multiple transitions between local equilibria of nuclear reactions and non-equilibrium phases \citep{magkotsios2011,sieverding2023}. The right two panels show the mass fractions and production factors of the final composition of the selected tracers, and it's clearly seen that the elemental abundances are easily changed by more than one order of magnitude due solely to the differences in the average ejection velocities.

\begin{figure}
    \centering
    \includegraphics[width=0.48\textwidth]{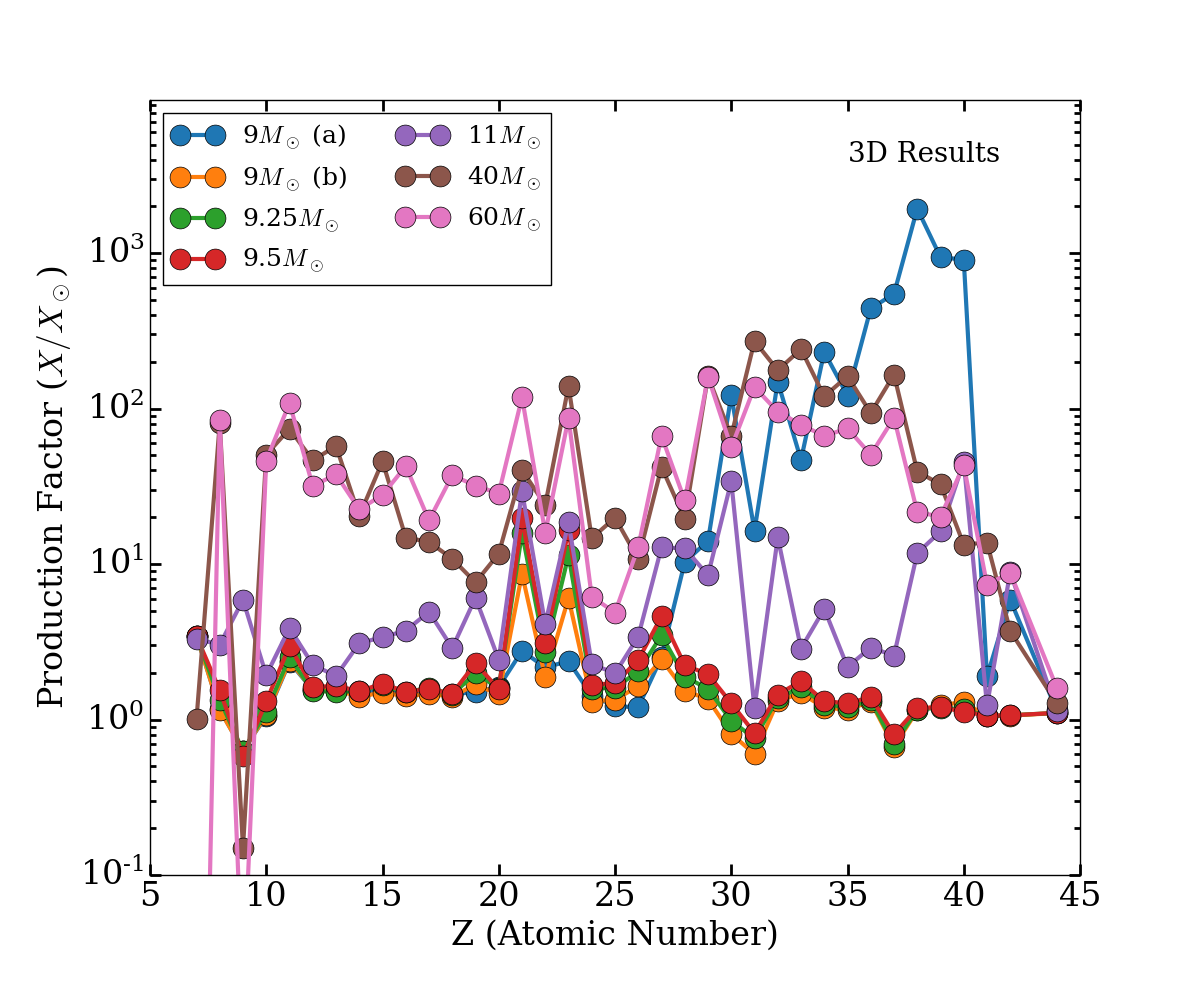}
    \includegraphics[width=0.48\textwidth]{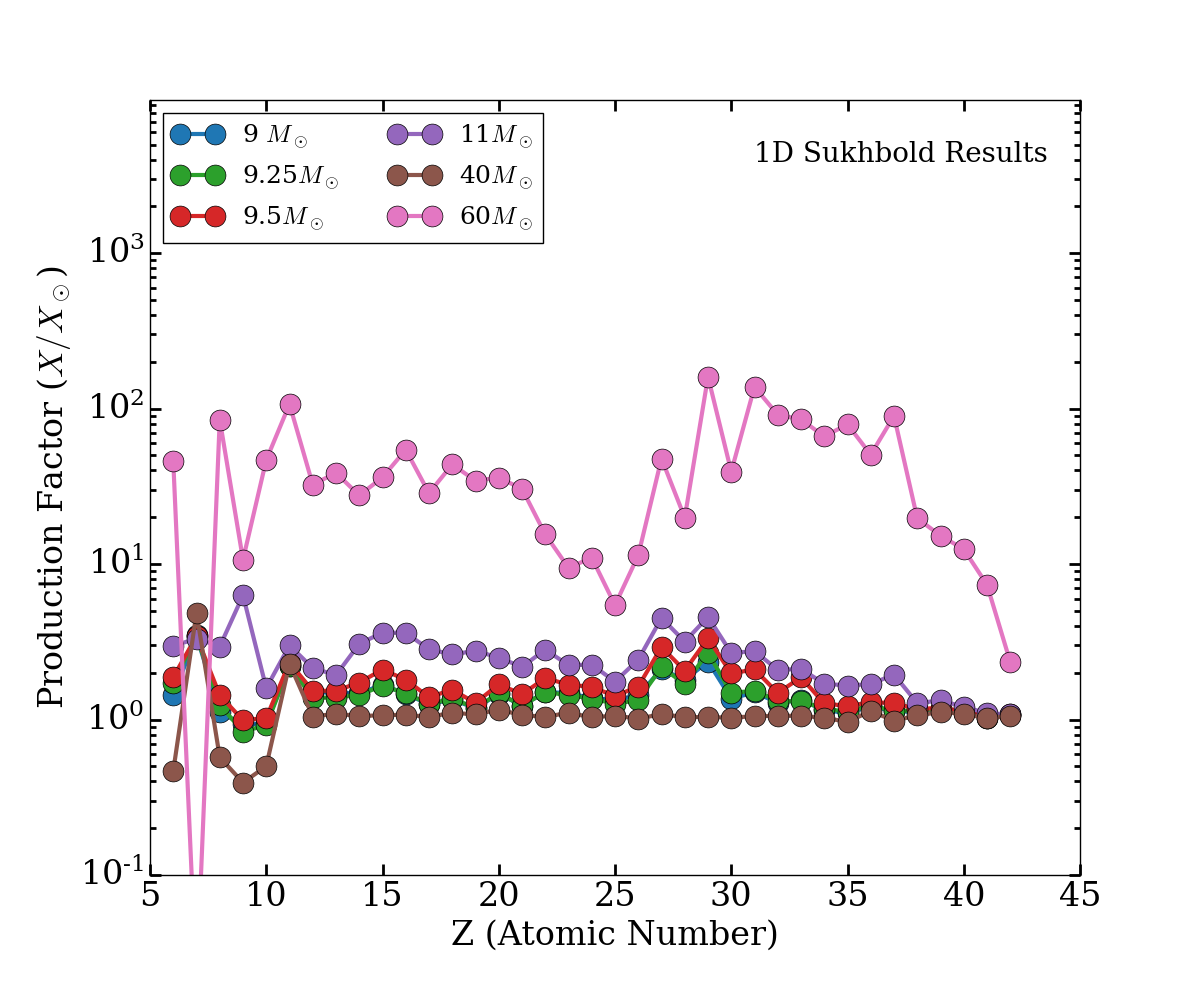}
    \caption{Comparison between 3D simulations done by this work (left) and 1D models from \citet{sukhbold2016} (right) of the same progenitors. Although the mass-function weighted result is close to the solar pattern, our 3D simulations show larger variation between different models due to the multi-dimensional effects, such as explosion asymmetry, non-monotonic tracer histories, and stochastic electron fraction evolutions,that can't be reproduced by 1D simulations. {The 40 $M_\odot$ model is special, because in 1D it doesn't explode. The 1D production factor shown here is the pre-SN winds.}}
    \label{fig:comparison}
\end{figure}

The stochastic $Y_e$ evolution, complex peak temperature distributions, and non-monotonic ejecta trajectories and thermal histories are some of the major multi-dimensional effects  of relevance to nucleosynthesis. Figure \ref{fig:comparison} compares the final elemental abundances calculated by this work using 3D simulations and by \citet{sukhbold2016} using 1D models of the same progenitor models. Significantly larger variations is seen in the 3D results, as a result of all the complexities mentioned above. In addition, the 3D simulations also have larger variations in explosion properties (explosion energy, ejected mass, etc.) compared to the 1D models in \citet{sukhbold2016} which are imposed by a simple ad hoc algorithm. The most significant difference is perhaps between the 3D and 1D 40 $M_\odot$ models. In our simulations, the 40 $M_\odot$ model explodes with an energy $\sim1.6$ B \citep{burrows2023}, while the 1D model assumes that the CCSN would fail. In general, the larger variation in 3D calls for more long-term CCSN simulations, and this work is the first step towards the goal of a comprehensive 3D nucleosynthesis dataset.

\section{Nucleosynthesis Results}\label{sec:results}
\begin{figure}
    \centering
    \includegraphics[width=0.48\textwidth]{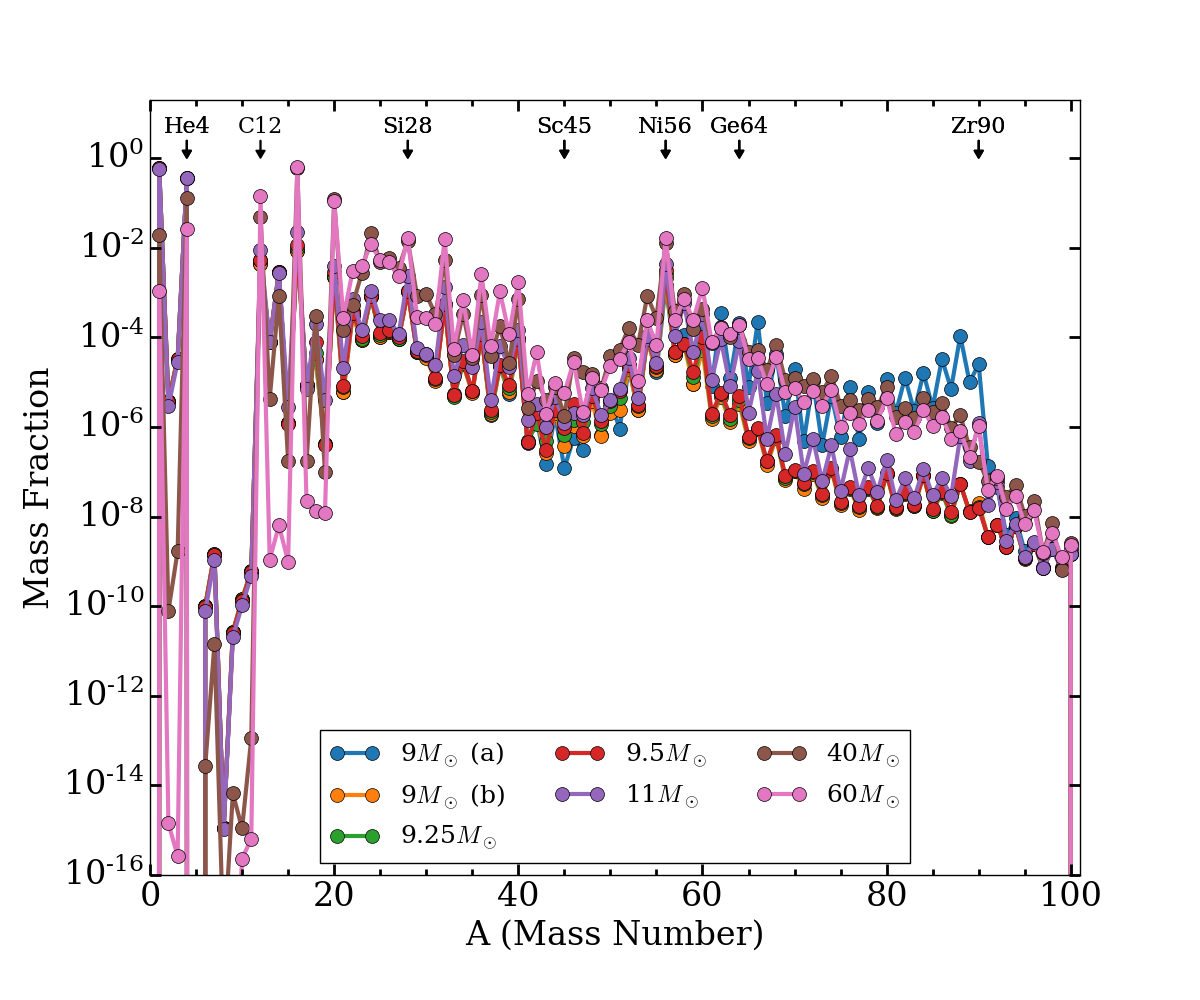}
    \includegraphics[width=0.48\textwidth]{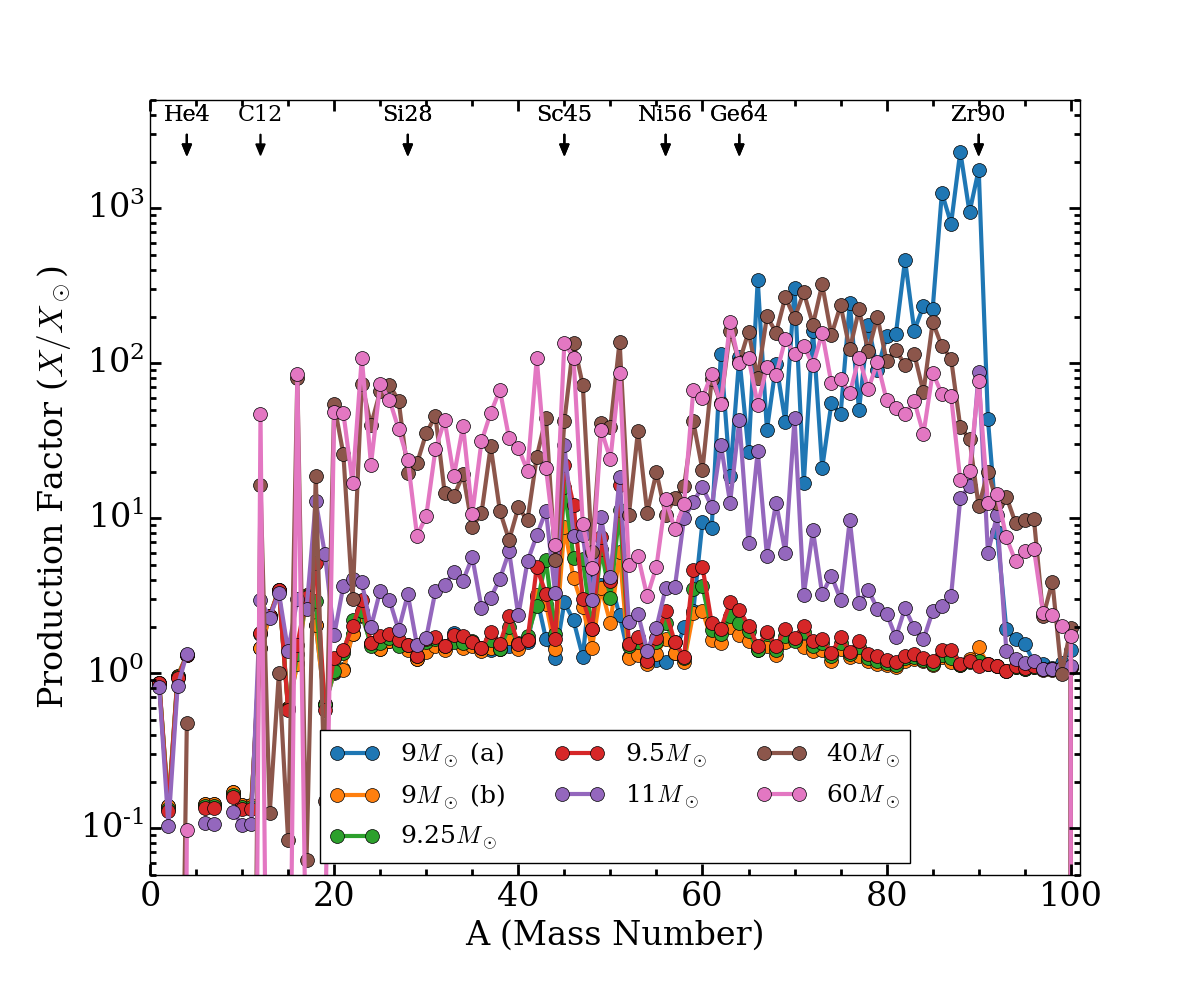}
    \caption{Mass abundances and production factors of ejecta at the end of each simulation. Isotopes with the same mass number $A$ are summed together. The abundances shown here include contributions from both pre-CCSN and explosion phases. The 9(b), 9.25, and 9.5 $M_\odot$ models show similar production curves peaking at $^{45}$Sc, which is typical for proton-rich ejecta. This is because these models don't have neutron-rich ejecta. The 9(a) model, on the other hand, experiences a more vigorous explosion and faster ejection compared to model 9(b), which leads to much more neutron-rich matter. As a result, this model is able to produce significantly more heavy elements up to the peak at about $^{90}$Zr. The other three models (11, 40, and 60 $M_\odot$) show random neutron-rich phases and they are able to produce some heavier elements. They also show larger variations in the abundances of elements lighter than the iron group compared to the less massive models.}
    \label{fig:prod-factor}
\end{figure}

Figure \ref{fig:prod-factor} summarizes the mass fractions and production factors ($X/X_{\odot}$) at the end of each simulation. Only S16 models are shown in this figure because they have complete pre-CCSN nucleosynthesis results. The abundances shown here include contributions from both pre-CCSN phases and the CCSN explosions. The 9(b), 9.25, and 9.5 $M_\odot$ models show similar production curves peaking at $^{45}$Sc, which is typical for proton-rich ejecta. These models haven't experienced a neutron-rich phase. On the other hand, the 9(a) model explodes more vigorously and boasts faster ejection compared to the 9(b) model, and this leads to the  ejection of more neutron-rich matter. As a result, this model is able to produce significantly more heavy elements up to the peak at about $^{90}$Zr. The other three models (11, 40, and 60 $M_\odot$) show random neutron-rich phases and they are able to produce some heavier elements. They also show larger variations in the abundances of elements lighter than the iron group compared to that seen in the less massive models. This is because the multi-dimensional effects are stronger in more massive models due to their longer delay to explosion and more vigorous post-shock convection at ``ignition."

\begin{figure}
    \centering
    \includegraphics[width=0.48\textwidth]{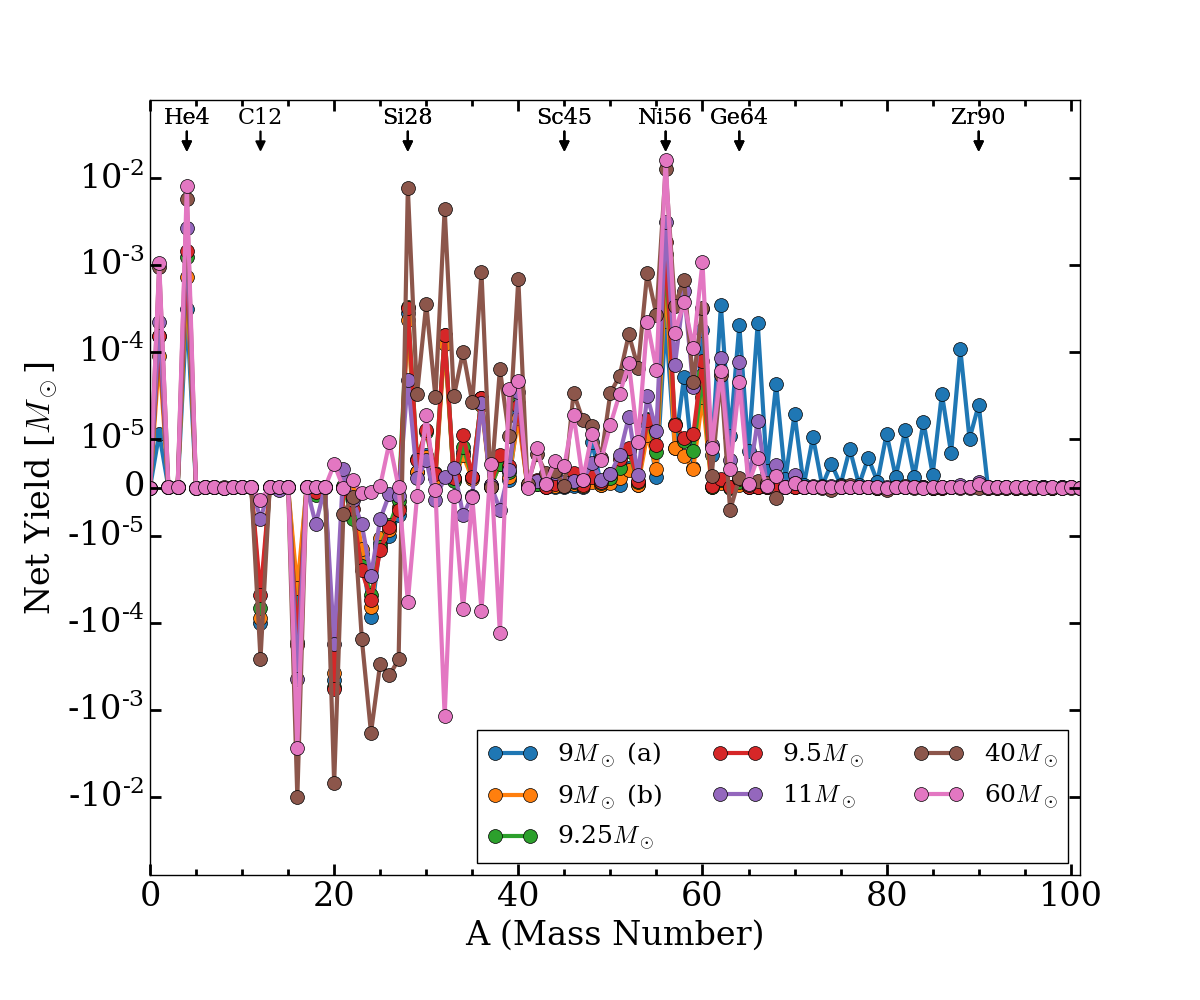}
    \includegraphics[width=0.48\textwidth]{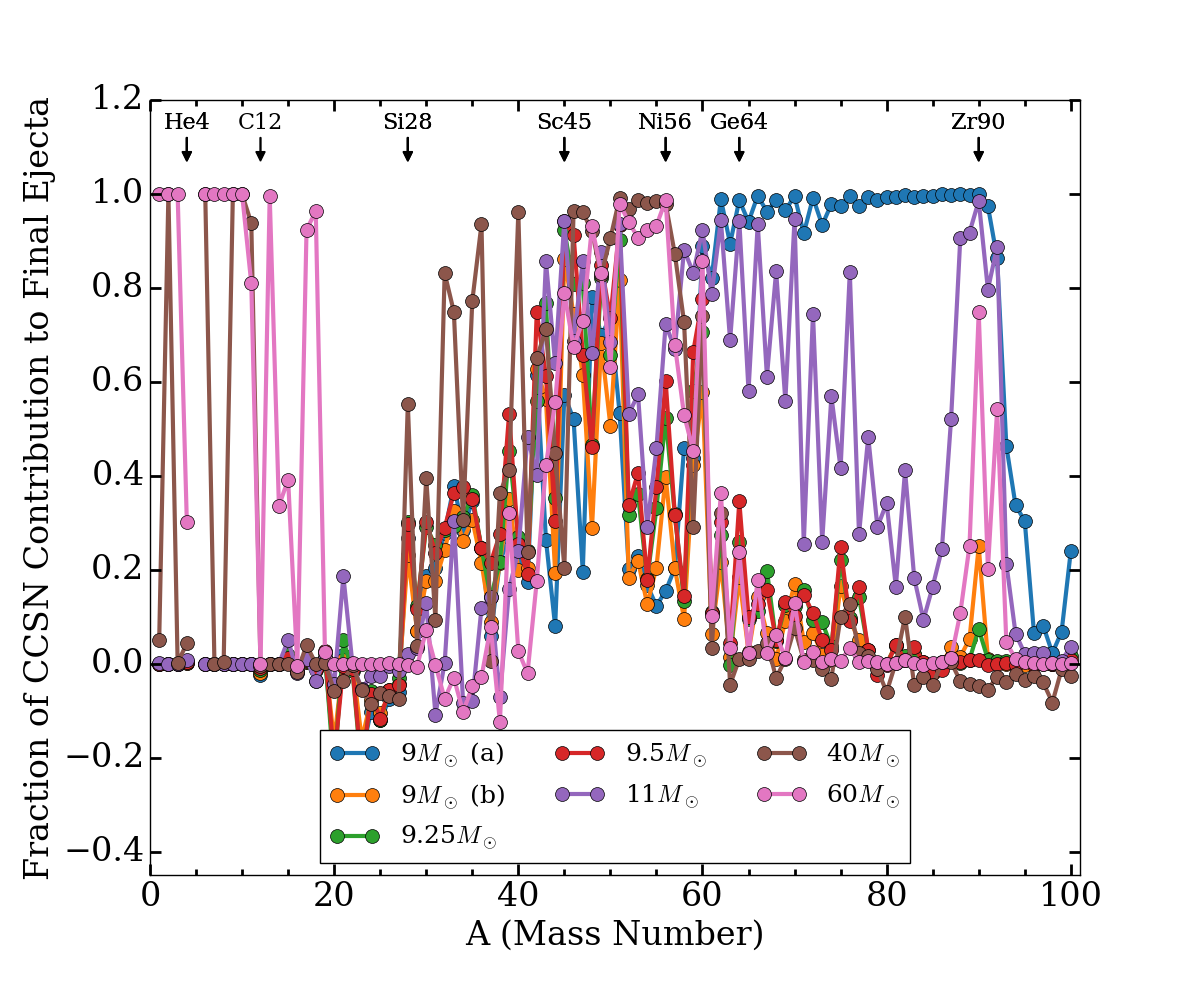}
    \caption{Net yield and the fractional contribution of explosive nucleosynthesis. Isotopes with the same mass number $A$ are summed together. The CCSN contribution fraction shown here is the ratio between the net yield and the final ejected amount. For isotopes lighter than $^{20}$Ne, for most models the CCSN contribution is negligible compared to the that of the pre-CCSN phase. The most massive models (40 and 60 $M_\odot$) have ejected almost their entire hydrogen and helium shells before explosion. Thus, the CCSN contribution dominates some lighter elements, but the absolute amount is very low, which can be seen in the left panel of Figure \ref{fig:net-yield} and in both panels of Figure \ref{fig:prod-factor}. The CCSN explosion component starts to make a significant contribution at $^{20}$Ne by destroying about 20\% of the ejected neon mass. The net yields become positive in most models at around $^{28}$Si. The CCSN contribution drops back to zero at around $^{64}$Ge in models without neutron-rich phases, but for models that experience some neutron-rich periods the CCSN can significantly contribute to the heavier elements, or at least to the $^{90}$Zr peak.}
    \label{fig:net-yield}
\end{figure}

Figure \ref{fig:net-yield} shows the net yields and the fractional contribution of the explosion itself to the total ejecta. These quantities are calculated using
\begin{equation}
\begin{split}
    &M_{\text{net}} = M_{\text{final, freeze-out}}+(M_{\text{final, non freeze-out}}-M_{\text{progenitor, non freeze-out}}),\\
    &f_{\text{CCSN}} = \frac{M_{\text{net}}}{M_{\text{total}}}\, ,
\end{split}
\end{equation}
which assumes that the initial abundance is zero if the matter parcel ever reaches NSE (the freeze-out part).

For isotopes lighter than $^{20}$Ne, the CCSN contribution is negligible compared to pre-CCSN phases in most models. The most massive models (40 and 60 $M_\odot$) have ejected almost their entire hydrogen and helium shell before core collapse; thus, the CCSN contribution dominates some lighter elements, but the absolute amount is low. This can be seen in the left panel of Figure \ref{fig:net-yield} and in both panels of Figure \ref{fig:prod-factor}. The CCSN explosions start to have a significant effect on the final abundances at $^{20}$Ne by destroying about 20\% of the pre-CCSN neon mass. The net yields become positive in most models at around $^{28}$Si, and the production of most iron group isotopes is dominated by the explosion component itself. The CCSN contribution drops back to about zero at around $^{64}$Ge in models without neutron-rich phases, but for models that have some neutron-rich ejecta the explosion component can significantly contribute to heavier elements, or at least to elements near the $^{90}$Zr peak. 

\begin{figure}
    \centering
    \includegraphics[width=0.48\textwidth]{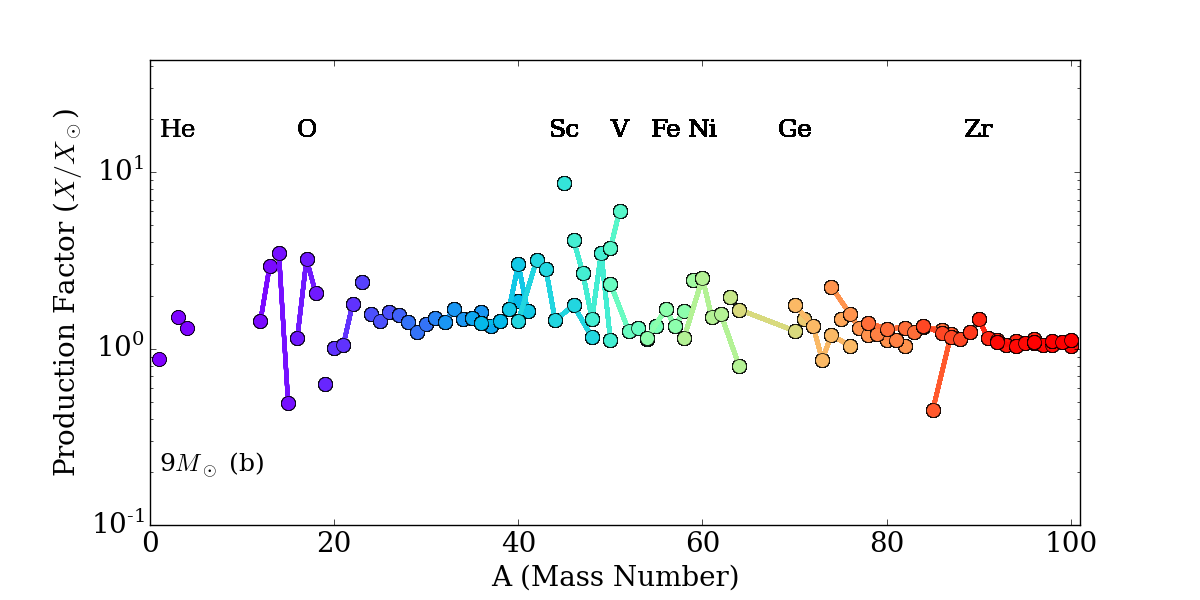}
    \includegraphics[width=0.48\textwidth]{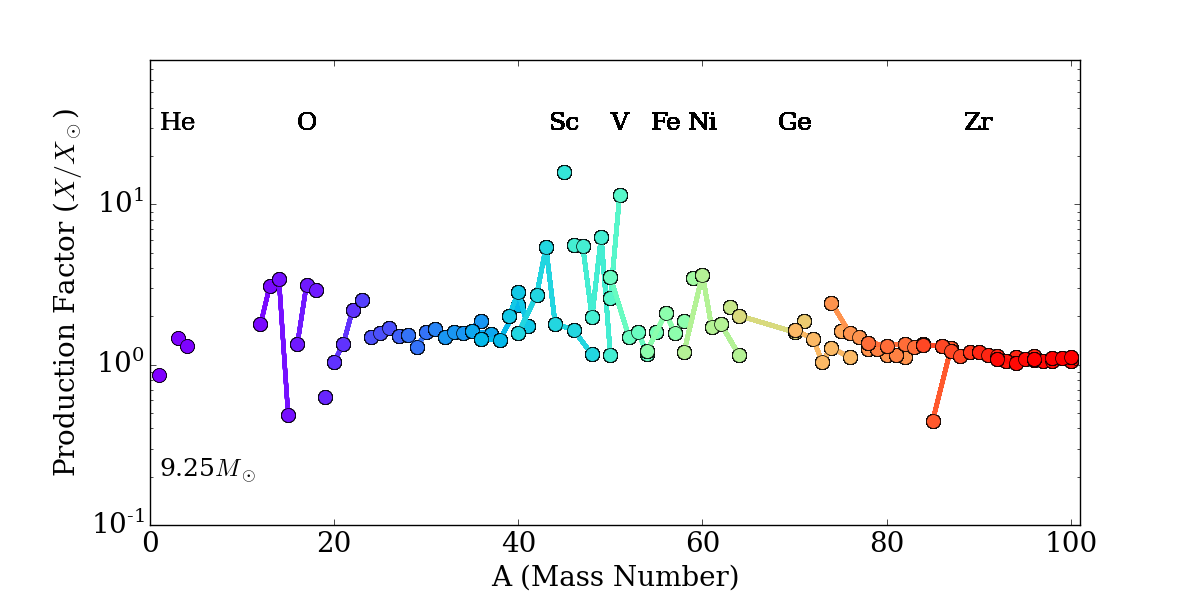}
    \includegraphics[width=0.48\textwidth]{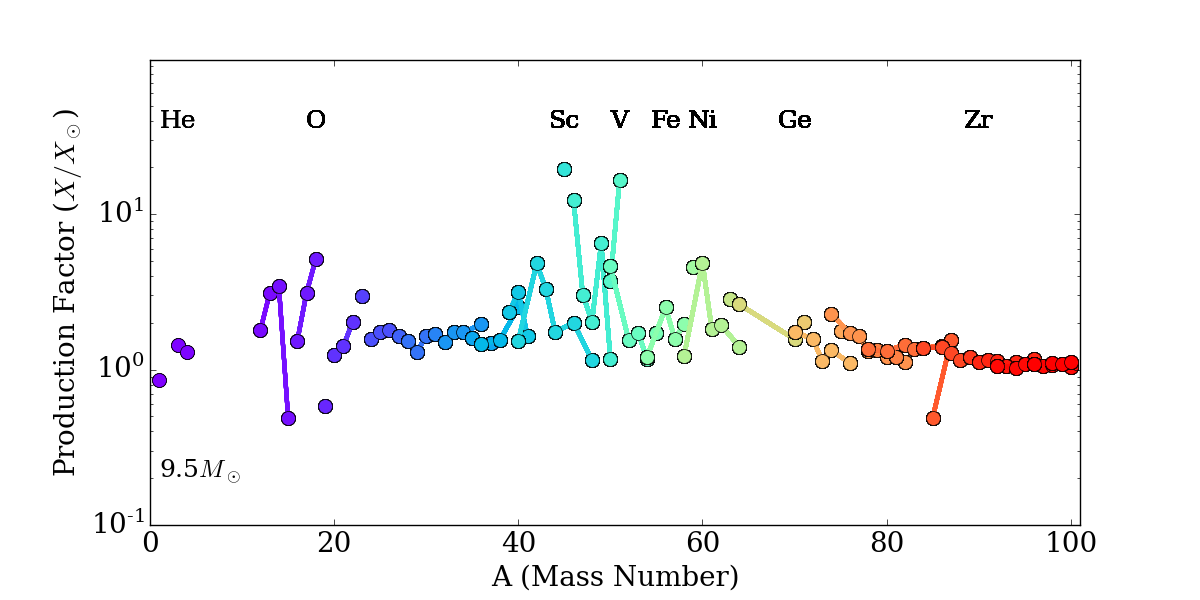}
    \includegraphics[width=0.48\textwidth]{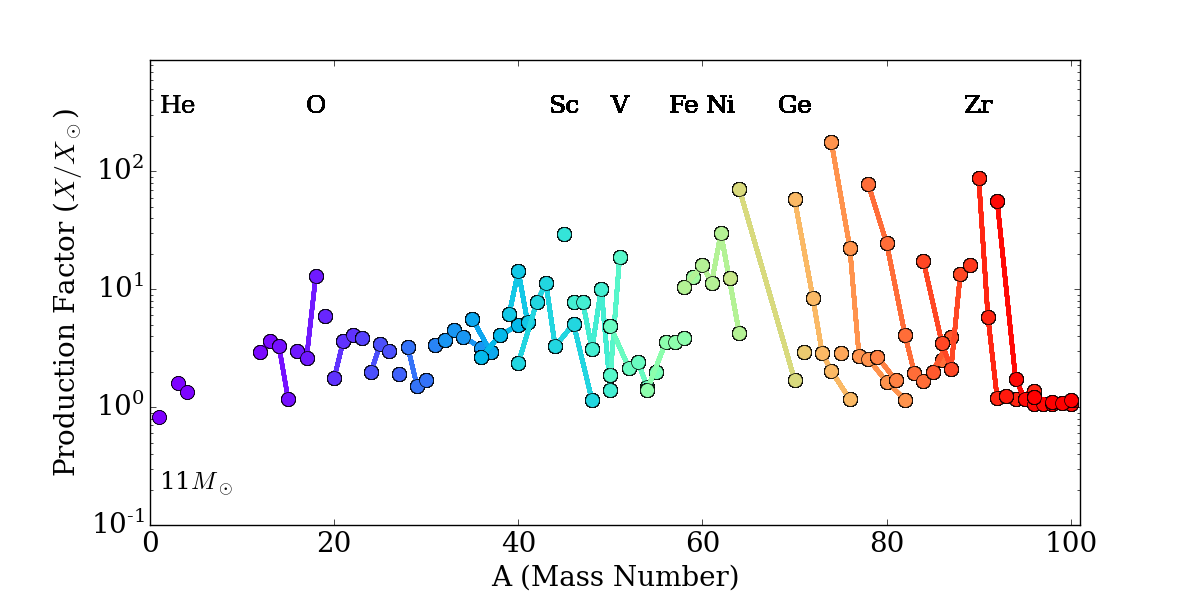}
    \includegraphics[width=0.48\textwidth]{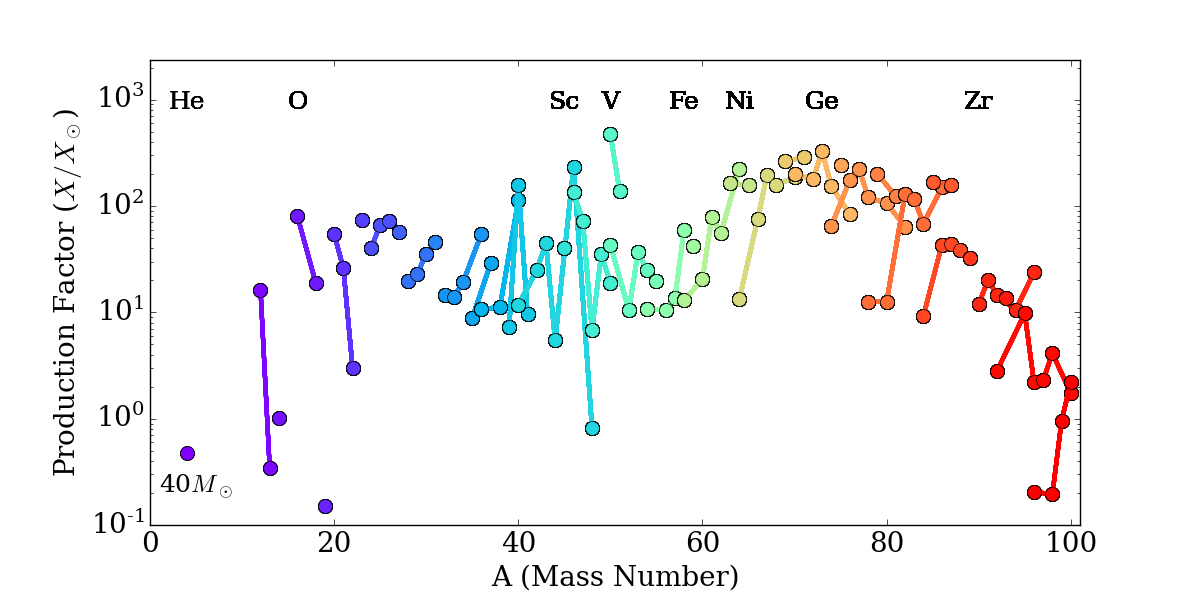}
    \includegraphics[width=0.48\textwidth]{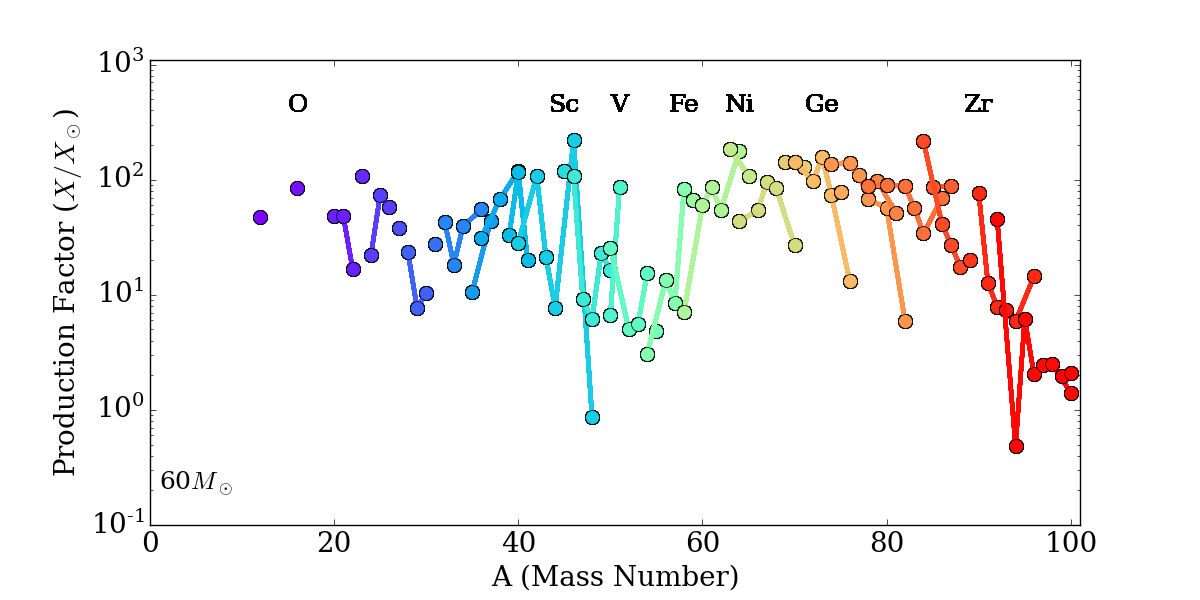}
    \caption{Production factors of stable isotopes after all decays have finished for the 9 $M_\odot$ (b), 9.25 $M_\odot$, 9.5 $M_\odot$, 11 $M_\odot$, 20 $M_\odot$, and 60 $M_\odot$ models. Models without neutron-rich components show a peak around scandium which is typical for proto-rich ejecta, while a small amount of neutron-rich ejecta will lead to the production of elements heavier than the iron group. None of our models show a very flat production factor curve as opposed to previous 1D works (e.g., \citet{sukhbold2016}, see Figure \ref{fig:comparison}).
    }
    \label{fig:stable}
\end{figure}

Figure \ref{fig:stable} shows the production factors of stable isotopes after all decays have finished. Models without neutron-rich phases show a peak around scandium, which is typical for proto-rich ejecta, while a small amount of neutron-rich ejecta leads to the production of elements heavier than the iron group. None of our 3D models show a very flat production factor curve as opposed to the 1D models of the same progenitors done by \citet{sukhbold2016} (See Figure \ref{fig:comparison}).

\subsection{Lighter $\alpha-$nuclei}
The small 19-isotope network includes $\alpha$-nuclei with $A\leq40$. We find that the final abundances of such $\alpha$-nuclei are weakly influenced by the choice of pre-CCSN networks. The small network is able to reproduce the abundances of such $\alpha$-nuclei within $\pm$50\%, which is an acceptable error compared to the uncertainties due to the multi-dimensional complexities. Therefore, we include both the S16 and the S18 models in the discussion of this subsection. 

\begin{figure}
    \centering
    \includegraphics[width=0.48\textwidth]{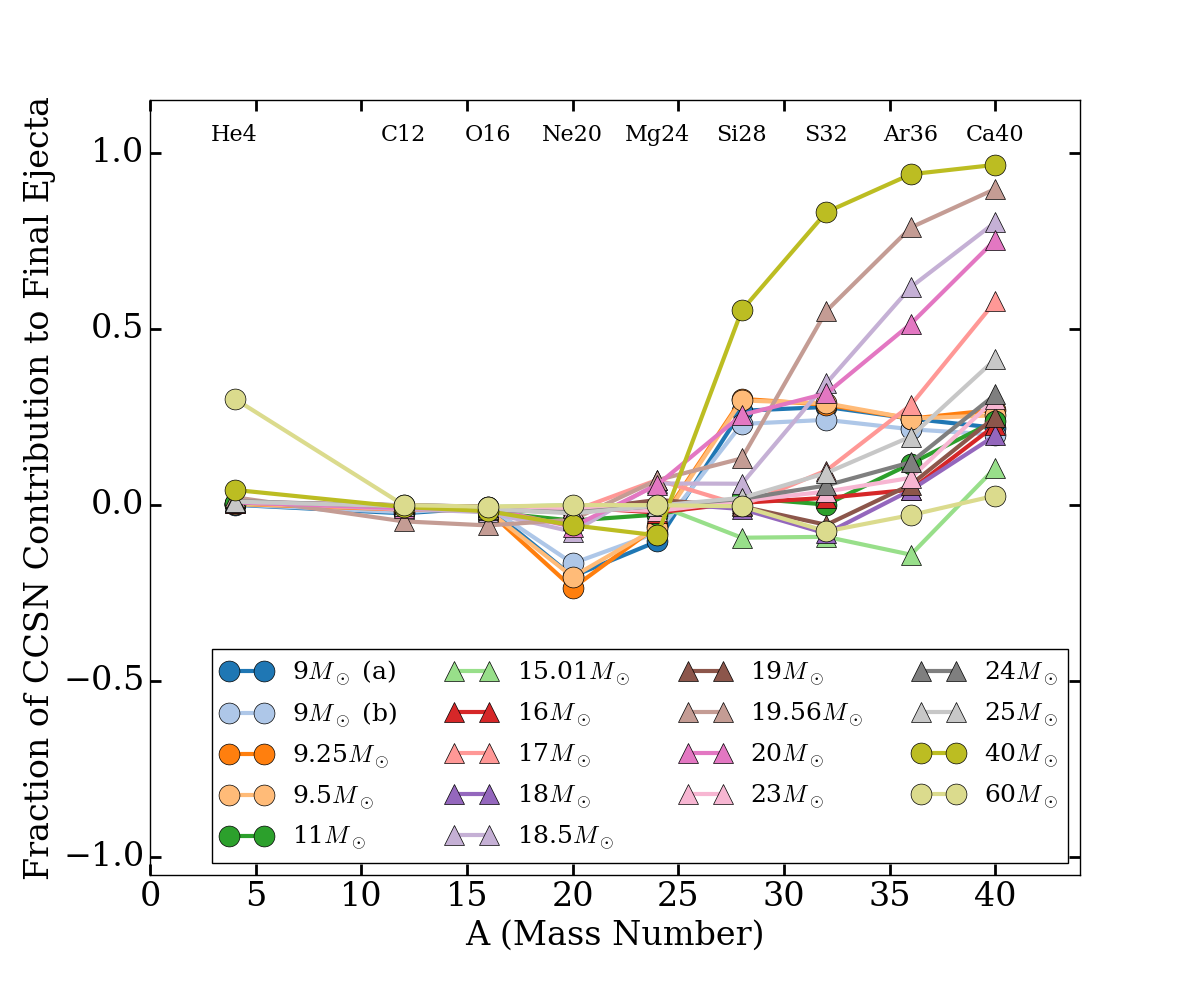}
    \includegraphics[width=0.48\textwidth]{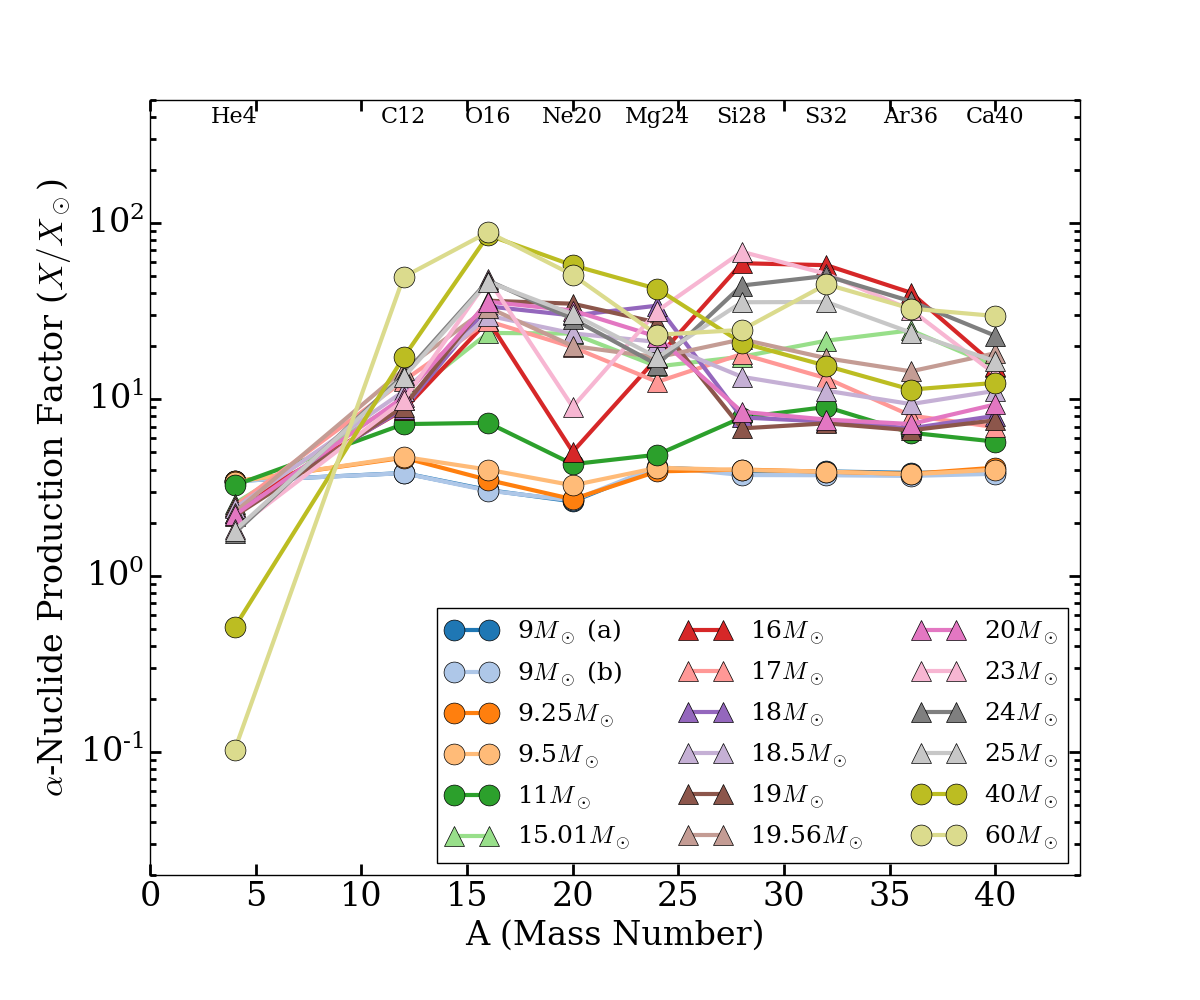}
    \caption{Fractions of explosive contributions and production factors of lighter $\alpha$-nuclei at the end of each simulation. S16 models are marked as circles, while S18 models are triangles. Elements lighter than $^{28}$Si are likely to be destroyed during the CCSN explosion, while $^{28}$Si and heavier elements will be produced. It is clear that the CCSN contributes more to the production of $^{36}$Ar and $^{40}$Ca if the explosion is more energetic (e.g., the 19.56 $M_\odot$ model), but the progenitor has a significant contribution to these elements even for the strongest explosions.}
    \label{fig:alpha}
\end{figure}

Figure \ref{fig:alpha} shows the fractional contribution of he explosion and the production factors of $A\leq40$ $\alpha$-nuclei at the end of each simulation. As described before, elements lighter than $^{28}$Si are likely to be destroyed during the CCSN explosion, while $^{28}$Si and heavier isotopes will be produced. It is also clear that the explosion component contributes more to the production of $^{36}$Ar and $^{40}$Ca if the explosion is more energetic, but the pre-explosion progenitor makes a significant contribution to these elements even for the strongest explosions (e.g., the 19.56 $M_\odot$, which has $E>2$ Bethes)\footnote{The solar-metallicity 40 $M_\odot$ progenitor is special not only because it explodes energetically (E=1.6 B), but because it has ejected a significant fraction of its envelope prior to explosion due to its high luminosity. This is why the Ar and Ca are almost all produced by the CCSN explosion itself.}. Recently, \citet{braun2023} measured the abundances of Ne, Mg, Si, S, Ar, Ca, and Fe for six supernova remnants and found that there is usually a large deviation between the observed Ar and Ca abundances and those predicted using 1D models. We see similar behavior in our 3D simulations, in which the Ne, Si, S, and Fe roughly follow the observed trends in \citet{braun2023}, while Ar and Ca are largely underproduced. This indicates that the multi-dimensional effects in CCSNe may not yet be able to explain the discrepancy. Further improvements in both progenitor models and the CCSN simulations may be required to resolve this issue. {Uncertainties in nuclear reaction rates may also play a role here.}

\subsection{Radioactive Isotopes}
In this section, we focus on the radioactive isotopes, especially on $^{44}$Ti, $^{56}$Ni, and $^{57}$Ni. $^{44}$Ti and $^{56}$Ni are included by the 19-isotope network, and the $^{57}$Ni is dominated by matter that has reached NSE, so the differences between the two networks used by S16 and S18 progenitors are small. Therefore, we include both S16 and S18 models in this subsection. However, it is important to note that the 19-isotope network generally produced much more ``pseudo-$^{44}$Ti" in the outer stellar envelopes which is never seen using the large network. Thus, we don't include the progenitor contribution of $^{44}$Ti in any model.

\begin{table}
    \centering
    \begin{tabular}{c|ccccc}
    ZAMS Mass [$M_\odot$]  &$^{44}$Ti (Freeze-Out) [$10^{-6}M_\odot$]  &$^{44}$Ti (Non Freeze-Out) [$10^{-6}M_\odot$] &$^{56}$Ni [$10^{-2}M_\odot$] &$^{56}$Fe (Envelope) [$10^{-2}M_\odot$] \\
    \hline
    9(a)    &0.61 &0.37 &0.168  &0.922  \\
    9(b)    &2.47 &0.32 &0.612  &0.922\\
    9.25    &6.26 &0.74 &1.04   &0.944\\
    9.5     &4.44 &0.77 &1.47   &0.967\\
    11      &25.8 &1.83 &2.92   &1.10\\
    40      &32.3 &13.4 &16.5   &0.222\\
    60      &18.6 &10.8 &9.04   &0.036\\
    \hline
    15.01   &3.76 &12.6 &5.42   &0.0\\ 
    16      &6.74 &5.86 &6.06   &0.0\\
    17      &4.35 &4.47 &4.65   &0.0\\
    18      &5.45 &7.54 &8.58   &0.0\\
    18.5    &5.66 &13.8 &12.2   &0.0\\
    19      &14.2 &6.23 &7.74   &0.0\\
    19.56   &15.2 &19.5 &25.6   &0.0\\
    20      &4.01 &11.3 &8.42   &0.0\\
    23      &12.6 &14.9 &8.77   &0.0\\
    24      &6.37 &16.6 &12.5   &0.0\\
    25      &11.2 &20.7 &10.4   &0.0\\
    \end{tabular}
    \caption{This table summarizes the production of $^{44}$Ti and $^{56}$Ni in different components of the ejecta at the end of each simulation. The amount of $^{56}$Fe in the outer stellar envelope produced during the pre-CCSN phase is also shown. As indicated in Figure \ref{fig:ni56-ti44-t}, although the $^{56}$Ni yield in most models approaches its asymptotic value, the amount of $^{44}$Ti is still quickly increasing at the end of the simulation. By then, the shock temperature has dropped below 1.5 GK and explosive nucleosynthesis of $^{44}$Ti is terminated. Thus, the late-time contribution of $^{44}$Ti is due mostly to the freeze-out component of the ejecta. Hence, $^{44}$Ti production will eventually be dominated by the freeze-out part (e.g., in the 11 $M_\odot$ model), but this final phase will take many extra seconds in most models. $^{56}$Ni production in 9(a) is four times lower than that in 9(b) because 9(a) has more neutron-rich ejecta and favors the production of $^{58}$Ni and $^{60}$Ni instead of $^{56}$Ni. For low mass progenitors, the outer envelope contributes about $10^{-2}M_\odot$ of $^{56}$Fe, which is comparable to more than the $^{56}$Ni produced during the explosion. This iron component will have a very different spatial distribution compared to $^{56}$Ni and $^{44}$Ti. This is discussed in Section \ref{sec:morphology}. Unfortunately, the S18-derived progenitors with the 19-isotope network don't provide the correct $^{56}$Fe abundances in the outer envelopes, and it is unclear if this component will play an important role in the final iron abundances of more massive models. {In general, the iron in the outer envelopes is of order $10^{-2}M_\odot$ for each model.}}
    \label{tab:ti44-ni56}
\end{table}

\begin{figure}
    \centering
    \includegraphics[width=0.48\textwidth]{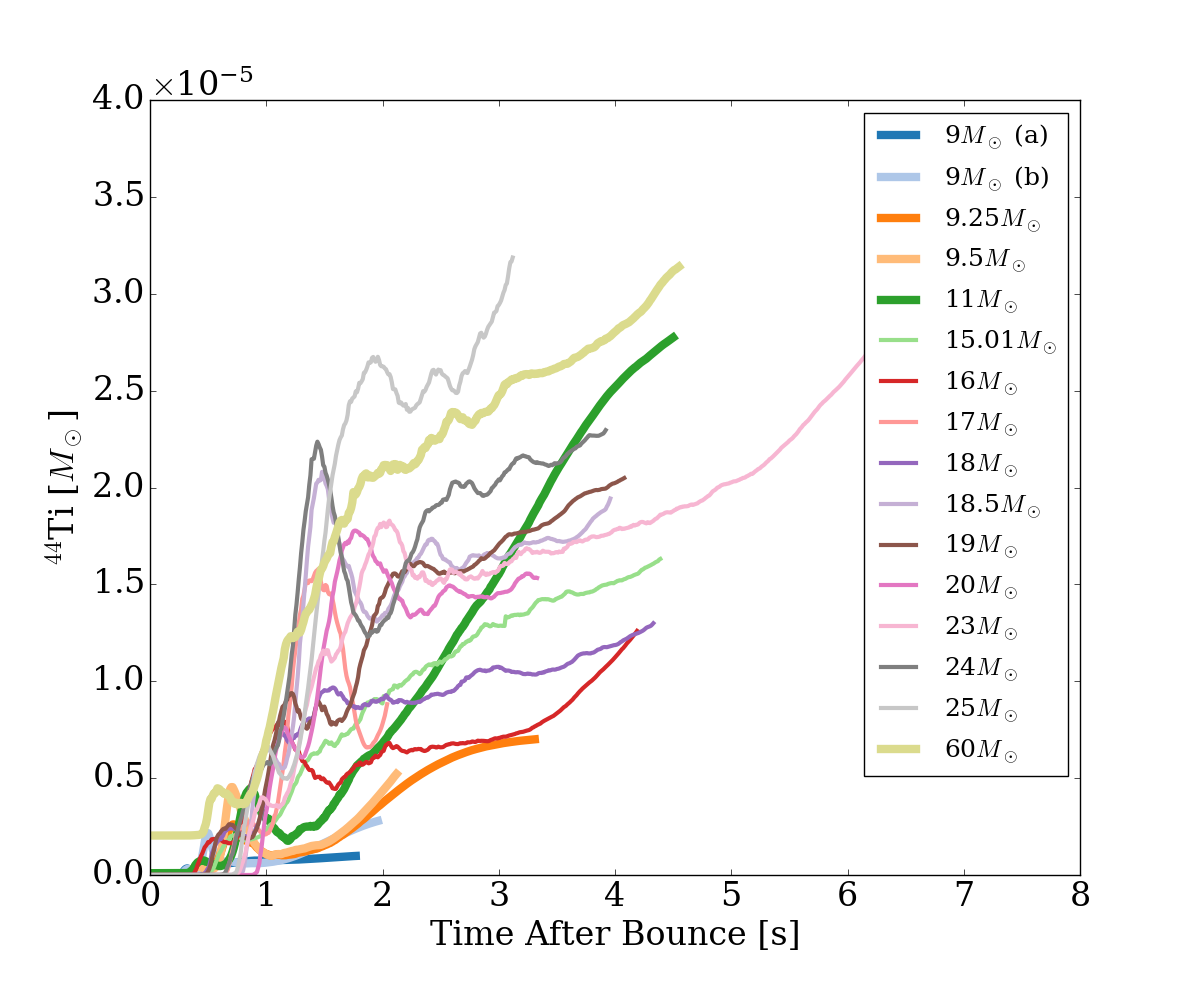}
    \includegraphics[width=0.48\textwidth]{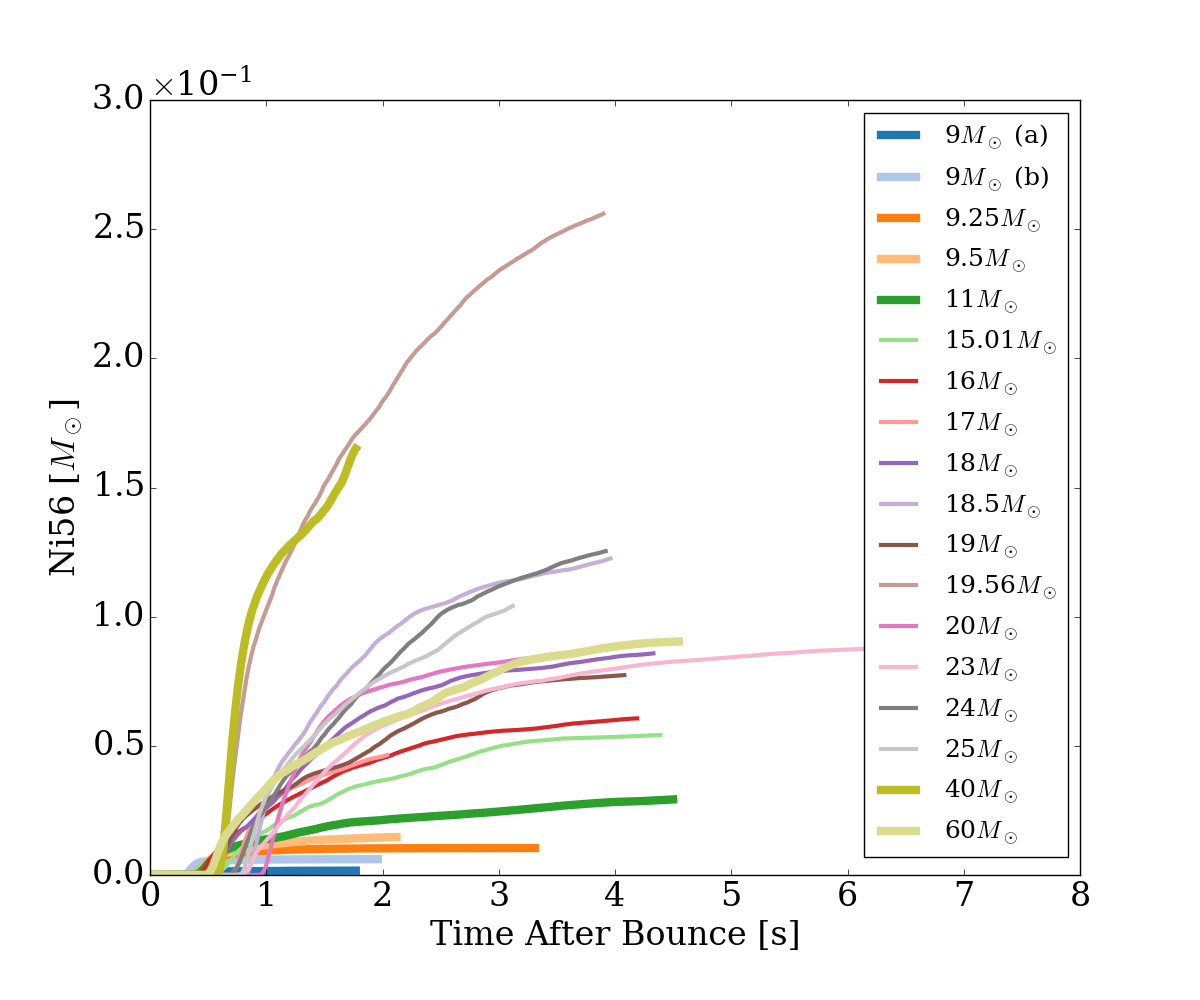}
    \caption{Temporal evolution of the $^{44}$Ti and $^{56}$Ni abundances. S16 models are marked as thicker lines. Although the $^{56}$Ni yield in most models approaches its asymptotic value, the amount of $^{44}$Ti is still quickly increasing. At the end of each simulation, the shock temperature has dropped below 1.5 GK and the explosive nucleosynthesis of $^{44}$Ti is terminated. Thus, the late-time contribution of $^{44}$Ti is mostly due to the freeze-out component of the ejecta. $^{44}$Ti production will eventually be dominated by the freeze-out part (e.g., as in the 11 $M_\odot$ model), but this final phase will take many extra seconds to reach in most models.} 
    \label{fig:ni56-ti44-t}
\end{figure}

Table \ref{tab:ti44-ni56} summarizes the production of $^{44}$Ti and $^{56}$Ni. The amount of $^{56}$Fe produced in pre-CCSN phases is also included. ``(Non) Freeze-Out" means that the matter has (not) reached NSE ($T>0.6$ MeV) before its ejection, while ``Envelope" represents the matter that experiences no supernova nucleosynthesis, most of which is exterior to our simulation grid. As shown in Figure \ref{fig:ni56-ti44-t}, although the $^{56}$Ni yield in most models approaches asymptotic values, the amount of $^{44}$Ti is still quickly increasing at the end of each simulation. At that time, the shock temperature has dropped below $\sim$1.5 GK and the explosive nucleosynthesis of $^{44}$Ti is almost terminated. Thus, the later-time contribution of $^{44}$Ti is mostly due to the freeze-out component of the ejecta. The $^{44}$Ti production will eventually be dominated by the freeze-out part (see, for example, the 11 $M_\odot$ model), but this final phase will take many extra seconds to achieve in most models.

\begin{figure}
    \centering
    \includegraphics[width=0.48\textwidth]{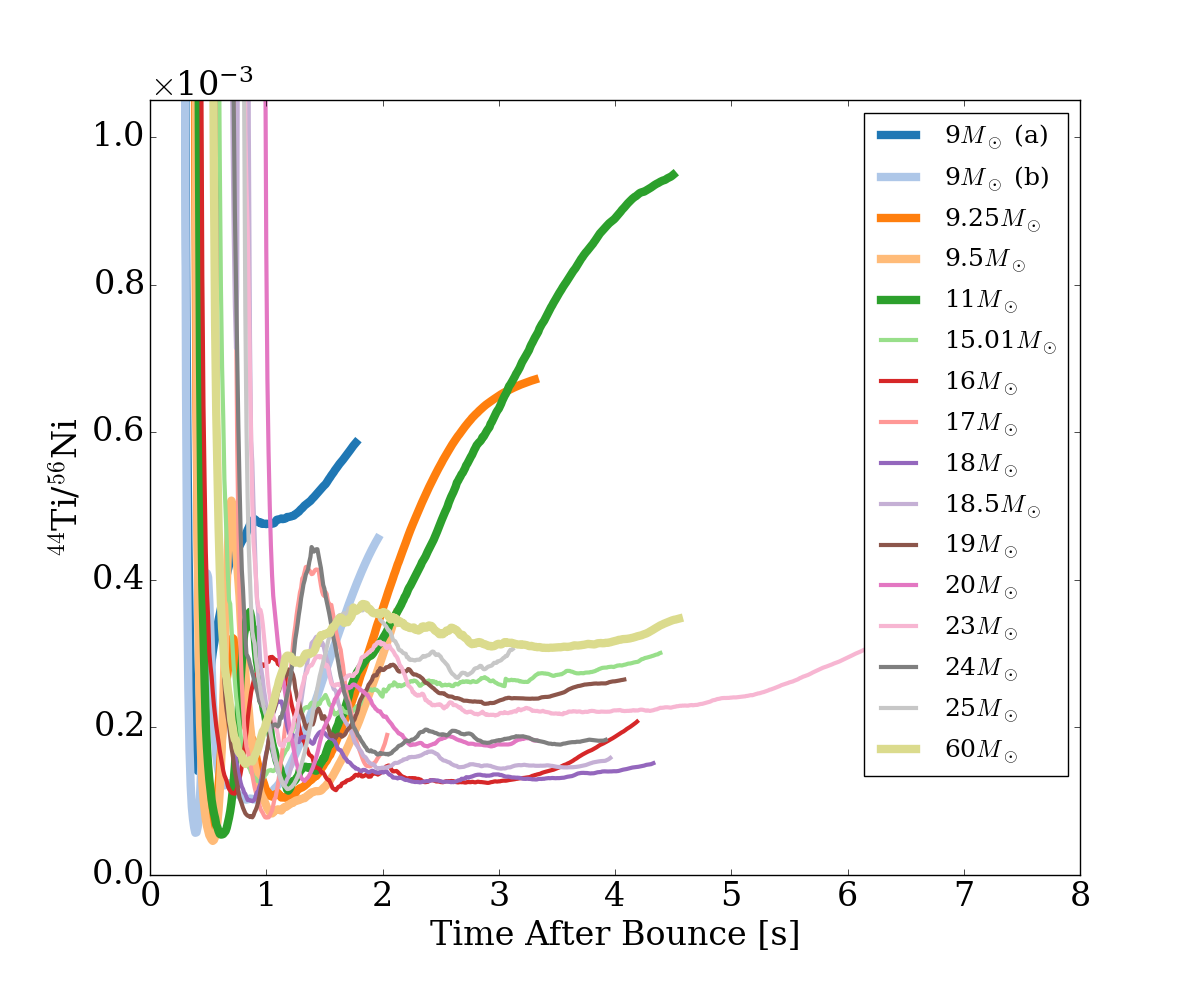}
    \includegraphics[width=0.48\textwidth]{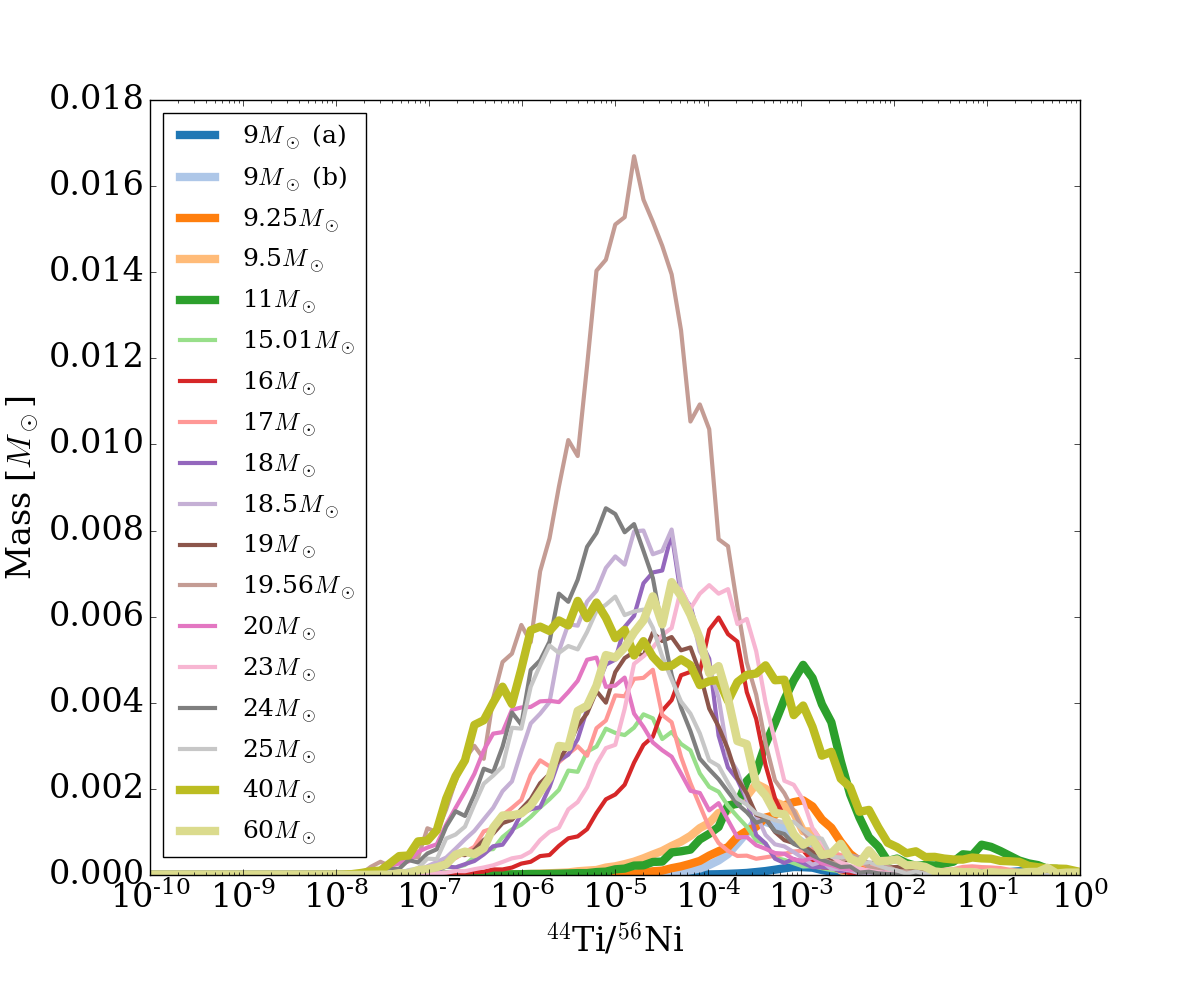}
    \caption{The $^{44}$Ti/$^{56}$Ni ratio as a function of time for each simulation, together with the histogram of the distribution of the ejecta mass as a function of this ratio at the end of each simulation. S16 models are marked as thicker lines. It is clear that $^{44}$Ti/$^{56}$Ni ratios in less massive models are quickly increasing with time by the end of the simulation and the distributions are peaked at around $10^{-3}$. $^{44}$Ti/$^{56}$Ni ratios in more massive models seem to stay at low values, but this is because $^{44}$Ti is produced in a much longer timescale compared to $^{56}$Ni, especially in more massive models. Notice that in Figure \ref{fig:ni56-ti44-t} the amount of $^{44}$Ti is still increasing at the end of all simulations. The late phase synthesis of $^{44}$Ti is explained by \citet{magkotsios2011} and \citet{sieverding2023}. The non-monotonic temperature histories of the ejecta allow multiple transitions between local equilibria of nuclear reactions and non-equilibrium phases, which enhances the production of $^{44}$Ti.}
    \label{fig:ni56-ti44}
\end{figure}

\begin{figure}
    \centering
    \includegraphics[width=0.48\textwidth]{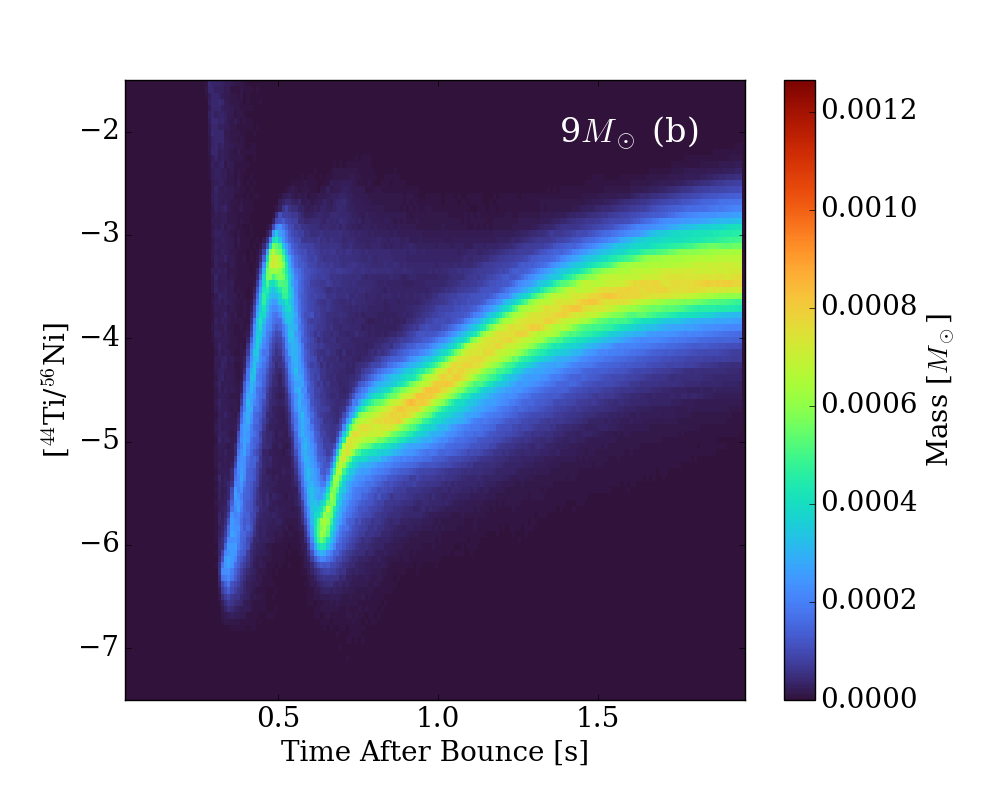}
    \includegraphics[width=0.48\textwidth]{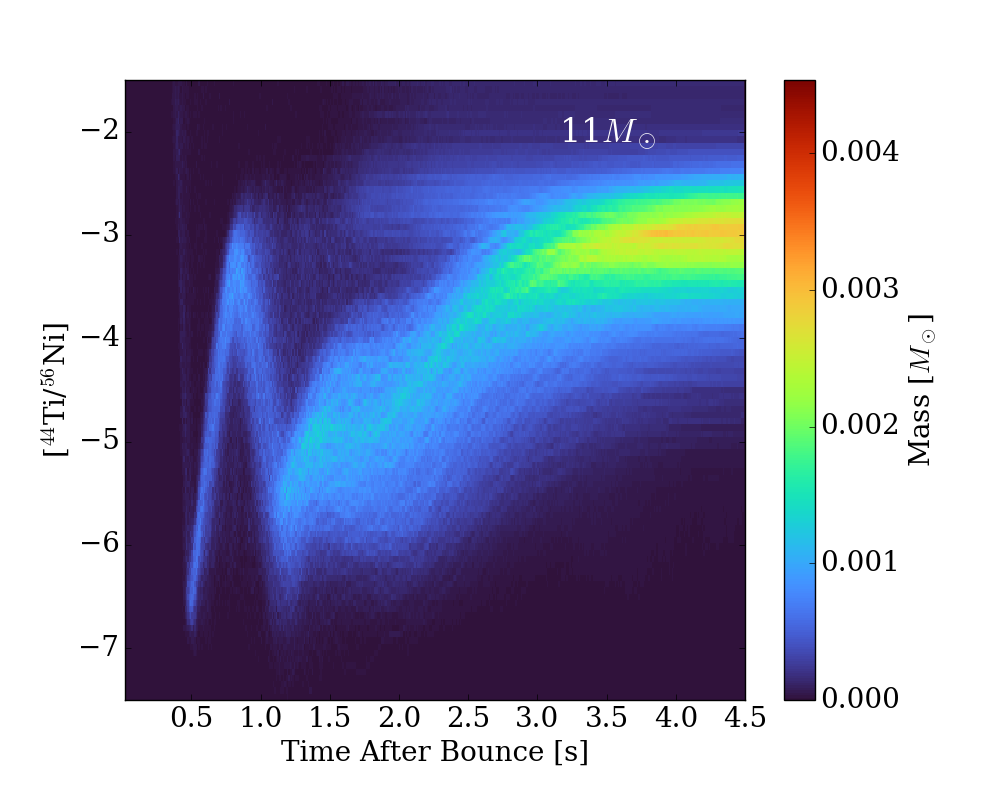}
    \includegraphics[width=0.48\textwidth]{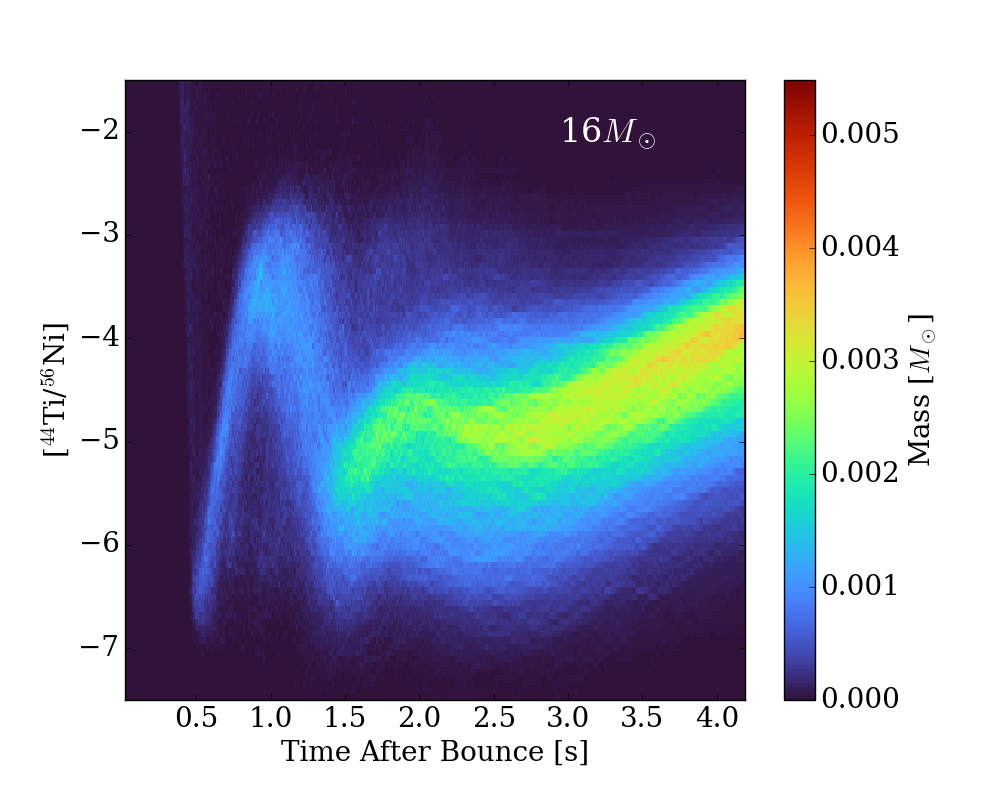}
    \includegraphics[width=0.48\textwidth]{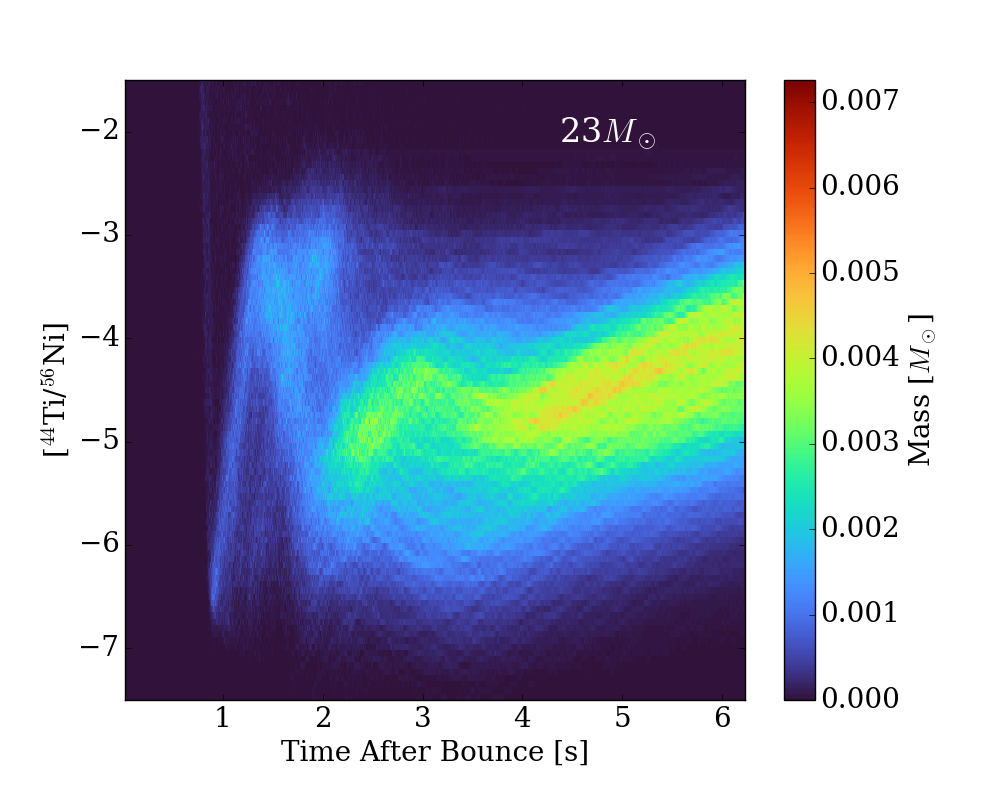}
    \caption{Distributions of $^{44}$Ti/$^{56}$Ni as a function of time for the freeze-out component after it leaves NSE. The first peak is due to the production of $^{44}$Ti, but the ratio then drops to $\sim10^{-6}$ when $^{56}$Ni starts to be synthesized. Afterwards, the ratio starts to increase slowly, and it generally takes many seconds to reach its asymptotic value. More massive models have wider bands on this plot because long-lasting accretion makes the ejecta motion more complicated. As a result, they need more time to finish the synthesis of $^{44}$Ti. None of our 3D simulations are able to reach the termination stage of $^{44}$Ti production, even for our longest 6.3-second model.}
    \label{fig:ni56-ti44-ejected-t}
\end{figure}

{Observed high ratios\footnote{{The measured values in supernova remnants vary from $\sim10^{-4}$ to a few times $10^{-3}$, with large uncertainties \citep{weinberger2020}.}} of $^{44}$Ti/$^{56}$Ni \citep{weinberger2020} have long been recognized as a problem for theoretical CCSN models \citep{the2006}.}
Figure \ref{fig:ni56-ti44} shows the $^{44}$Ti/$^{56}$Ni ratio as a function of time for each simulation, together with the histogram of the distribution of the ejecta mass as a function of this ratio at the end of each simulation. It is clear that the $^{44}$Ti/$^{56}$Ni ratio in less massive models is quickly increasing with time and that the distributions are peaked at around $10^{-3}$. $^{44}$Ti/$^{56}$Ni ratios in more massive models seem to stay at low values, but this is because $^{44}$Ti is produced on a much longer timescale compared to $^{56}$Ni, especially in the more massive models. Notice that in Figure \ref{fig:ni56-ti44-t} the amount of $^{44}$Ti is still increasing at the end of all simulations. The late phase synthesis of $^{44}$Ti is discussed by \citet{magkotsios2011} and \citet{sieverding2023}. The non-monotonic temperature histories of the ejecta allow multiple transitions between local equilibria of nuclear reactions and non-equilibrium phases, which enhances the production of $^{44}$Ti. Figure \ref{fig:ni56-ti44-ejected-t} shows the temporal evolution of the $^{44}$Ti/$^{56}$Ni ratios of the freeze-out component after it leaves NSE. The first peak is due to the production of $^{44}$Ti, but the ratio then drops to $\sim10^{-6}$ when $^{56}$Ni starts to be synthesized. Then the ratios start to increase slowly, and generally take many seconds to reach their asymptotic values. More massive models have wider bands on this plot because their long-lasting accretion makes the ejecta motion more complicated. As a result, they need much longer to finish the synthesis of $^{44}$Ti. None of our 3D simulations are able to reach the termination stage of $^{44}$Ti production, even for our longest 6.3-second model.

In short, $^{44}$Ti production is sensitive to the complex ejecta histories in 3D simulations at late post-bounce phases, and generally takes much longer to reach the asymptotic state than $^{56}$Ni. {This timescale difference between the production of $^{44}$Ti and $^{56}$Ni is related to the morphology differences, as discussed in Section \ref{sec:morphology}.} Long-term 3D CCSN simulations with full neutrino radiation transport are required for the study of $^{44}$Ti production, because the ejecta motion is strongly dependent upon the details of inflow and outflow motions, which are sensitive to neutrino heating at late times. {Due to these complexities, it remains to be seen whether simplified models, such as using spherical symmetric inner boundary conditions \citep{stockinger2020} and approximate heating terms based on 1D simulations \citep{bollig2021}, can accurately predict $^{44}$Ti yields.}

\begin{figure}
    \centering
    \includegraphics[width=0.48\textwidth]{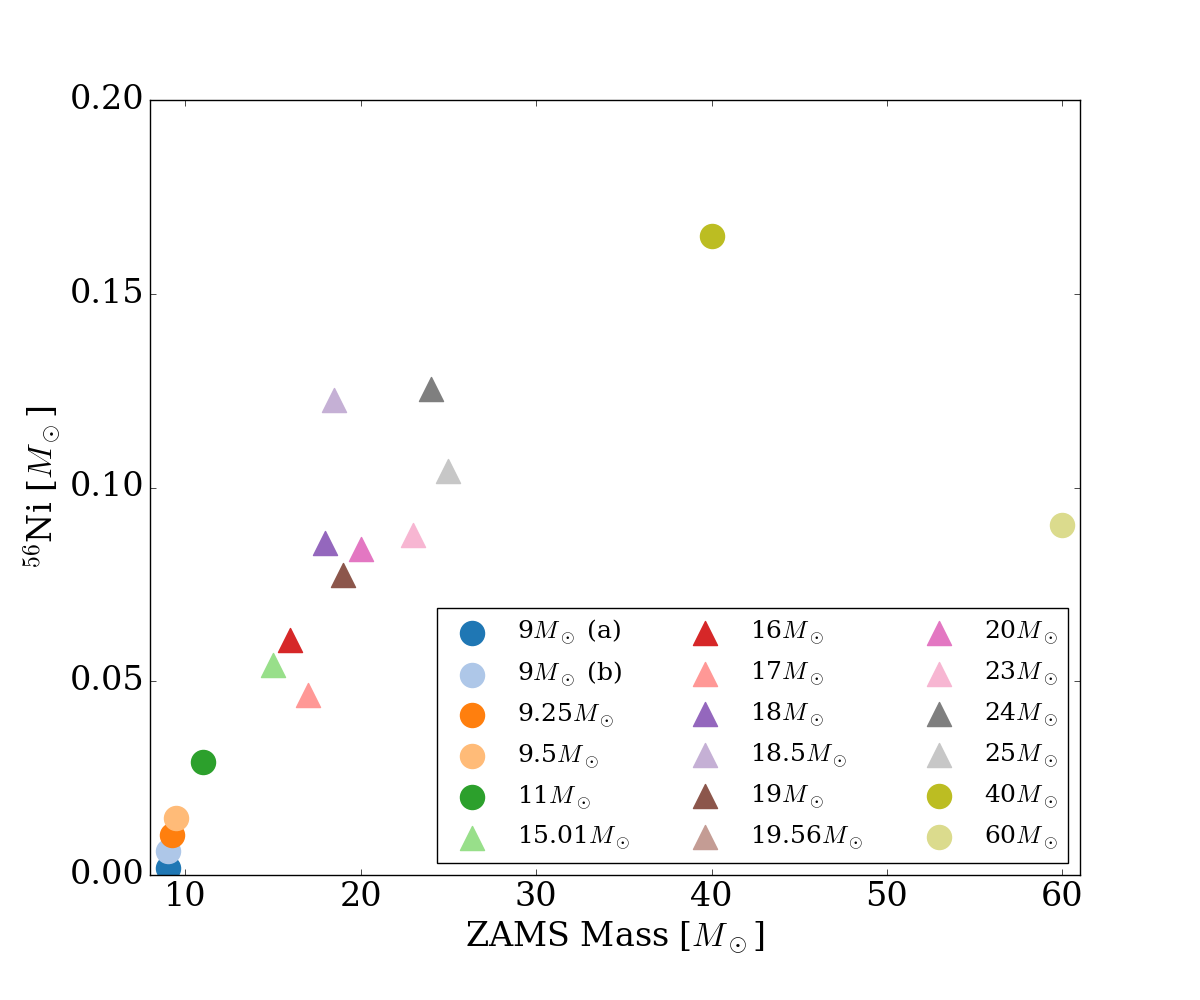}
    \includegraphics[width=0.48\textwidth]{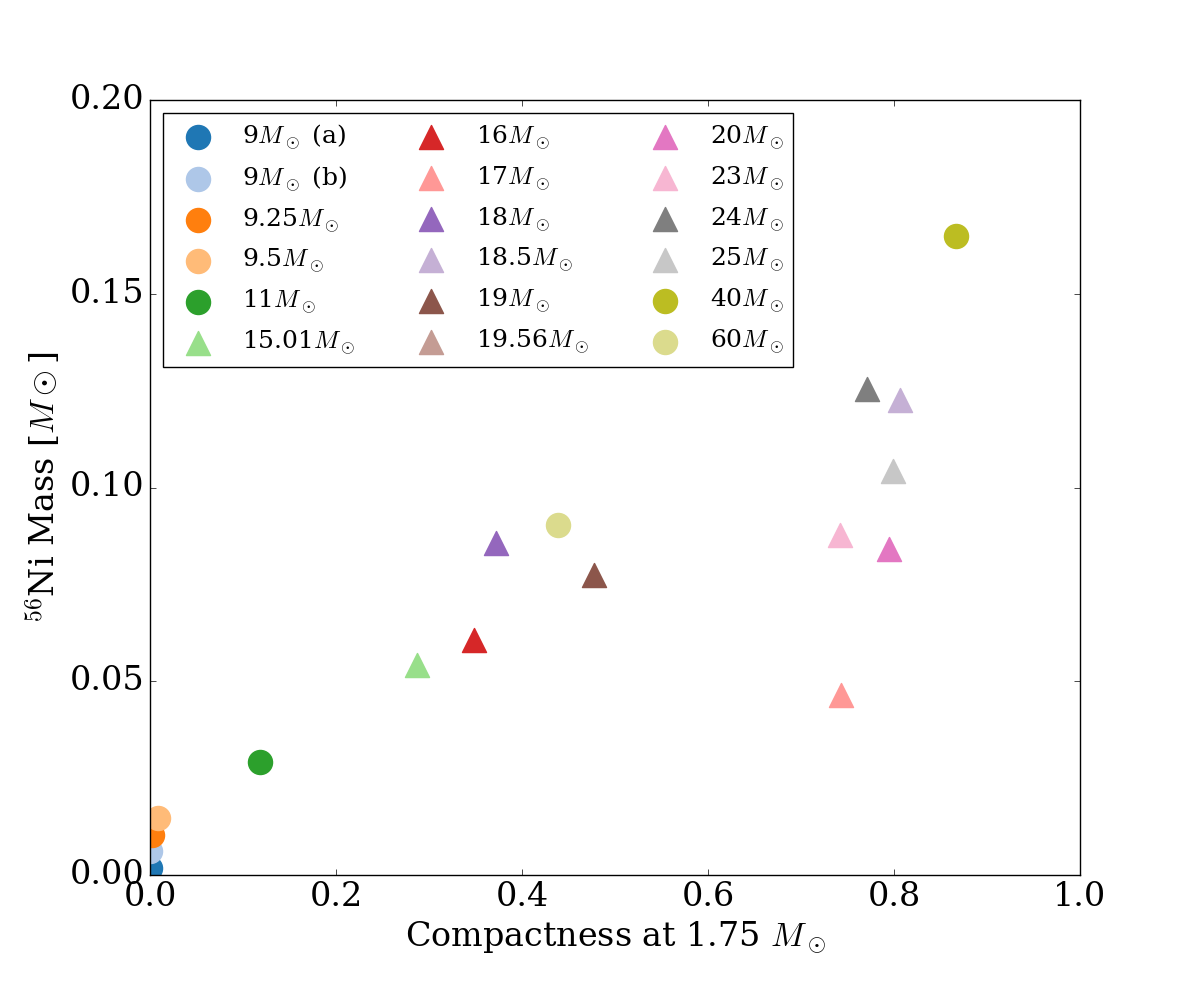}
    \caption{Relations between $^{56}$Ni yield and progenitor ZAMS mass (left) and compactness parameter (right). S16 models are marked as circles, while S18 models are triangles. Although more massive models seem to have higher $^{56}$Ni production, the relation between nickel yield and ZAMS mass is non-monotonic. However, an almost linear relation is found between $^{56}$Ni and the compactness parameter which directly represents the initial progenitor structures. The 17 $M_\odot$ model on the right panel seems to be an outlier because the simulation is the shortest one among all massive models and the $^{56}$Ni is still produced quickly. It is very likely that the 17 $M_\odot$ model will reach the linear relation when it is continued further.}
    \label{fig:ni56-E-cpt}
\end{figure}

On the other hand, most our simulations have reached the asymptotic stages of $^{56}$Ni production. Therefore, we are able to study the relation between $^{56}$Ni yield and the progenitor properties. Figure \ref{fig:ni56-E-cpt} shows the amount of $^{56}$Ni as a function of progenitor ZAMS mass and the compactness parameter at 1.75 $M_\odot$\footnote{The compactness parameter is defined to be $\frac{M/M_\odot}{R(M)/1000\text{km}}$.}. Although more massive models seem to have higher $^{56}$Ni production, the relation between nickel yield and ZAMS mass is non-monotonic. However, an almost linear relation is found between $^{56}$Ni and the compactness parameter, which directly represents the progenitor structure. The 17 $M_\odot$ model on the right panel seems to be an outlier because the simulation is the shortest duration one among all our massive models and $^{56}$Ni is still being produced. It is very likely that the 17 $M_\odot$ model will reach the linear relation shown in Figure \ref{fig:ni56-E-cpt} when it is continued further.

\begin{figure}
    \centering
    \includegraphics[width=0.48\textwidth]{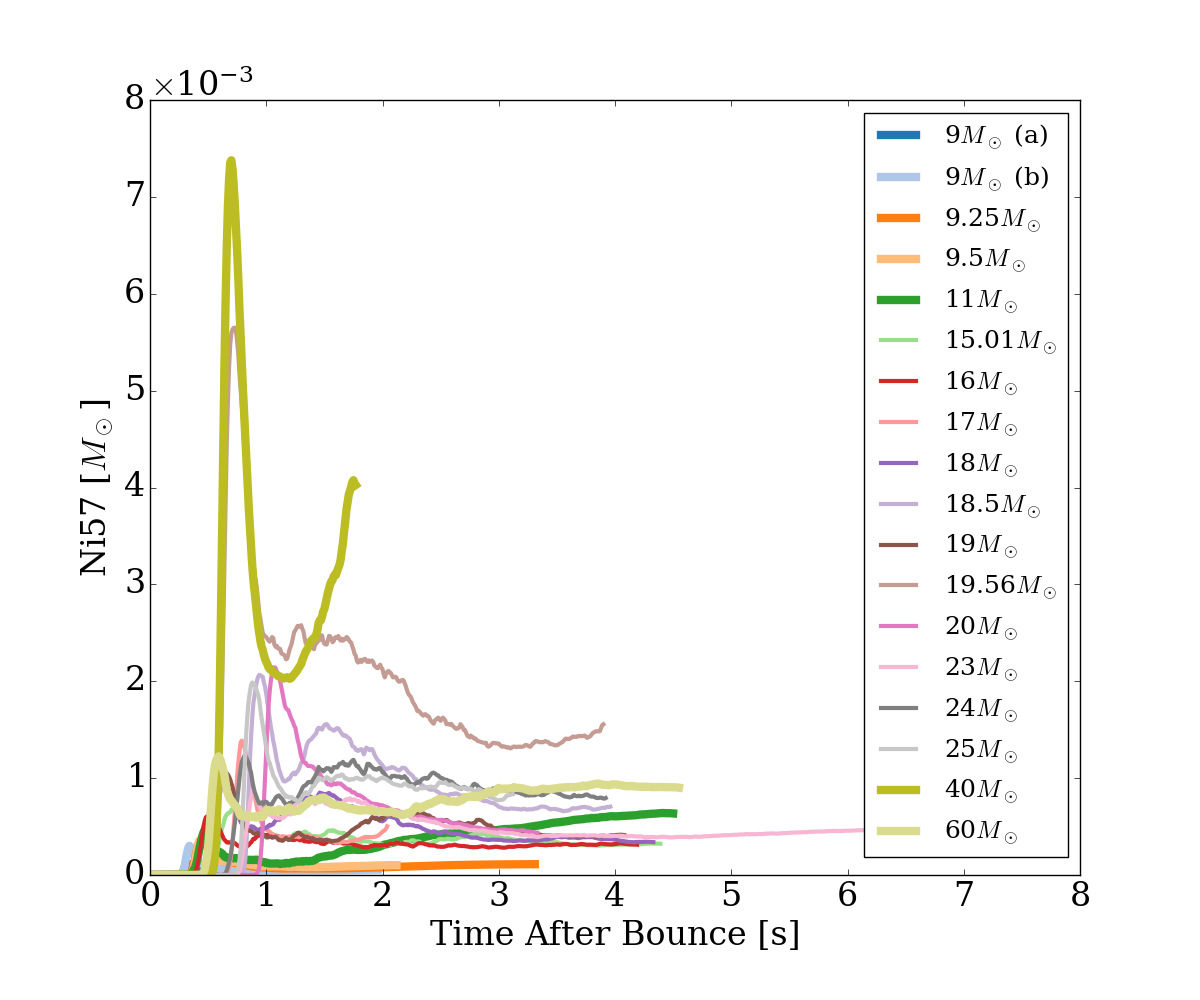}
    \includegraphics[width=0.48\textwidth]{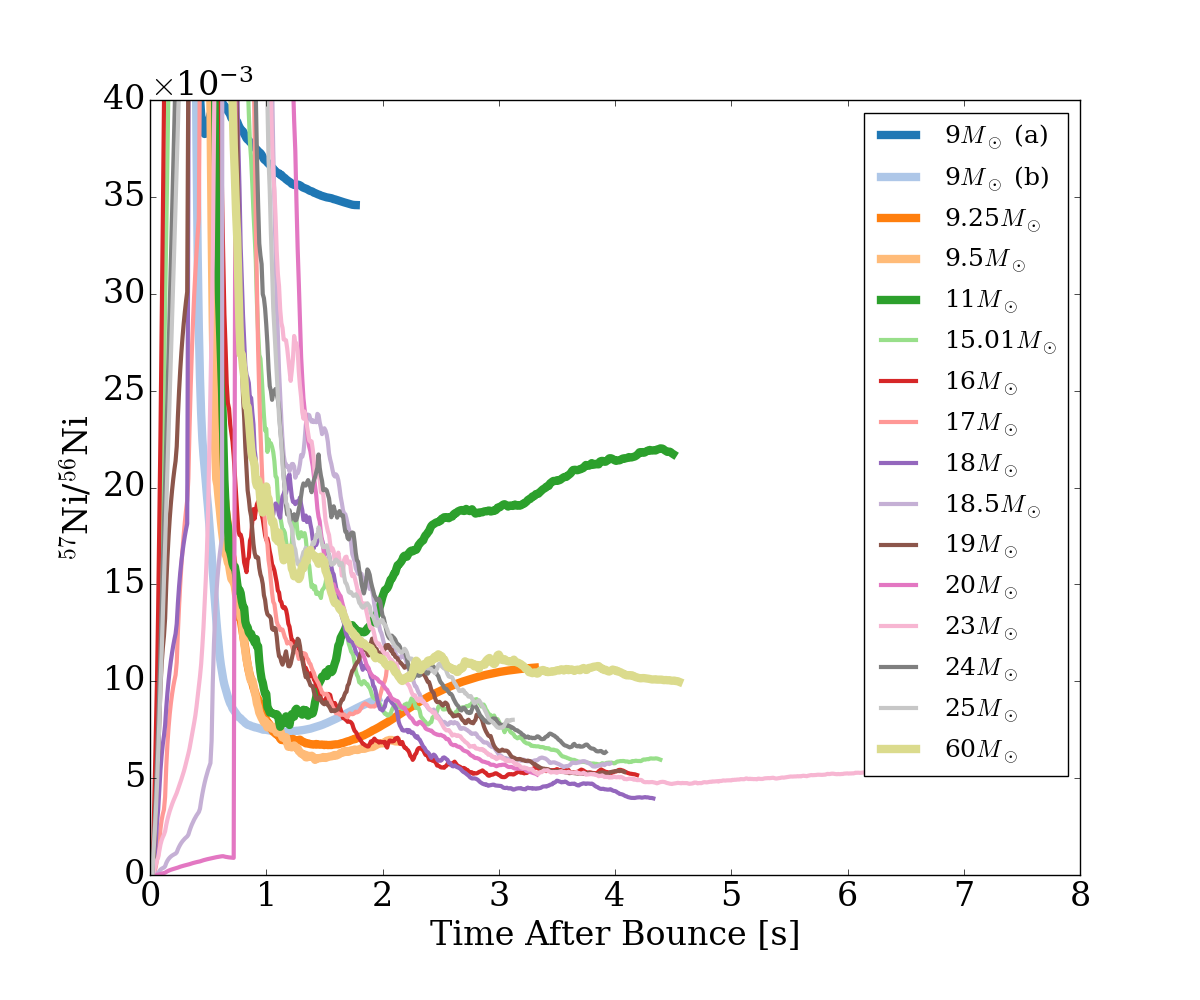}
    \caption{$^{57}$Ni yield and $^{57}$Ni/$^{56}$Ni ratio as a function of time. The S16 models are marked as thicker lines. Models without any neutron-rich phases have final $^{57}$Ni/$^{56}$Ni ratios around $5\times10^{-3}$, while stochastic neutron-rich phases can easily increase this number by a factor of a few. Faster outflows also lead to higher $^{57}$Ni production, which means that later neutrino-driven winds may have higher $^{57}$Ni abundances. However, since the mass outflow rates are decaying almost exponentially with time \citep{wang2023}, it is unclear how much more $^{57}$Ni will be contributed from the later winds.}
    \label{fig:ni57}
\end{figure}

$^{57}$Ni is another interesting radioactive isotope produced by CCSN explosions. It plays a role in gamma-ray observations has been observed in SN1987a \citep{kurfess1992,diehl2023}. Figure \ref{fig:ni57} shows the temporal evolution of $^{57}$Ni abundances and $^{57}$Ni/$^{56}$Ni ratios for all models. Models without any neutron-rich phases have final $^{57}$Ni/$^{56}$Ni ratios around $5\times10^{-3}$, while stochastic neutron-rich phases can easily increase this number by a factor of a few. Faster outflows also lead to higher $^{57}$Ni production, which means that later neutrino-driven winds may have higher $^{57}$Ni abundances. However, since the mass outflow rates are decaying almost exponentially with time \citep{wang2023}, it is unclear how much more $^{57}$Ni will be contributed from the later winds. This is another example that calls for more long-term 3D CCSN simulations.

\subsection{Heavier Elements}
Elements heavier than the iron group are also synthesized in CCSN explosions via several different processes. Since the S18 progenitors have only the small 19-isotope network which doesn't contain such heavy elements, we include in this subsection only the simulations starting from the S16 progenitors.

The rapid neutron-capture process (r-process) accounts for the production of about half of the heavy elements in general. Although the strong r-process requiring extreme neutron-rich conditions seems impossible in CCSNe, the weak r-process {(also referred to as the $\alpha$-process \citep{woosley1992})} is clearly seen in some of our simulations and it produces many heavy elements up to the $^{90}$Zr peak. Table \ref{tab:r-process} summarizes the production of some weak r-process isotopes. The percentage in the bracket represents the fractional contribution from the explosive phase. Models without neutron-rich ejecta show negligible influence on such weak r-process isotopes, but CCSN explosions with neutron-rich phases contribute significantly to such isotopes, even if the neutron phase is very short (e.g., the 9(b) and 60 $M_\odot$ models, see Figure \ref{fig:ye-t}). The production of these isotopes will be dominated by the CCSN explosion if the neutron-rich phase is longer (e.g., the 9(a) and 11 $M_\odot$ models). We have mentioned in previous sections that the $Y_e$ evolution seems stochastic. Thus, the production of such r-process elements will be stochastic as well. 

Another process that may contribute to the r-process isotopes is the $\nu p$-process \citep{frohlich2006,Pruet2006}. In this process, free protons are converted into neutrons by absorbing anti-electron type neutrinos, and are then captured by heavy nuclei. This process is included by default in our nucleosynthesis calculations. To study the effects of the $\nu p$-process, we turn off the neutrino interactions in SkyNet and compare the results. Although some tracers can be influenced, {their masses are too low to have} significant effects on the $\nu p$-process and the final abundances of the heavy elements. This is probably because the ejecta are on average not proton-rich enough and the entropies are on average too low compared to those required by the $\nu p$-process. Although models such as the 9(b) model have matter with $Y_e>0.65$, such high-$Y_e$ ejecta are ejected later in the neutrino-driven winds with very high radial velocities and there is not much time for the $\nu p$-process to happen. The neutrino luminosities are lower at later time, which may also suppress the $\nu p$-process.

However, it is still too early to rule out the possibility of $\nu p$-process in CCSNe due to the following caveats: (a) The $Y_e$ evolution seems stochastic and the high $Y_e$ required by $\nu p$-process may not always occur as late as in the 9(b) case. (b) This comparison is done using only the S16 models, which covers only a small range of progenitor ZAMS masses. More complex ejecta motions due to long-lasting accretion in the more massive models may increase the chances for the interactions between neutrinos and high $Y_e$ ejecta, as opposed to the fast ejection seen in low-mass models.

In addition to the r-process isotopes, proton-rich nuclei (p-nuclei) are also produced during CCSN explosions. 
Table \ref{tab:p-process} summarizes the yields and the fractional contribution of CCSN for some selected p-nuclei. The amount of lighter p-nuclei (such as $^{74}$Se and $^{84}$Sr) are significantly influenced by the explosion in all our simulations, while large variations of the CCSN contribution fractions are seen for heavier p-nuclei. More p-nuclei are produced when the model experiences neutron-rich phases, {which is a result of the $\alpha-$process.} 

In short, the fractional contribution of CCSNe to the production of heavier elements varies from 0\% to 100\% in the different models provided in this paper. This large variation is due to the intrinsic complexities of 3D CCSN explosions, such as the stochastic $Y_e$ evolution and the complicated ejecta histories. Clearly, more long-term 3D simulations are required to fully understand the production of heavy elements in CCSNe.

\begin{table}
    \centering
    \begin{tabular}{c|cccc}
    ZAMS Mass [$M_\odot$]   &$^{87}$Ru [$M_\odot$] &$^{88}$Sr [$M_\odot$]  &$^{89}$Y [$M_\odot$] &$^{90}$Zr [$M_\odot$]\\
    \hline
    9(a)    &5.2$\times10^{-5}$ (99.9\%)    &8.1$\times10^{-4}$ (100.0\%)   &7.4$\times10^{-5}$ (99.9\%)    &1.8$\times10^{-4}$ (99.9\%)\\
    9(b)    &4.5$\times10^{-8}$ (5.9\%)     &3.9$\times10^{-7}$ (1.4\%)     &9.7$\times10^{-8}$ (5.5\%)     &1.5$\times10^{-7}$ (25.1\%)\\
    9.25    &4.8$\times10^{-8}$ (3.5\%)     &4.1$\times10^{-7}$ (0.0\%)     &9.6$\times10^{-8}$ (0.2\%)     &1.3$\times10^{-7}$ (7.2\%)\\
    9.5     &6.1$\times10^{-8}$ (3.6\%)     &4.2$\times10^{-7}$ (0.2\%)     &1.0$\times10^{-7}$ (0.0\%)     &1.2$\times10^{-7}$ (0.7\%)\\
    11      &1.8$\times10^{-7}$ (56.3\%)    &5.8$\times10^{-6}$ (90.6\%)    &1.6$\times10^{-6}$ (92.1\%)    &1.1$\times10^{-5}$ (98.6\%)\\
    40      &1.0$\times10^{-5}$ (2.0\%)     &2.3$\times10^{-5}$ (--3.7\%)    &2.6$\times10^{-6}$ (--3.3\%)    &1.3$\times10^{-6}$ (--5.6\%)\\
    60      &2.4$\times10^{-6}$ (1.2\%)     &4.5$\times10^{-6}$ (10.8\%)    &9.3$\times10^{-7}$ (32.0\%)    &5.8$\times10^{-6}$ (76.2\%)\\
    \end{tabular}
    \caption{This table summarizes the yields of selected weak r-process nuclei. The percentage in brackets indicates the fraction that is contributed by nucleosynthesis the explosion. A negative percentage means that the element is destroyed in our CCSN simulations. This table shows that the CCSN contribution to the production of weak r-process elements can vary from $\sim$0\% to $\sim$100\% of the total ejected amount, depending on the details of the ejecta electron fraction. In \citet{wang2023}, we have shown that the electron fractions of neutrino-driven winds are a stochastic function of time. The same is true here, and a stochastic neutron-rich period is enough to contribute significantly to the production of weak r-process elements (e.g., see the 11 $M_\odot$ model).}
    \label{tab:r-process}
\end{table}

\begin{table}
    \centering
    \begin{tabular}{c|cccc}
    ZAMS Mass [$M_\odot$]   &$^{74}$Se [$M_\odot$] &$^{84}$Sr [$M_\odot$]  &$^{92}$Mo [$M_\odot$] &$^{94}$Mo\\ 
    \hline
    9(a)    &4.7$\times10^{-6}$ (99.8\%)    &8.0$\times10^{-8}$ (97.1\%)    &2.3$\times10^{-7}$ (96.6\%)    &1.8$\times10^{-8}$ (72.5\%)    \\
    9(b)    &1.9$\times10^{-8}$ (53.2\%)    &2.9$\times10^{-9}$ (22.2\%)    &7.8$\times10^{-9}$ (2.5\%)     &5.0$\times10^{-9}$ (0.2\%)     \\
    9.25    &2.2$\times10^{-8}$ (57.2\%)    &3.0$\times10^{-9}$ (22.1\%)    &8.0$\times10^{-9}$ (2.5\%)     &5.2$\times10^{-9}$ (--0.1\%)   \\
    9.5     &2.1$\times10^{-8}$ (54.1\%)    &3.2$\times10^{-9}$ (24.6\%)    &7.9$\times10^{-9}$ (0.1\%)     &5.3$\times10^{-9}$ (0.2\%)     \\
    11      &1.9$\times10^{-6}$ (99.2\%)    &4.7$\times10^{-8}$ (85.0\%)    &5.0$\times10^{-7}$ (98.2\%)    &7.3$\times10^{-9}$ (17.4\%)    \\
    40      &9.8$\times10^{-7}$ (30.1\%)    &3.6$\times10^{-8}$ (--32.2\%)   &3.5$\times10^{-8}$ (92.5\%)   &2.4$\times10^{-9}$ (48.6\%)    \\
    60      &9.0$\times10^{-7}$ (38.2\%)    &3.6$\times10^{-7}$ (--11.1\%)   &2.5$\times10^{-7}$ (99.9\%)   &5.8$\times10^{-10}$ (90.6\%)   \\
    \end{tabular}
    \caption{This table summarizes the yields of selected p-nuclei. The percentage in brackets shows the fraction that is contributed during CCSN explosions. A negative percentage means that the element is destroyed during the CCSN explosion. Similar to the weak r-process nuclei, the CCSN contribution to p-nuclei also varies between $\sim$0\% to $\sim$100\% of the total ejected amount. The p-nuclei are produced mostly by the photodisintegration of other heavy nuclei. Therefore, models with neutron-rich ejecta will also produce more p-nuclei.}
    \label{tab:p-process}
\end{table}

\subsection{Other Isotopes}
In this section, we briefly discuss the production of a few other interesting isotopes, {including the gamma-ray emitters $^{26}$Al and $^{60}$Fe \citep{diehl2021} and the extremely neutron-rich isotope $^{48}$Ca.} {The production of these isotopes has larger uncertainties because our calculations are either limited by the small 19-isotope pre-SN network, or are influenced by intrinsic 3D complexities.}

$^{26}$Al is a radioactive isotope which is interesting for gamma-ray observations. The net production of $^{26}$Al varies between $10^{-7}M_\odot$ and $10^{-5}M_\odot$ in our simulations. We see that the net production of $^{26}$Al in our CCSN simulations is dominated by the explosive nucleosynthesis component, while the freeze-out part contributes only a negligible amount. The net yield of $^{26}$Al also seems to have a strong correlation with the explosion energy. However, a more significant fraction of $^{26}$Al is contributed by the pre-CCSN nucleosynthesis phase. In the low-mass S16 models, the CCSN contribution to the final yield of $^{26}$Al is only around 10\%, probably because of their relatively lower explosion energies. The CCSN contribution in the 60 $M_\odot$ model is about 40\%. The 40$M_\odot$ model has a relative high explosion energy ($\sim$1.6 B), but our nucleosynthesis calculation is carried out only to $1.76$s due to the formation of a black hole \citep{burrows2023}, at which time $^{26}$Al is still vigorously being synthesized. Some S18 models (e.g., the 17, 18.5, and 19.56 $M_\odot$ models) produce significantly more $^{26}$Al than the S16 models, probably due to their higher explosion energies. However, the 19-isotope network doesn't cover $^{26}$Al so the pre-supernova envelope contribution of $^{26}$Al remains unknown for all the S18 models. The network differences for the two different progenitor suites may also lead to different explosive nucleosynthesis results, so $^{26}$Al production in the explosions of S18 models remains very uncertain.

$^{60}$Fe is another radioactive isotope useful for gamma-ray observations. The production of $^{60}$Fe depends strongly on the electron fraction of the ejecta. Only models with relatively longer neutron-rich phases are able to produce a significant amount of $^{60}$Fe, compared to those inherited from the progenitor phase. The exceptionally low-$Y_e$ 9(a) model produces $7\times10^{-5}M_\odot$ $^{60}$Fe, which is more than 95\% of the total ejected amount. The 11 $M_\odot$ model, which experiences a relative long stochastic neutron-rich phase, produces $4\times10^{-6}M_\odot$ $^{60}$Fe during the CCSN explosion, which is roughly 1/3 of the ejected amount. $^{60}$Fe production during the CCSN explosion in other simulations is negligible and the ejected amount is dominated by the pre-CCSN abundances. The total ejected mass of $^{60}$Fe in S16 models varies from $5\times10^{-7}M_\odot$ to $4\times10^{-4}M_\odot$. The S18 models are very uncertain due to the small network used in the pre-CCSN nucleosynthesis calculations.

$^{48}$Ca is a very neutron-rich isotope whose production sites are not very clear. In most of our simulations, $^{48}$Ca is not produced by CCSN explosions because they don't experience a  low enough $Y_e$. However, the 9(a) model produces $7\times10^{-5}M_\odot$ $^{48}$Ca, due to its exceptionally neutron-rich conditions. This preserves the possibility that $^{48}$Ca is (partly) produced by CCSN explosions.

\subsection{Morphology}\label{sec:morphology}
\begin{figure}
    \centering
    \includegraphics[width=0.48\textwidth]{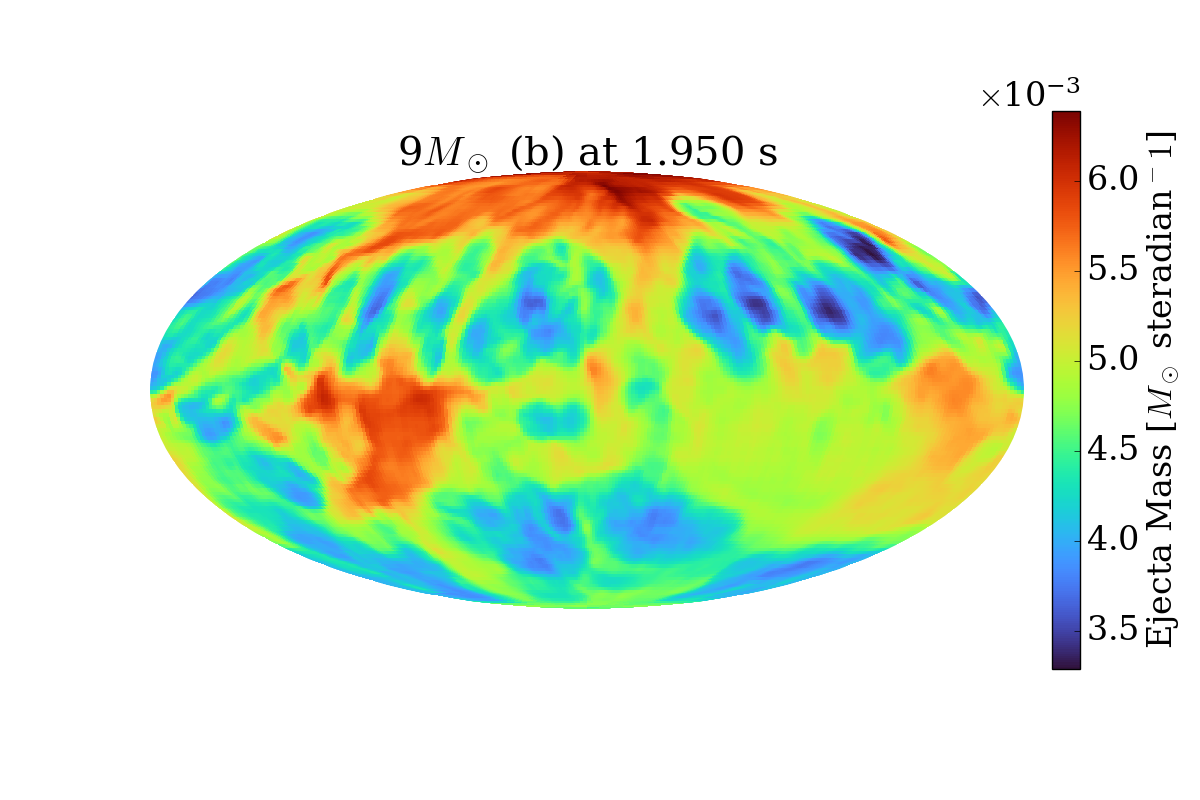}
    \includegraphics[width=0.48\textwidth]{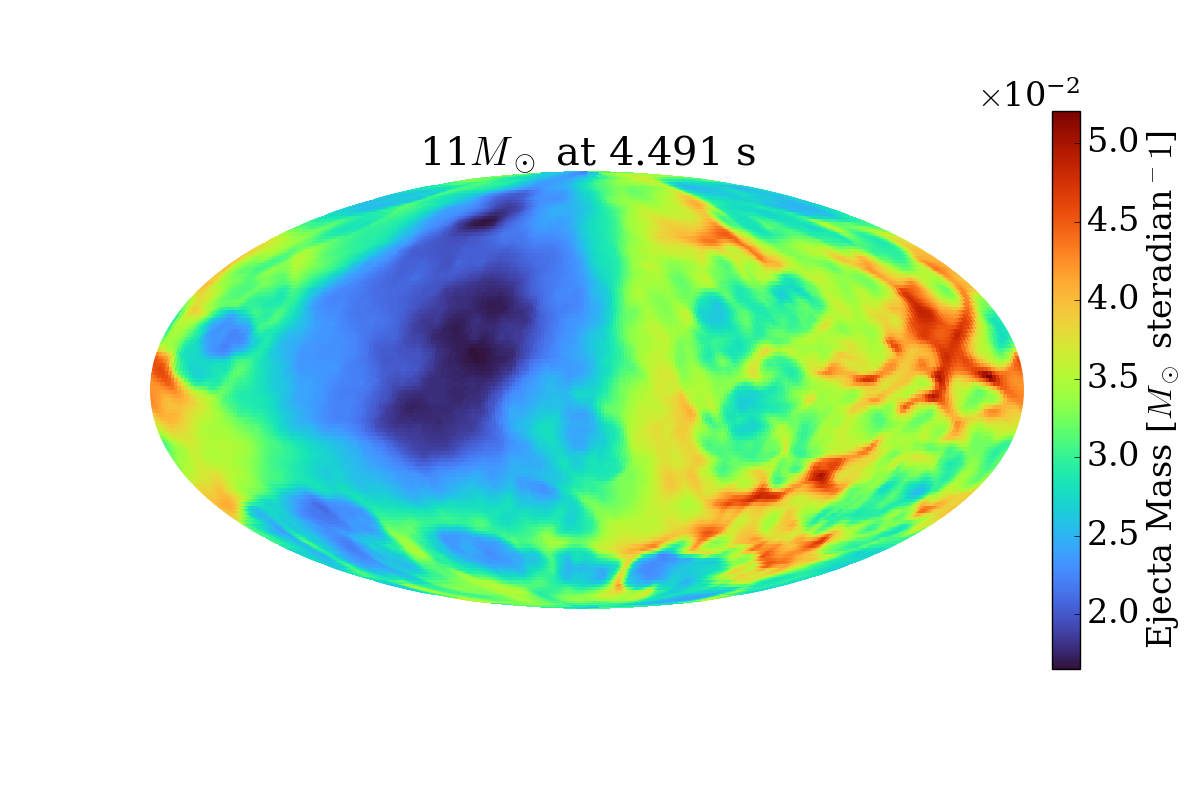}
    \includegraphics[width=0.48\textwidth]{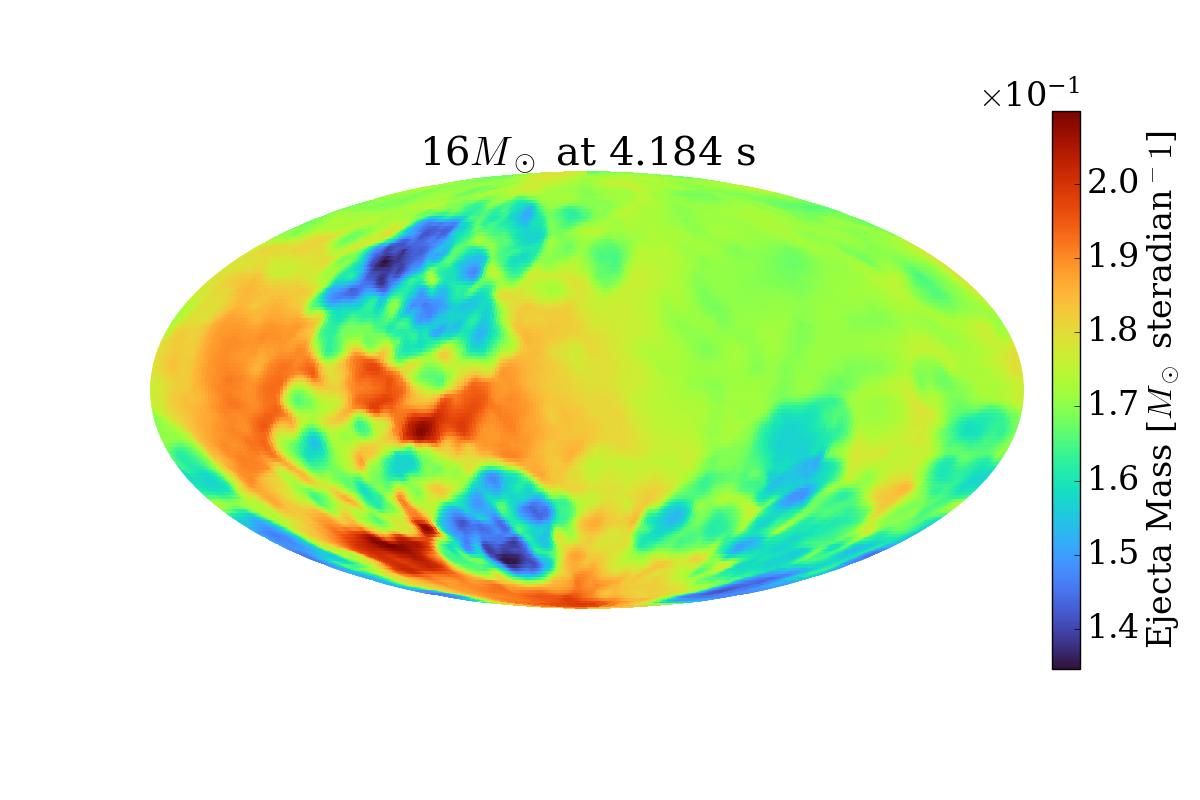}
    \includegraphics[width=0.48\textwidth]{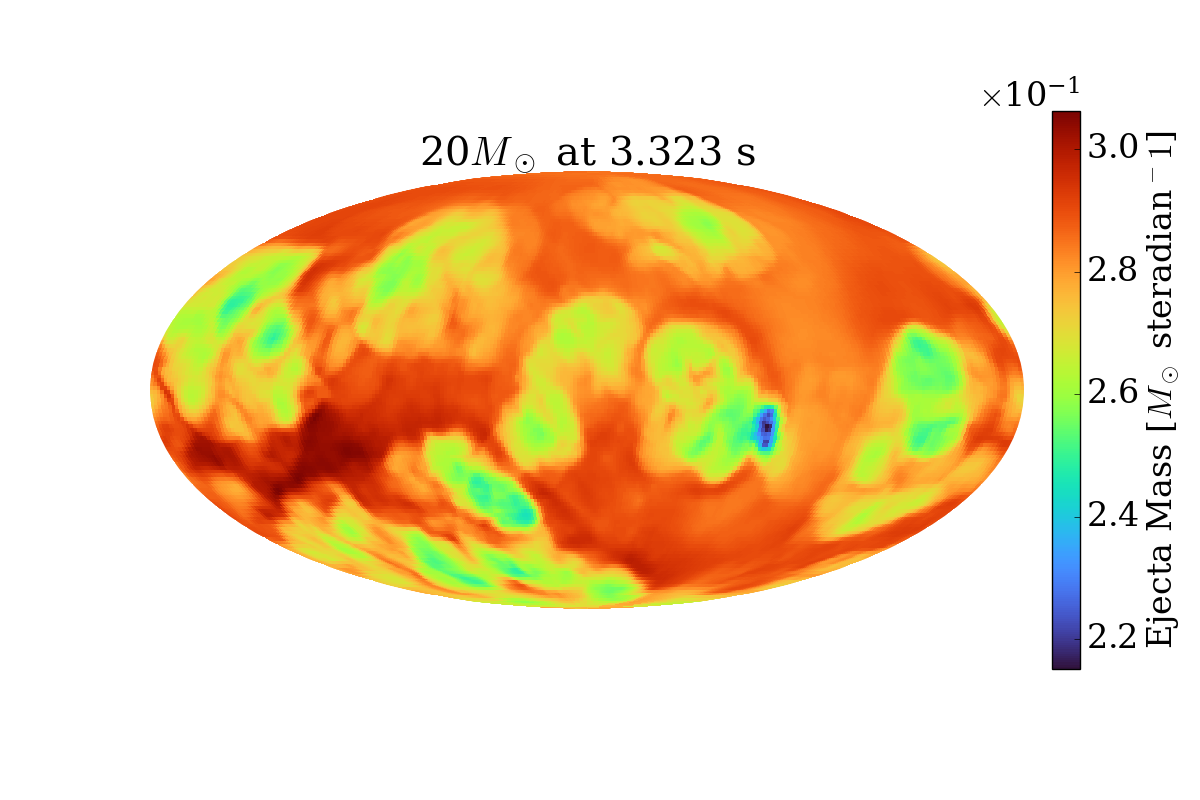}
    \includegraphics[width=0.48\textwidth]{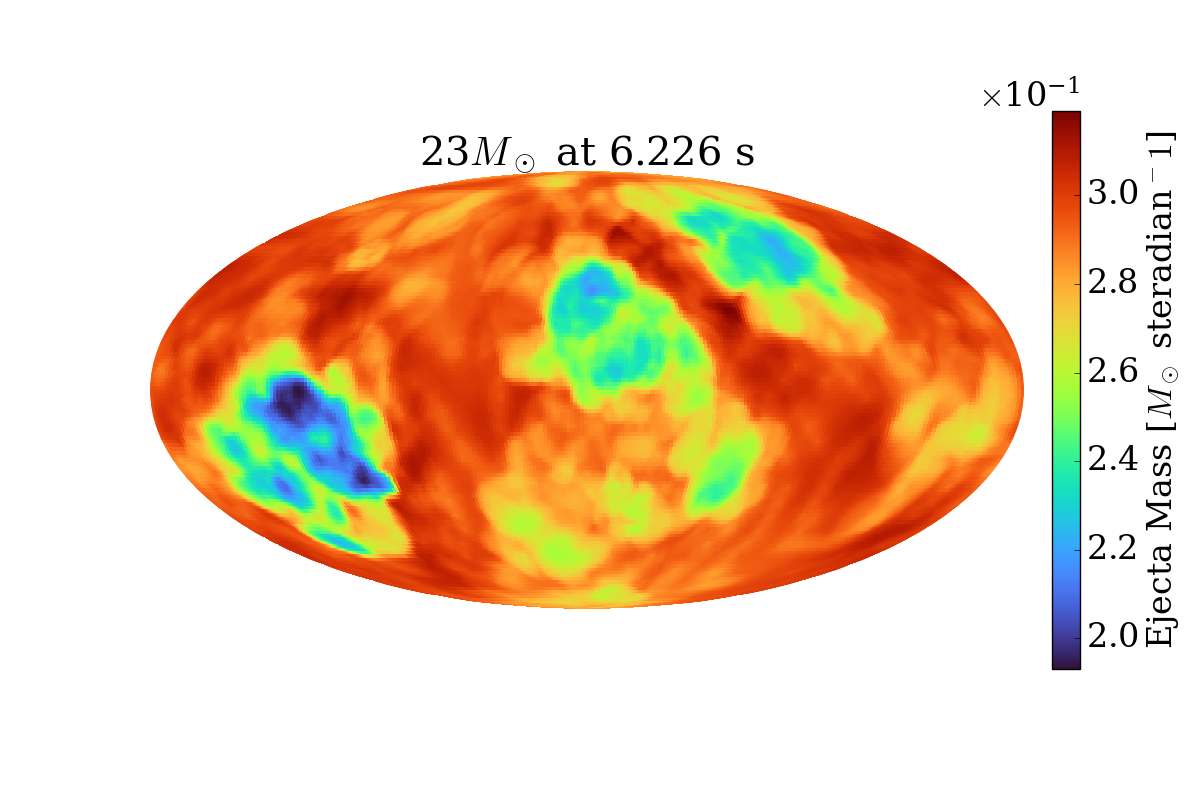}
    \includegraphics[width=0.48\textwidth]{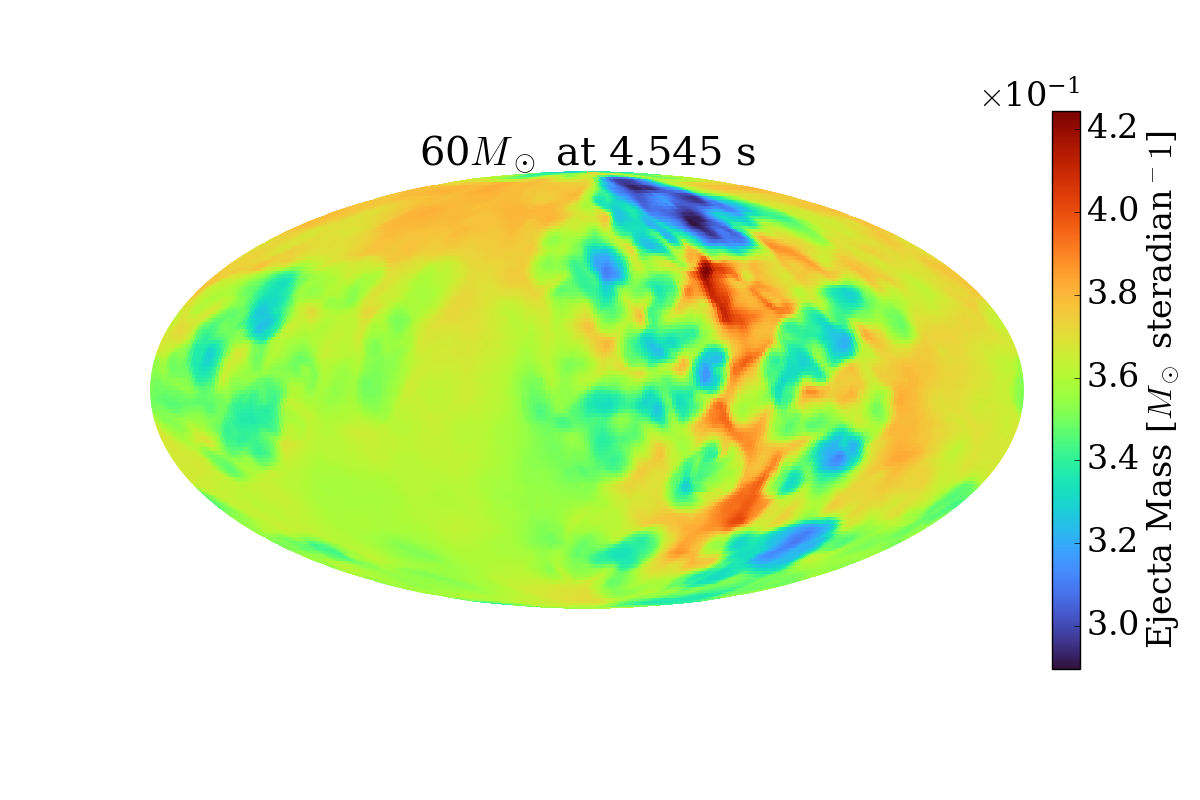}
    \caption{Angular mass distributions of unbound matter at the end of several selected simulations. The 9(b), 11, 16, 20, 23, and 60 $M_\odot$ models are shown here. The large amount of unshocked matter is not included by this plot. Strong dipoles can be seen in some models (e.g., the 11, 16, and 60 $M_\odot$ models), which means that the explosion is very asymmetric and the shock velocities along different directions vary significantly. Low mass holes can be seen on this plot as well. These holes are the bubbles formed by the neutrino-heated matter. These bubbles have lower density, higher entropy and higher velocity compared to the matter directly ejected by the shock. Most of the freeze-out component and part of the explosive component are contained in such bubbles.}
    \label{fig:ejecta-dist}
\end{figure}

\begin{figure}
    \centering
    \includegraphics[width=0.48\textwidth]{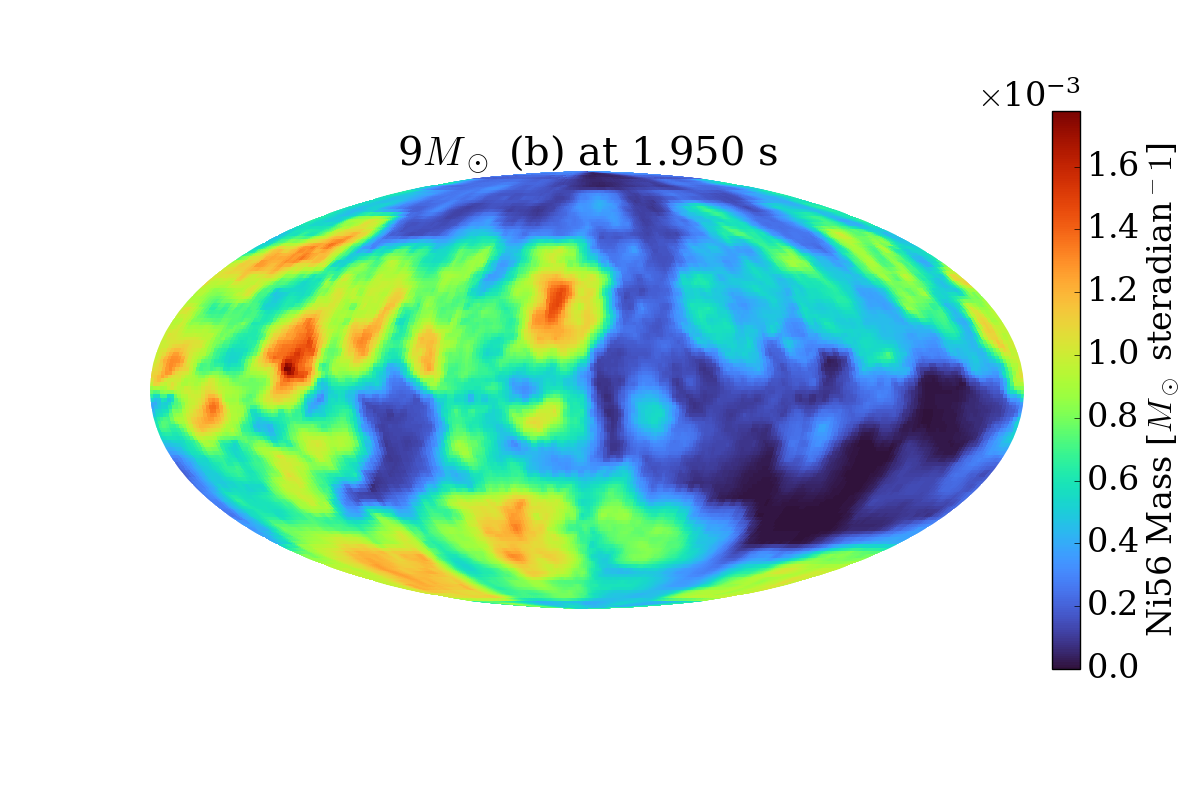}
    \includegraphics[width=0.48\textwidth]{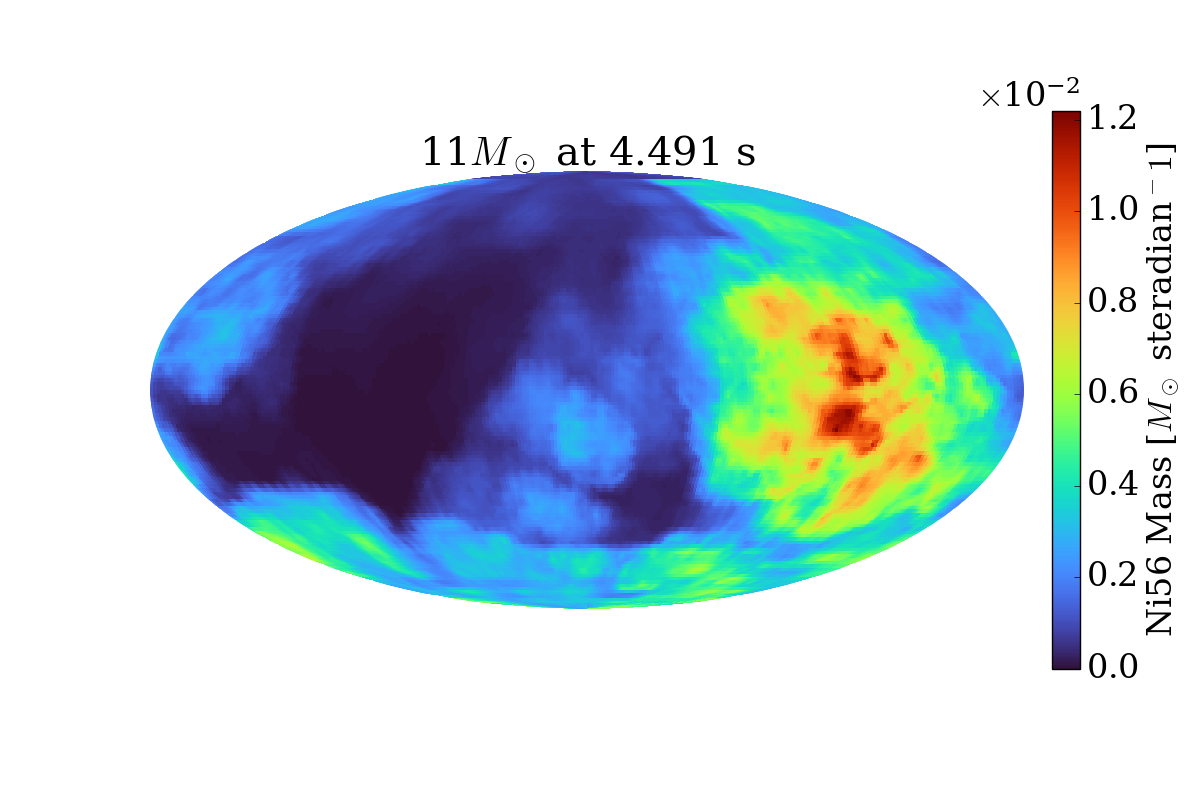}
    \includegraphics[width=0.48\textwidth]{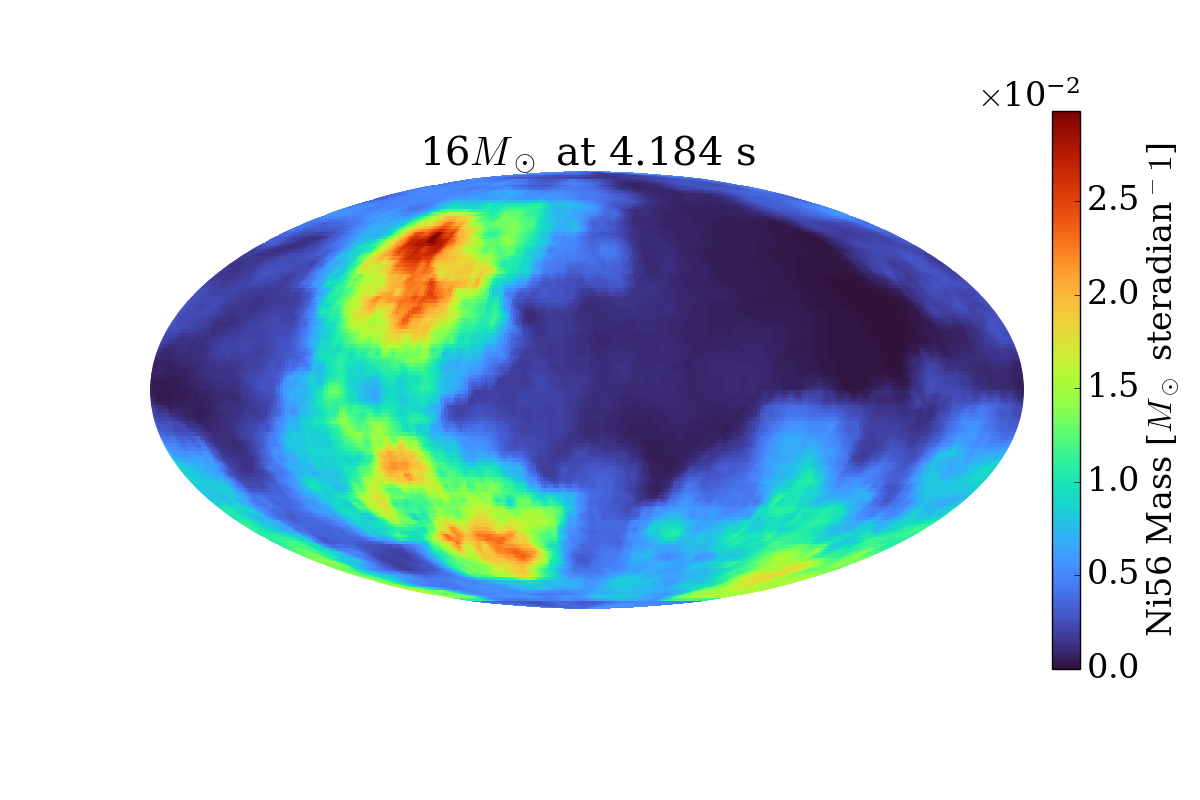}
    \includegraphics[width=0.48\textwidth]{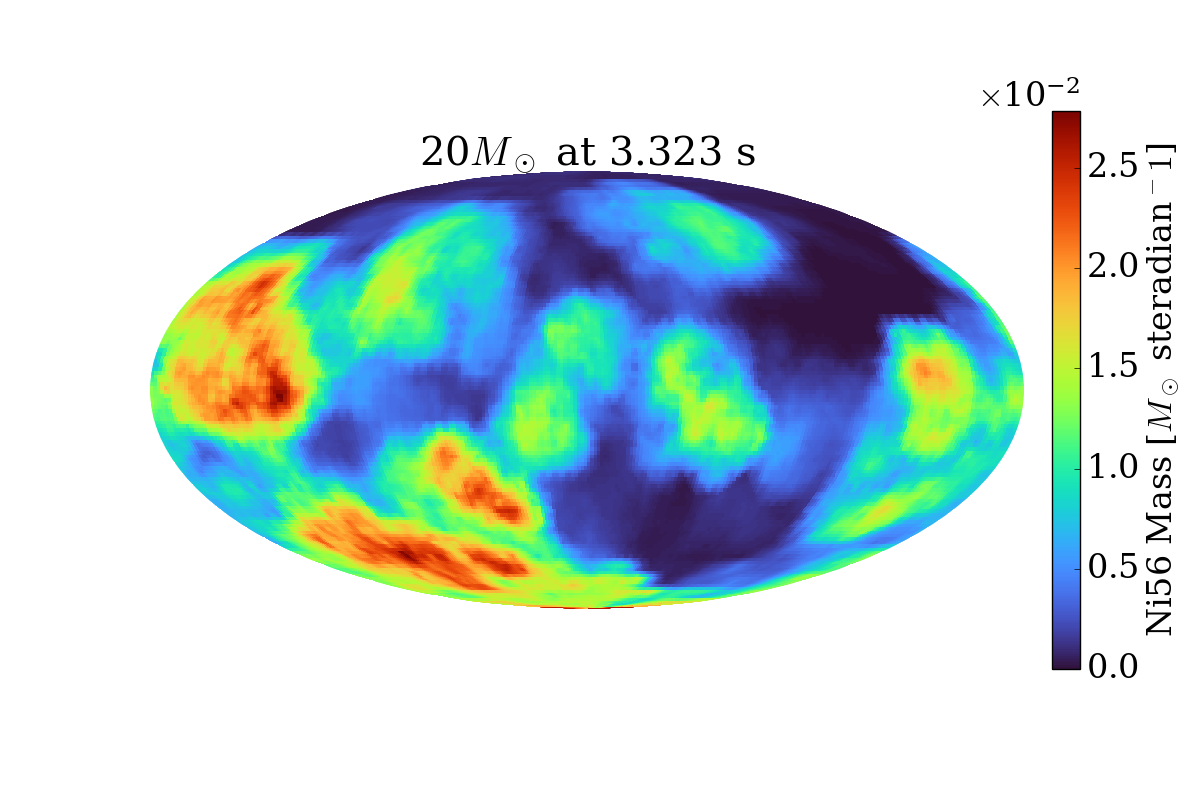}
    \includegraphics[width=0.48\textwidth]{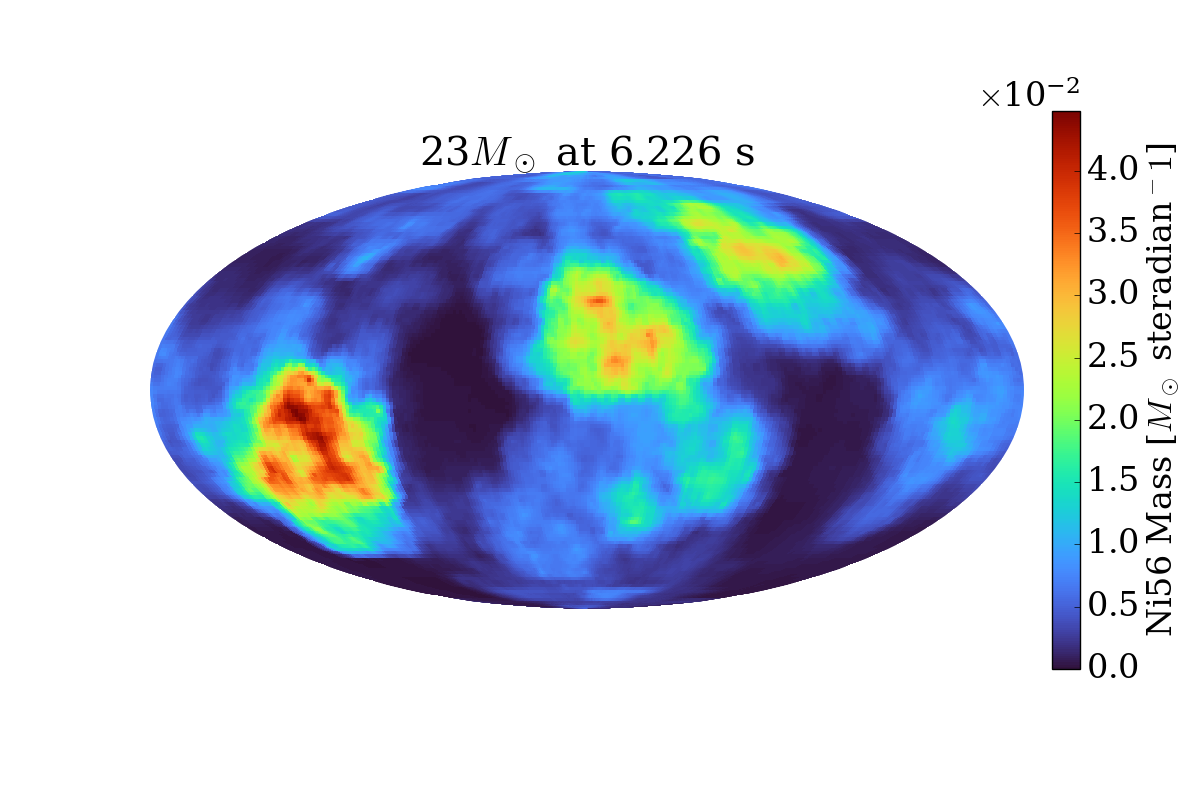}
    \includegraphics[width=0.48\textwidth]{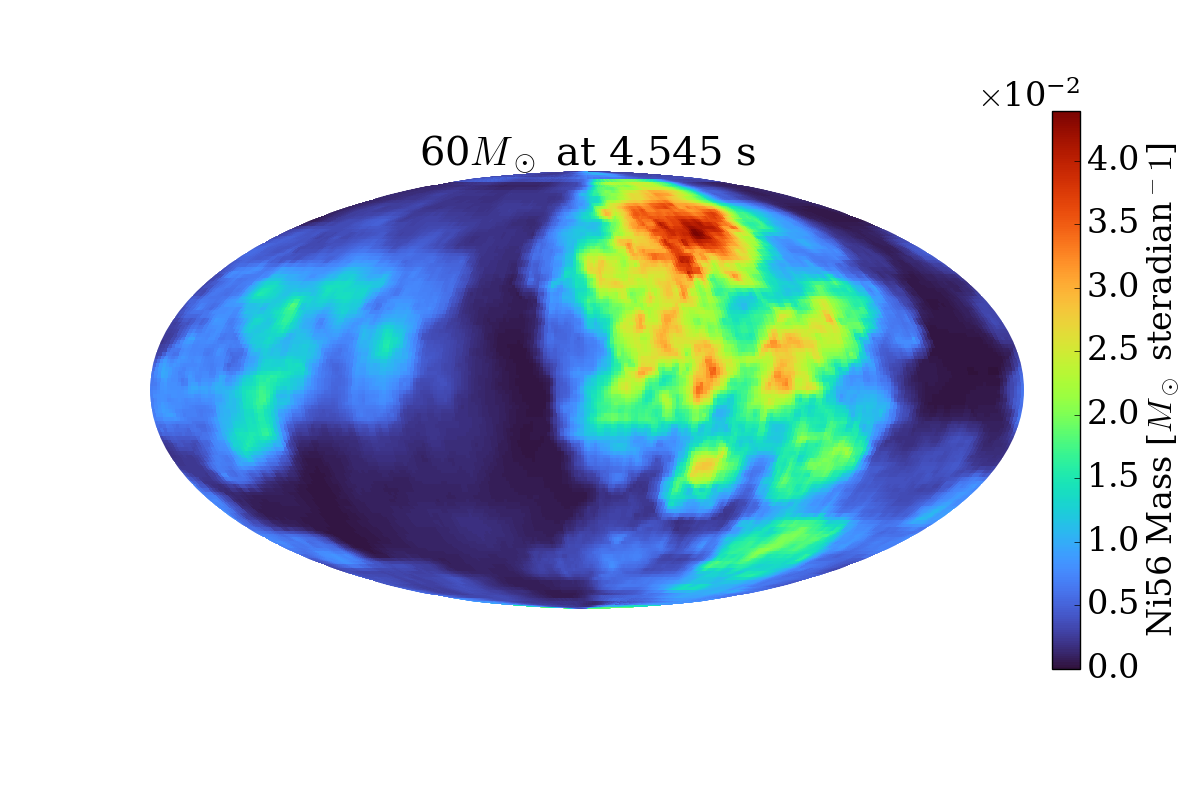}
    \caption{Same as Figure \ref{fig:ejecta-dist}, but with $^{56}$Ni angular distributions. It is clear that the nickel is mostly distributed in the bubbles, which are anti-correlated on relatively small scales with the distribution of the total unbound mass. The dipole moment is still aligned with the total mass distribution. The typical angular size of these bubbles is 15--30 degrees, and is still evolving with time due to interaction with the infalling matter.}
    \label{fig:ni56-dist}
\end{figure}

In addition to the abundances, it is also interesting to study how these elements are distributed spatially in CCSN explosions. Figure \ref{fig:ejecta-dist} shows the angular mass distribution of unbound matter at the end of the simulation in a few selected models. The large amount of unshocked matter is not included in these plots. Strong dipoles can be seen in some models (e.g., the 11, 16, and 60 $M_\odot$ models), which means that the explosion is very asymmetrical and the shock velocities along different directions vary significantly. Low mass holes can be seen on these plots as well. Such holes are the bubbles formed by the neutrino-heated matter. These bubbles have lower density, higher entropy and higher velocity compared to their surroundings. Most of the freeze-out component and part of the explosive component are contained by such bubbles. For example, Figure \ref{fig:ni56-dist} shows the angular distribution of $^{56}$Ni and it is clear that $^{56}$Ni is mostly distributed in these bubbles, which is anti-correlated with the total ejecta mass distribution on relatively small scales. The dipolar moment of $^{56}$Ni, however, has the same direction as that of the total ejecta mass, because they are both related to the explosion asymmetry. The typical angular size of these larger bubbles is  15$-$30 degrees, and is still evolving with time due to continuing interaction with the infalling matter.

Iron and $^{44}$Ti in the Cassiopeia A (Cas A) supernova remnant show very different spatial distributions \citep{hwang2004,grefenstette2014,grefenstette2017}. Since iron is thought to be mostly derived from $^{56}$Ni, this indicates different spatial distributions of $^{56}$Ni and $^{44}$Ti, which deviate from the predictions of traditional CCSN nucleosynthesis models. Previous CCSN simulations almost always produce $^{56}$Ni and $^{44}$Ti with similar spatial distributions, which is also seen in most of our 3D simulations. However, the 9(a) model shows very different $^{44}$Ti and $^{56}$Ni angular distributions, which leads to an interesting possibility that may be related to Cas A. Although $^{44}$Ti is formed mostly in the freeze-out component, $^{56}$Ni can be formed by both freeze-out and explosive nucleosynthesis. If the freeze-out component is more neutron-rich than those in most current CCSN simulations {for the first few hundred milliseconds}, it will suppress the formation of $^{56}$Ni, but is still able to produce some amount of $^{44}$Ti {at later times}. In this case, $^{44}$Ti follows the distribution of the freeze-out component, while $^{56}$Ni follows the distribution of the explosive nucleosynthesis component. This may lead to uncorrelated spatial distributions of the two isotopes. Because the $^{56}$Ni produced in the freeze-out component is suppressed, this possibility also leads to a high $^{44}$Ti to $^{56}$Ni ratio. However, it is unclear if it is normal to have a lot of neutron-rich ejecta in CCSN explosions. The 9(a) is a special model due to its earlier explosion and sensitivity to initial perturbations, and other models may not show similar behavior, even with initial perturbations. There is another factor that may influence the distribution of iron in supernova remnants. In the S16 progenitors, we see that there is $\sim10^{-2}M_\odot$ $^{56}$Fe in the outer envelopes, which is more than the $^{56}$Ni produced in our lowest mass models. If the outer envelope eventually shows a similar angular distribution to that of the unbound mass shown in Figure \ref{fig:ejecta-dist}, this $^{56}$Fe distribution will be anti-correlated with the bubbles which contain most of the $^{44}$Ti. 

It is worth noting that the morphologies we describe here are still evolving as the blast propagates. When the shock passes through interfaces in outer stellar envelopes, hydrodynamic instabilities will occur and significantly change the morphology of the ejecta. The structures seen at the end of our simulations will serve as the initial seeds of such instabilities. Therefore, morphological comparisons directly with the observed supernova remnants require our models to be continued by hydrodynamic simulations to at least the time of shock breakout. A detailed morphology study is beyond the scope of this paper, but we plan to do so in future work.

\section{Conclusion}\label{sec:conclusion}
The nucleosynthetic outcome is one of the most fundamental topics in core-collapse supernova  theory. In this paper, we presented a nucleosynthetic analysis of 18 long-term 3D CCSN simulations spanning a large progenitor ZAMS mass range from 9 to 60 $M_\odot$; this constitutes an unprecedentedly large 3D CCSN nucleosynthesis dataset. We study the complex ejecta conditions in 3D simulations and their consequences for nucleosynthesis, show the temporal evolution of a few important isotopes and the timescale of their production, calculate the elemental abundances of the ejecta and compare them with the pre-CCSN contribution, and discuss the spatial distributions of some species.

Multi-dimensional effects cause very complicated ejecta motions, which lead to large variations in the ejecta conditions. In our simulations, we see stochastic temporal evolution of ejecta electron fractions. Although most ejecta are proton-rich, there are some random neutron-rich phases in a few models. The mass of such neutron-rich ejecta is small, but it's enough to strongly influence the nucleosynthetic outcome. In addition, the peak temperature achieved by each layer in the progenitor model has a strong influence on the final nucleosynthetic abundances. A large range of peak temperatures is seen at smaller radii when the matter is able to enter the turbulent region above the PNS before its ejection. At larger radii, a power-law relation between peak temperature and radius is seen when the peak temperature drops below $\sim$2GK. In previous 1D models, a commonly-used assumption is that the explosion energy is deposited instantly, which leads approximately to a $T\propto R^{-3/4}$ relation. This $-3/4$ power-law is only seen in our 3D models with lower ZAMS masses. In some of our more massive models, the explosion energy is still increasing at several seconds after bounce, which continuously heats the blast and leads to a shallower slope in the power-law region. Another important factor in 3D nucleosynthetic calculations is the time ejecta spend in certain temperature ranges. We see that it is very common for ejecta to experience non-monotonic thermal histories, which leads to multiple transitions between local equilibria of nuclear reactions and non-equilibrium phases. The more massive models usually experience larger effects caused by such non-monotonic histories, as they generally have more manifest multi-dimensional effects. Stochastic $Y_e$ evolution, complex peak temperature distributions, and non-monotonic ejecta trajectories and thermal histories are some of the major complexities in the ejecta conditions in 3D. As a result, the nucleosynthesis results in our 3D simulations experience significantly larger variations between models than seen in \citet{sukhbold2016}. 

In our simulations, lighter $\alpha$-nuclei are mostly contributed by pre-explosion progenitor nucleosynthesis, while the CCSN contribution starts to be non-negligible for elements heavier than Si. There is a correlation between the net yields of $^{36}$Ar and $^{40}$Ca and the explosion energy, but the progenitor contribution still plays an important role in the their final abundances. X-ray analyses performed by \citet{braun2023} show that 1D nucleosynthesis models usually underpredict Ar and Ca abundances in supernova remnants. The same trend is seen in our 3D simulations, which may indicate that the discrepancy can't be solved by multi-dimensional effects in CCSN explosions. Further improvements in both progenitor models and CCSN simulations are likely required to solve this issue.

$^{56}$Ni is perhaps the most important isotope produced during CCSNe. Most of our simulations are able to reach the asymptotic stages of nickel production, and we are thus able to study its progenitor dependence. We find that although the nickel yield is not a monotonic function of progenitor ZAMS mass, there is a power-law relation between $^{56}$Ni and the compactness parameter which describes the structure of progenitor models. In our models, $^{56}$Ni is seen to be produced by both $\alpha$-rich freeze-out and explosive nucleosynthesis. Another important isotope is $^{44}$Ti, whose production is dominated by the freeze-out component in our simulations. We see that $^{44}$Ti is still being vigorously synthesized by the end of our simulations. The production of $^{44}$Ti is enhanced by the multiple transitions between local equilibria of nuclear reactions and non-equilibrium phases due to complex ejecta motions \citep{sieverding2023}, and the $^{44}$Ti/$^{56}$Ni ratio increases fast even at 6 seconds after bounce in some models. This means that 3D CCSN simulations carried out to $\sim$10 seconds after bounce are essential for an understanding of the production of $^{44}$Ti. We have also looked at another radioactive isotope $^{57}$Ni. Its production is sensitive to the stochastic $Y_e$ evolution, and the abundances in some models are still changing at simulation's end. This is another topic that calls for longer-term {($\sim$10 s)} 3D CCSN simulations.

Elements heavier than the iron group can be produced whenever the model experiences a neutron-rich phase. Although a strong r-process requiring extreme neutron-rich conditions does not seem possible in neutrino-driven CCSNe, the weak r-process is clearly seen and it produces many heavy elements up to the $^{90}$Zr peak. There are always more proton-rich nuclei when the neutron-rich condition is met. This is because the p-nuclei are mostly formed via photodisintegration of other heavy elements \citep{woosley1978}, whose abundances are enhanced by the weak r-process. The $\nu p$-process is another proposed process that may synthesize heavy elements. In general, we don't see significant effects of $\nu p$-process on the final abundances of the heavy elements. This is probably because the ejecta are on average not proton-rich enough and the entropies are on average too low compared to those required by the $\nu p$-process. High-$Y_e$ ejecta can reach $Y_e=0.65$ in some models (e.g., 9(b)), but these parcels are ejected later in the neutrino-driven wind phase with very high radial velocities and they don't have enough time for $\nu p$-process to occur. The neutrino luminosities are lower at later time, which may also suppress the $\nu p$-process. However, it is still too early to rule out the possibility of $\nu p$-process in CCSNe due to the following caveats: (a) The $Y_e$ evolution seems stochastic and the high $Y_e$ condition required by $\nu p$-process may not always occur as late as in the 9(b) case. (b) This comparison is done using only the S16 model set, which covers only a small range of progenitor ZAMS masses. More complex ejecta motions due to long-lasting, post-explosion accretion in the more massive models may increase the chances for interaction between neutrinos and high $Y_e$ ejecta, unlike in the case of fast ejection seen in low-mass models.

A few other interesting isotopes are also discussed in this paper. $^{26}$Al is a radioactive isotope which is interesting for gamma-ray astronomy. The net production of $^{26}$Al varies between $10^{-7}M_\odot$ and $10^{-5}M_\odot$ in different models. We see that the net production of $^{26}$Al in our CCSN simulations is dominated by the explosive nucleosynthetic component, while the freeze-out contribution provides only a negligible amount. The net yield of $^{26}$Al seems to be strongly correlated with the explosion energy. However, a more significant contribution of $^{26}$Al comes from the pre-CCSN nucleosynthesis phase. In some models, the pre-CCSN contribution is 10 times that produced during the explosion. But more energetic explosions may possibly produce more than half of the total $^{26}$Al. $^{60}$Fe is another radioactive isotope useful for gamma-ray astronomy. The production of $^{60}$Fe depends strongly on the electron fraction of the ejecta. Only models with relatively longer neutron-rich phases (e.g., the 9(a) and 11 $M_\odot$ models) are able to produce a significant amount of $^{60}$Fe compared to the component inherited from the progenitor. $^{48}$Ca is a very neutron-rich isotope whose production sites are not known. In most of our simulations, $^{48}$Ca is not produced by CCSN explosions because we don't achieve a low enough $Y_e$. However, the 9(a) model produces $7\times10^{-5}M_\odot$ of $^{48}$Ca, due to its exceptionally neutron-rich conditions. This preserves the possibility that $^{48}$Ca is (partly) produced in CCSN explosions.

We also briefly discussed the spatial distribution of the synthesized elements. Most $^{56}$Ni is located in the bubbles formed by the neutrino-heated matter, which is anti-correlated with the distribution of the total unbound mass. In most of our simulations, $^{44}$Ti is distributed similarly to $^{56}$Ni. However, in the most neutron-rich model (the 9(a) model), we see very different $^{44}$Ti and $^{56}$Ni distributions. Although $^{44}$Ti is formed mostly in the freeze-out component, $^{56}$Ni can be formed by both freeze-out and explosive nucleosynthesis. If the freeze-out component is more neutron-rich than those in most of our current 3D CCSN simulations (as in the 9(a) model) {for the first few hundred milliseconds}, it will suppress the formation of $^{56}$Ni, but will still be able to produce some $^{44}$Ti {at later times}. In this case, $^{44}$Ti follows the distribution of the freeze-out component, while $^{56}$Ni follows the distribution of the explosive nucleosynthesis component, which are uncorrelated. This possibility also leads to a high $^{44}$Ti to $^{56}$Ni ratio. However, it is unclear if it is normal to have a lot of neutron-rich ejecta in CCSN explosions. It is also worth noting that in the less massive models there is more $^{56}$Fe in the outer envelopes than the $^{56}$Ni produced. This can also influence the final inferred iron distribution.

This paper provides the largest and longest 3D CCSN simulation set for nucleosynthetic studies generated. With this set, we find that 3D CCSN models are much more complex than previous 1D models. Multi-dimensional effects are able to qualitatively change the elemental abundances of the CCSN ejecta, and significantly larger variations are seen in the 3D nucleosynthesis results. As a result, a larger number of longer-term {($\sim$10 s)} simulations are required to study the temporal evolution and model dependence in detail, and this paper is just the first step in this direction.

It is important to list some limitations of this work. First, our calculations are all based upon progenitors calculated by Kepler in \citet{sukhbold2016} and \citet{sukhbold2018}. It is unclear if these progenitors cover the same range of structures and properties as experienced in real stars. In addition, the S18 progenitors have only a 19-isotope network, so the behavior of models with ZAMS mass between 12 and 27 $M_\odot$ is unknown for most of the isotopes. Second, the nuclear reaction rates are all taken from the JINA Reaclib \citep{cyburt2010}, for which there are still large uncertainties in some rates (e.g., \citet{bliss2020}). Therefore, we make public not only our final nucleosynthesis results, but also the tracer trajectories, so that people can re-do the calculations with different reactions and rates. Third, although there are in general convective regions in progenitor stars, most of our simulations are not done with initial perturbations. This is due to the lack of knowledge of the convective and mixing asymmetries in progenitors in the last phases before core collapse. The perturbed 9(a) model uses an arbitrarily chosen spherical harmonic velocity field following \citet{muller2015}, but whether this captures the correct behavior of initial perturbations is unknown. 
We have done a simple 1D vs 3D progenitor comparison of the 18 $M_\odot$ model in this work, but many more three-dimensional progenitors are required for a more comprehensive study of the effects of initial perturbations.

\section*{Data Availability}
All the tracer trajectories (masses, positions, temperatures, $Y_e$, and local neutrino spectra), together with the final ejecta abundances of each model, can be downloaded at \url{https://doi.org/10.5281/zenodo.10498615}. The progenitor models from \citet{sukhbold2016} and \citet{sukhbold2018} used by this work will be available for download as well, including the full nucleosynthesis results in pre-CCSN phases of the S16 progenitors (Woosley, private communication). Other data from this set of 3D CCSN simulations can be made available upon reasonable request.

\section*{Acknowledgments}
We thank Stan Woosley for sharing the nucleosynthesis results of the co-processed adaptive network for the progenitors in \citet{sukhbold2016} and for allowing us to make them public. We also thank Matthew Coleman, David Vartanyan, Christopher White, Yongzhong Qian, Bronson Messer, Alexander Friedland, Amol Patwardhan, Alexander Heger, and Thomas Janka for their helpful insights and discussions on this paper.  We acknowledge support from the U.~S.\ Department of Energy Office of Science and the Office of Advanced Scientific Computing Research via the Scientific Discovery through Advanced Computing (SciDAC4) program and Grant DE-SC0018297 (subaward 00009650), support from the U.~S.\ National Science Foundation (NSF) under Grants AST-1714267 and PHY-1804048 (the latter via the Max-Planck/Princeton Center (MPPC) for Plasma Physics), and support from NASA under award JWST-GO-01947.011-A.  A generous award of computer time was provided by the INCITE program, using resources of the Argonne Leadership Computing Facility, a DOE Office of Science User Facility supported under Contract DE-AC02-06CH11357. We also acknowledge access to the Frontera cluster (under awards AST20020 and AST21003); this research is part of the Frontera computing project at the Texas Advanced Computing Center \citep{stanzione2020} under NSF award OAC-1818253. Finally, the authors acknowledge computational resources provided by the high-performance computer center at Princeton University, which is jointly supported by the Princeton Institute for Computational Science and Engineering (PICSciE) and the Princeton University Office of Information Technology, and our continuing allocation at the National Energy Research Scientific Computing Center (NERSC), which is supported by the Office of Science of the U.~S.\ Department of Energy under contract DE-AC03-76SF00098.

\bibliography{sample631}{}
\bibliographystyle{aasjournal}




\end{document}